\pdfoutput=1
\documentclass[prd,showpacs,letterpaper,10pt,aps,notitlepage,nofootinbib,superscriptaddress]{revtex4-1}

\usepackage{amsfonts,amsmath,graphicx}
\usepackage{slashed}
\usepackage{hyperref}
\usepackage{placeins}
\usepackage{mathrsfs}
\usepackage{rotating}

\hypersetup{colorlinks, linkcolor = [rgb]{0,0.0,0.75}, citecolor = [rgb]{0,0.0,0.75}, urlcolor = [rgb]{0,0.0,0.75}}

\newcommand{\nb}{\phantom{0}}
\newcommand{\wm}{\phantom{-}}

\begin{document}

\title{$\Lambda_b \to p \,\ell^-\, \bar{\nu}_\ell$ and $\Lambda_b \to \Lambda_c \,\ell^-\, \bar{\nu}_\ell$ form factors from lattice QCD \\ with relativistic heavy quarks}

\author{William Detmold}
\affiliation{Center for Theoretical Physics, Massachusetts Institute of Technology, Cambridge, MA 02139, USA}
\author{Christoph Lehner}
\affiliation{Physics Department, Brookhaven National Laboratory, Upton, NY 11973, USA}
\author{Stefan Meinel\:}
\email{smeinel@email.arizona.edu}
\affiliation{Department of Physics, University of Arizona, Tucson, AZ 85721, USA}
\affiliation{RIKEN BNL Research Center, Brookhaven National Laboratory, Upton, NY 11973, USA}

\begin{abstract}
Measurements of the $\Lambda_b \to p \,\ell^- \bar{\nu}_\ell$ and $\Lambda_b \to \Lambda_c \,\ell^- \bar{\nu}_\ell$ decay rates can
be used to determine the magnitudes of the CKM matrix elements $V_{ub}$ and $V_{cb}$, provided that the relevant hadronic form factors are known.
Here we present a precise calculation of these form factors using lattice QCD with 2+1 flavors of dynamical domain-wall fermions. The $b$ and $c$
quarks are implemented with relativistic heavy-quark actions, allowing us to work directly at the physical heavy-quark masses.
The lattice computation is performed for six different pion masses and two different lattice spacings, using gauge-field configurations
generated by the RBC and UKQCD collaborations. The $b \to u$ and $b \to c$ currents are renormalized with a mostly nonperturbative method.
We extrapolate the form factor results to the physical pion mass and the continuum limit, parametrizing the $q^2$-dependence
using $z$-expansions. The form factors are presented in such a way as to enable the correlated propagation of both statistical and systematic uncertainties
into derived quantities such as differential decay rates and asymmetries. Using these form factors, we present predictions for the
$\Lambda_b \to p \,\ell^- \bar{\nu}_\ell$ and $\Lambda_b \to \Lambda_c \,\ell^- \bar{\nu}_\ell$ differential and integrated decay rates.
Combined with experimental data, our results enable determinations of $|V_{ub}|$, $|V_{cb}|$, and $|V_{ub}/V_{cb}|$ with theory uncertainties
of $4.4\%$, $2.2\%$, and $4.9\%$, respectively.
\end{abstract}

\maketitle

\FloatBarrier
\section{Introduction}
\FloatBarrier

To date, all direct determinations of the CKM matrix element magnitudes $|V_{ub}|$ and $|V_{cb}|$ were performed using measurements of $B$ meson
semileptonic or leptonic decays at $e^+ e^-$ colliders. For both $|V_{ub}|$ and $|V_{cb}|$, there are tensions between the most precise extractions
from exclusive and inclusive semileptonic $B$ decays. The 2014 Review
of Particle Physics lists \cite{Agashe:2014kda}
\begin{equation}
 \begin{array}{ll} |V_{ub}|_{\rm excl.} = (3.28 \pm 0.29 )\times 10^{-3}, & |V_{cb}|_{\rm excl.} = (39.5\pm 0.8)\times 10^{-3}, \\[1ex]
 |V_{ub}|_{\rm incl.} = (4.41 \pm 0.15 ^{+0.15}_{-0.17} )\times 10^{-3}, \hspace{2ex} & |V_{cb}|_{\rm incl.} = (42.2 \pm 0.7 )\times 10^{-3}. \label{eq:VubVcbRPP2014}
 \end{array}
\end{equation}
The exclusive results in Eq.~(\ref{eq:VubVcbRPP2014}) are from the decays $B \to \pi \ell \bar{\nu}$ and $B \to D^* \ell \bar{\nu}$ (where $\ell=e,\mu$)
and use hadronic form factors from lattice QCD \cite{Bailey:2008wp, Bernard:2008dn}.
The discrepancy between the exclusive and inclusive results is a long-standing puzzle in flavor physics \cite{Kowalewski:2010zz, Mannel:2010zz, Ricciardi:2014aya}, and
right-handed currents beyond the Standard Model have been considered as a possible explanation \cite{Chen:2008se, Crivellin:2009sd, Buras:2010pz, Crivellin:2014zpa}.
New lattice QCD calculations of the $B \to \pi$ form factors published recently yield somewhat higher values of $|V_{ub}|_{\rm excl.}=(3.72\pm0.16)\times10^{-3}$
\cite{Lattice:2015tia} and $|V_{ub}|_{\rm excl.}=(3.61\pm0.32)\times10^{-3}$ \cite{Flynn:2015mha}, but the latest analysis of $B \to D^* \ell \bar{\nu}$ using
lattice QCD gives $|V_{cb}|_{\rm excl.}=(39.04\pm 0.75)$ \cite{Bailey:2014tva} and slightly increases the exclusive-inclusive tension. Moreover, the
current experimental results for the ratios of the $B\to D^{(*)} \tau \bar{\nu}$ and $B\to D^{(*)} \ell \bar{\nu}$ ($\ell=e,\mu$) branching fractions
differ from the Standard-Model expectation with a combined significance of 3.4$\sigma$ \cite{Lees:2012xj}.

On the experimental front, new results are expected from the future Belle II detector at the SuperKEKB $e^+e^-$ collider, and in the near future also from
LHCb at the Large Hadron Collider. The LHCb Collaboration is currently analyzing the ratio of branching fractions of the baryonic $b\to u$ and $b \to c$
decays $\Lambda_b \to p\, \mu^-\bar{\nu}_\mu$ and $\Lambda_b \to \Lambda_c\, \mu^-\bar{\nu}_\mu$, with the aim of determining $|V_{ub}/V_{cb}|$ for the first
time at a hadron collider. These decays were chosen over the more conventional $B \to \pi \mu \bar{\nu}$ and $B \to D \mu \bar{\nu}$ decays because, with the
LHCb detector, final states containing protons are easier to identify than final states with pions \cite{Adinolfi:2012qfa}. Note that the production rate of
$\Lambda_b$ baryons at the LHC is remarkably high, equal to approximately $1/2$ times the production rate of $\bar{B}^0$ mesons \cite{Aaij:2014jyk}.
 
The extraction of $|V_{ub}|$ and $|V_{cb}|$ (or their ratio) from the measured $\Lambda_b \to p\, \mu^-\bar{\nu}_\mu$ and $\Lambda_b \to \Lambda_c\, \mu^-\bar{\nu}_\mu$
branching fractions requires knowledge of the form factors describing the $\Lambda_b \to p$ and $\Lambda_b \to \Lambda_c$ matrix elements
of the relevant $b\to u$ and $b \to c$ currents in the weak effective Hamiltonian. These form factors have been studied using
sum rules and quark models \cite{Cardarelli:1997sx, Dosch:1997zx, Huang:1998rq, Carvalho:1999ia, Huang:2004vf, Pervin:2005ve, Ke:2007tg, Wang:2009hra, Azizi:2009wn, Khodjamirian:2011jp, Gutsche:2014zna, Gutsche:2015mxa}.
Nonperturbative QCD calculations of the $\Lambda_b \to p$ and $\Lambda_b \to \Lambda_c$ form factors can be performed using lattice gauge theory.
The first lattice QCD calculation of $\Lambda_b \to p$ form factors, published in Ref.~\cite{Detmold:2013nia}, employed static $b$ quarks (i.e., 
leading-order heavy-quark effective theory) to simplify the analysis. The static limit reduces the number of independent $\Lambda_b \to p$ form factors
to two \cite{Mannel:1990vg, Hussain:1990uu, Hussain:1992rb}, but introduces systematic uncertainties of order $\Lambda_{\rm QCD}/m_b$
and $|\mathbf{p}^\prime| /m_b$ in the  $\Lambda_b \to p\, \mu^-\bar{\nu}_\mu$ differential decay rate (where $\mathbf{p}^\prime$
is the momentum of the proton in the $\Lambda_b$ rest frame). Here we present a new lattice calculation which improves upon Ref.~\cite{Detmold:2013nia}
by replacing the static $b$ quarks by relativistic $b$ quarks, eliminating this systematic uncertainty. In addition to the six form factors describing
the hadronic part of the decay $\Lambda_b \to p\, \mu^-\bar{\nu}_\mu$ in fully relativistic QCD, we also compute the six analogous form factors for
$\Lambda_b \to \Lambda_c\, \mu^-\bar{\nu}_\mu$ (note that early lattice studies of $\Lambda_b \to \Lambda_c$ form factors in the quenched approximation
can be found in Refs.~\cite{Bowler:1997ej, Gottlieb:2003yb}). Preliminary results from the present work were shown in Ref.~\cite{Meinel:2014wua}.

In Sec.~\ref{sec:FFdefinitions} we provide the definitions of the form factors employed here. The lattice actions and parameters, as well as the matching
of the $b\to u$ and $b \to c$ currents from the lattice renormalization scheme to the continuum $\overline{\rm MS}$ scheme are discussed in Sec.~\ref{sec:actions}. This calculation
is based on the same lattice gauge-field ensembles as Ref.~\cite{Detmold:2013nia}; the ensembles include 2+1 flavor of dynamical domain-wall fermions
and were generated by the RBC and UKQCD Collaborations \cite{Aoki:2010dy}. Section \ref{sec:ratios} explains our method for extracting
the form factors from ratios of three-point and two-point correlation functions and removing excited-state contamination by extrapolating to infinite source-sink separation.
Our fits of the quark-mass, lattice-spacing, and momentum-dependence of the form factors are discussed in Sec.~\ref{sec:ccextrap}. The form factors in the physical
limit are presented in terms of $z$-expansion \cite{Bourrely:2008za} parameters and their correlation matrices. Two different sets of parameters, referred to
as the ``nominal parameters'' and the ``higher-order parameters'' are given. The nominal parameters are used to obtain the central values and statistical
uncertainties of the form factors (and of derived quantities), while the higher-order parameters are used to calculate systematic uncertainties.
In Sec.~\ref{sec:dGamma} we then present predictions for the $\Lambda_b \to p \,\ell^- \bar{\nu}_\ell$ and $\Lambda_b \to \Lambda_c \,\ell^- \bar{\nu}_\ell$
differential and integrated decay rates using our form factors. Combined with experimental data, our results for the $\Lambda_b \to p \,\mu\, \bar{\nu}_\mu$
and $\Lambda_b \to \Lambda_c \,\mu\, \bar{\nu}_\mu$ decay rates in the high-$q^2$ region will allow determinations of $|V_{ub}|$ and $|V_{cb}|$ with theory
uncertainties of $4.4\%$ and $2.2\%$, respectively.

\FloatBarrier
\section{\label{sec:FFdefinitions}Definitions of the form factors}
\FloatBarrier

Allowing for possible right-handed currents beyond the Standard Model, the effective weak Hamiltonian for $b \to q \:\ell^-\, \bar{\nu}_\ell$
transitions (where $q=u,c$) can be written
as
\begin{equation}
\mathcal{H}_{\rm eff} = \frac{G_F}{\sqrt{2}}V_{qb}^L\: \left[ (1+\epsilon_q^{R})\bar{q} \gamma^\mu b - (1-\epsilon_q^{R})\:\bar{q} \gamma^\mu \gamma_5 b \right]\: \bar{\ell} \gamma_\mu (1-\gamma_5) \nu \label{eq:Heff}
\end{equation}
(in the Standard Model, $\epsilon_q^{R}=0$ and $V_{qb}^L=V_{qb}$). To calculate the differential decay rate and other observables, we therefore need
the hadronic matrix elements of the vector and axial vector currents, $\bar{q} \gamma^\mu b$ and $\bar{q} \gamma^\mu \gamma_5 b$.
In the following, we denote the final-state baryon by $X$ ($X=p,\Lambda_c$). Lorentz and discrete symmetries imply that the matrix elements
$\langle X | \overline{q} \,\gamma^\mu\, b | \Lambda_b \rangle$ and $\langle X | \overline{q} \,\gamma^\mu\gamma_5\, b | \Lambda_b \rangle$ can each be decomposed
into three form factors. In this work we primarily use a helicity-based definition of the $\Lambda_b \to X$
form factors, which was introduced in Ref.~\cite{Feldmann:2011xf} and is given by
\begin{eqnarray}
 \nonumber \langle X(p^\prime,s^\prime) | \overline{q} \,\gamma^\mu\, b | \Lambda_b(p,s) \rangle &=&
 \overline{u}_X(p^\prime,s^\prime) \bigg[ f_0(q^2)\: (m_{\Lambda_b}-m_X)\frac{q^\mu}{q^2} \\
 \nonumber && \phantom{\overline{u}_X \bigg[}+ f_+(q^2) \frac{m_{\Lambda_b}+m_X}{s_+}\left( p^\mu + p^{\prime \mu} - (m_{\Lambda_b}^2-m_X^2)\frac{q^\mu}{q^2}  \right) \\
 && \phantom{\overline{u}_X \bigg[}+ f_\perp(q^2) \left(\gamma^\mu - \frac{2m_X}{s_+} p^\mu - \frac{2 m_{\Lambda_b}}{s_+} p^{\prime \mu} \right) \bigg] u_{\Lambda_b}(p,s), \\
 \nonumber \langle X(p^\prime,s^\prime) | \overline{q} \,\gamma^\mu\gamma_5\, b | \Lambda_b(p,s) \rangle &=&
 -\overline{u}_X(p^\prime,s^\prime) \:\gamma_5 \bigg[ g_0(q^2)\: (m_{\Lambda_b}+m_X)\frac{q^\mu}{q^2} \\
 \nonumber && \phantom{\overline{u}_X \:\gamma_5\bigg[}+ g_+(q^2)\frac{m_{\Lambda_b}-m_X}{s_-}\left( p^\mu + p^{\prime \mu} - (m_{\Lambda_b}^2-m_X^2)\frac{q^\mu}{q^2}  \right) \\
 && \phantom{\overline{u}_X\:\gamma_5 \bigg[}+ g_\perp(q^2) \left(\gamma^\mu + \frac{2m_X}{s_-} p^\mu - \frac{2 m_{\Lambda_b}}{s_-} p^{\prime \mu} \right) \bigg]  u_{\Lambda_b}(p,s).
\end{eqnarray}
In these expressions, $q=p-p^\prime$ is the four-momentum transfer (whereas $\bar{q}$ is the $\bar{u}$ or $\bar{c}$ quark field),
and $s_\pm$ is defined as
\begin{equation}
 s_\pm =(m_{\Lambda_b} \pm m_X)^2-q^2.
\end{equation}
The form factors with subscripts $0$, $+$, $\perp$ describe the contractions of the above matrix elements with virtual polarization vectors $\epsilon^*_\mu$ that are, respectively,
time-like, longitudinal, and transverse to $q^\mu$. Consequently, this choice of form factors leads to particularly simple expressions for observables such as the differential
decay rate. Moreover, this choice simplifies the extraction of the form factors from correlation functions and clarifies the spin-parity quantum numbers of poles outside the physical kinematic region
$0 \leq q^2 \leq (m_{\Lambda_b}-m_X)^2$.

An alternate definition of the form factors that can be found in the literature (see, e.g., Ref.~\cite{Gutsche:2014zna}) is the following:
\begin{eqnarray}
 \langle X(p^\prime,s^\prime) | \overline{q} \,\gamma^\mu\, b | \Lambda_b(p) \rangle &=& \overline{u}_X(p^\prime,s^\prime) \left[ f_1^V(q^2)\: \gamma^\mu - \frac{f_2^V(q^2)}{m_{\Lambda_b}} i\sigma^{\mu\nu}q_\nu + \frac{f_3^V(q^2)}{m_{\Lambda_b}} q^\mu \right] u_{\Lambda_b}(p,s),  \label{eq:WeinbergFF1} \\
 \langle X(p^\prime,s^\prime) | \overline{q} \,\gamma^\mu\gamma_5\, b | \Lambda_b(p) \rangle &=& \overline{u}_X(p^\prime,s^\prime) \left[ f_1^A(q^2)\: \gamma^\mu - \frac{f_2^A(q^2)}{m_{\Lambda_b}} i\sigma^{\mu\nu}q_\nu + \frac{f_3^A(q^2)}{m_{\Lambda_b}} q^\mu \right]\gamma_5\: u_{\Lambda_b}(p,s),  \label{eq:WeinbergFF2}
\end{eqnarray}
where $\sigma^{\mu\nu}=\frac{i}{2}(\gamma^\mu\gamma^\nu-\gamma^\nu\gamma^\mu)$ and, as before, $q=p-p^\prime$. This choice decomposes the matrix elements
into form factors of the first and second class according to Weinberg's classification \cite{Weinberg:1958ut}. The second-class form factors $f_3^V$ and $f_2^A$ would
vanish in the limit $m_b=m_c$ (for $\Lambda_b \to \Lambda_c$) or $m_b=m_u$ (for $\Lambda_b \to p$) \cite{Sasaki:2008ha}. In the following,
we will refer to the form factors defined in Eqs.~(\ref{eq:WeinbergFF1}), (\ref{eq:WeinbergFF2}) as ``Weinberg form factors''. The helicity form factors are related to the Weinberg form factors as follows:
\begin{eqnarray}
 f_+(q^2)     &=& f_1^V(q^2) + \frac{q^2}{m_{\Lambda_b}(m_{\Lambda_b}+m_X)} f_2^V(q^2), \label{eq:FFR1} \\
 f_\perp(q^2) &=& f_1^V(q^2) + \frac{m_{\Lambda_b}+m_X}{m_{\Lambda_b}} f_2^V(q^2),  \\
 f_0(q^2)     &=& f_1^V(q^2) + \frac{q^2}{m_{\Lambda_b}(m_{\Lambda_b}-m_X)} f_3^V(q^2), \\
 g_+(q^2)     &=& f_1^A(q^2) - \frac{q^2}{m_{\Lambda_b}(m_{\Lambda_b}-m_X)} f_2^A(q^2), \\
 g_\perp(q^2) &=& f_1^A(q^2) - \frac{m_{\Lambda_b}-m_X}{m_{\Lambda_b}} f_2^A(q^2), \\
 g_0(q^2)     &=& f_1^A(q^2) - \frac{q^2}{m_{\Lambda_b}(m_{\Lambda_b}+m_X)} f_3^A(q^2). \label{eq:FFR6}
\end{eqnarray}
These relations also demonstrate the following endpoint constraints for the helicity form factors:
\begin{eqnarray}
 f_0(0) &=& f_+(0), \label{eq:FFC1} \\
 g_0(0) &=& g_+(0), \label{eq:FFC2} \\
 g_\perp(q^2_{\rm max}) &=& g_+(q^2_{\rm max}), \label{eq:FFC3}
\end{eqnarray}
where $q^2_{\rm max}=(m_{\Lambda_b}-m_X)^2$. At intermediate stages of our analysis of the lattice QCD data, it is beneficial to work
with both definitions of the form factors. However, we perform the chiral/continuum/kinematic extrapolations only in the helicity basis.

\FloatBarrier
\section{\label{sec:actions}Lattice actions and currents}
\FloatBarrier

This calculation is based on lattice gauge field ensembles generated by the RBC and UKQCD collaborations \cite{Aoki:2010dy}
with the Iwasaki gauge action \cite{Iwasaki:1983ck, Iwasaki:1984cj} and 2+1 flavors of dynamical domain-wall fermions \cite{Kaplan:1992bt, Furman:1994ky, Shamir:1993zy}.
We implement the light ($u$ or $d$) valence quarks with the same domain-wall action that was used in generating the ensembles. Our analysis uses six
different combinations of light-quark masses and lattice spacings as shown in Table \ref{tab:params}. These parameters are identical to those used in the earlier
calculation of $\Lambda_b \to p \ell\bar{\nu}$ form factors in Ref.~\cite{Detmold:2013nia}. However, instead of the static Eichten-Hill action \cite{Eichten:1989kb}
employed in Ref.~\cite{Detmold:2013nia}, we now use anisotropic clover actions for the heavy ($c$ and $b$) quarks \cite{ElKhadra:1996mp, Aoki:2001ra, Aoki:2003dg, Lin:2006ur}.
These actions have the form
\begin{equation}
S_Q = a^4 \sum_x \bar{Q} \left[ m_Q + \gamma_0 \nabla_0 - \frac{a}{2} \nabla^{(2)}_0 + \nu\sum_{i=1}^3\left(\gamma_i \nabla_i - \frac{a}{2} \nabla^{(2)}_i\right)
- c_E  \frac{a}{2} \sum_{i=1}^3 \sigma_{0i}F_{0i} - c_B \frac{a}{4} \sum_{i,\, j=1}^3 \sigma_{ij}F_{ij}  \right] Q \, , \label{eq:RHQ}
\end{equation}
where $Q$ is the lattice charm or bottom quark field, $\nabla_\mu$ and $\nabla_\mu^{(2)}$ are first- and second-order covariant lattice derivatives,
and $F_{\mu\nu}$ is a lattice expression for the field-strength tensor (all of which are defined as in Ref.~\cite{Aoki:2012xaa}).
By suitably tuning the parameters $\nu$, $c_E$, $c_B$ as functions of $a m_Q$, heavy-quark discretization errors proportional to powers of $a m_Q$ can be removed
to all orders. The remaining discretization errors are of order $a^2|\mathbf{p}|^2$, where $|\mathbf{p}|$ is
the typical magnitude of the spatial momentum of the heavy quark inside the hadron. As the continuum limit $a \to 0$ is approached, the values $\nu=1$ and $c_E=c_B=c_{\rm SW}$
corresponding to the standard clover-improved Wilson action are recovered. For the bottom quark, we use the parameters that were tuned nonperturbatively
by the RBC and UKQCD collaborations \cite{Aoki:2012xaa} using the condition that the action reproduces the correct spin-averaged $B_s$ meson mass and relativistic dispersion relation,
as well as the correct $B_s^*-B_s$ hyperfine splitting. For the charm quarks, we use the parameters from Ref.~\cite{Brown:2014ena}, where $a m_Q$ and $\nu$ were tuned nonperturbatively
to obtain the correct spin-averaged charmonium mass and relativistic dispersion relation, while $c_E$ and $c_B$ were set to mean-field improved tree-level predictions. Note that
after the parameters were tuned in this way, the calculated charmonium hyperfine splittings were also in agreement with experiment \cite{Brown:2014ena}. The values of all
heavy-quark action parameters used here are given in Table \ref{tab:HQparams}.

\begin{table}
\begin{tabular}{cccccccccccccccccccc}
\hline\hline
Set & \hspace{1ex} & $\beta$ & \hspace{1ex} & $N_s^3\times N_t\times N_5$ & \hspace{1ex} & $a m_5$ & \hspace{1ex} & $am_{s}^{(\mathrm{sea})}$
& \hspace{1ex} & $am_{u,d}^{(\mathrm{sea})}$   & \hspace{1ex} & $a$ (fm) & \hspace{1ex} & $am_{u,d}^{(\mathrm{val})}$ 
& \hspace{1ex} & $m_\pi^{(\mathrm{val})}$ (MeV) & \hspace{1ex} & $N_{\rm meas}$ \\
\hline
\texttt{C14} && $2.13$ && $24^3\times64\times16$ && $1.8$ && $0.04$ && $0.005$ && $0.1119(17)$ && $0.001$ && 245(4) && 2672 \\
\texttt{C24} && $2.13$ && $24^3\times64\times16$ && $1.8$ && $0.04$ && $0.005$ && $0.1119(17)$ && $0.002$ && 270(4) && 2676 \\
\texttt{C54} && $2.13$ && $24^3\times64\times16$ && $1.8$ && $0.04$ && $0.005$ && $0.1119(17)$ && $0.005$ && 336(5) && 2782 \\
\texttt{F23} && $2.25$ && $32^3\times64\times16$ && $1.8$ && $0.03$ && $0.004$ && $0.0849(12)$ && $0.002$ && 227(3) && 1907 \\
\texttt{F43} && $2.25$ && $32^3\times64\times16$ && $1.8$ && $0.03$ && $0.004$ && $0.0849(12)$ && $0.004$ && 295(4) && 1917 \\
\texttt{F63} && $2.25$ && $32^3\times64\times16$ && $1.8$ && $0.03$ && $0.006$ && $0.0848(17)$ && $0.006$ && 352(7) && 2782 \\
\hline\hline
\end{tabular}
\caption{\label{tab:params} Parameters of the lattice gauge field ensembles \cite{Aoki:2010dy} and light-quark propagators \cite{Detmold:2012vy, Detmold:2013nia}. The three groups
of data sets $\left\{ \mathtt{C14}, \mathtt{C24}, \mathtt{C54} \right\}$, $\left\{ \mathtt{F23}, \mathtt{F43} \right\}$, and  $\left\{ \mathtt{F63} \right\}$
correspond to three different ensembles of lattice gauge fields: one with a ``coarse'' lattice spacing $a\approx 0.11$ fm, and two with ``fine'' lattice spacings $a\approx 0.085$ fm
(we use the lattice spacing values determined in Ref.~\cite{Meinel:2010pv}). Within each group, the valence-quark masses $am_{u,d}^{(\mathrm{val})}$ used for the propagators
differ, resulting in different ``valence pion masses'' $m_\pi^{(\mathrm{val})}$; the number of light-quark propagators used in each data set is denoted as $N_{\rm meas}$.}
\end{table}

\begin{table}
\begin{center}
\small
\begin{tabular}{clll}
\hline\hline
Parameter              & \hspace{1.2ex} coarse    & \hspace{2ex} & \hspace{1.2ex} fine       \\
\hline
$a m_Q^{(b)}$          & $\wm8.45$   & & $\wm3.99$     \\
$\xi^{(b)}$            & $\wm3.1$    & & $\wm1.93$     \\
$c_{E,\,B}^{(b)}$      & $\wm5.8$    & & $\wm3.57$     \\
$a m_Q^{(c)}$          & $\wm0.1214$ & & $-0.0045$  \\
$\xi^{(c)}$            & $\wm1.2362$ & & $\wm1.1281$   \\
$c_{E}^{(c)}$          & $\wm1.6650$ & & $\wm1.5311$   \\
$c_{B}^{(c)}$          & $\wm1.8409$ & & $\wm1.6232$   \\
\hline\hline
\end{tabular}\vspace{-2ex}
\end{center}
\caption{\label{tab:HQparams} Parameters of the bottom and charm quark actions \cite{Aoki:2012xaa, Brown:2014ena}. }
\end{table}

We use a mostly nonperturbative method \cite{Hashimoto:1999yp, ElKhadra:2001rv} to match the $b\to q$ ($q=u,c$) vector and axial vector currents from the lattice scheme to
the continuum $\overline{\rm MS}$ scheme. The renormalized currents in the $\overline{\rm MS}$ scheme are written in terms of the lattice quark and gluon
fields as
\begin{eqnarray}
 V_0 &=& \sqrt{Z_V^{(qq)} Z_V^{(bb)}} \rho_{V_0} \left[ \bar{q} \gamma_0 b + 2a \left( c^R_{V_0}\, \bar{q} \gamma_0 \gamma_j \overrightarrow{\nabla}_j  b + c^L_{V_0}\, \bar{q} \overleftarrow{\nabla}_j \gamma_0 \gamma_j   b \right)  \right], \label{eq:V0} \\
 A_0 &=& \sqrt{Z_V^{(qq)} Z_V^{(bb)}} \rho_{A_0} \left[ \bar{q} \gamma_0 \gamma_5 b + 2a \left( c^R_{A_0}\, \bar{q} \gamma_0\gamma_5 \gamma_j \overrightarrow{\nabla}_j  b + c^L_{A_0}\, \bar{q} \overleftarrow{\nabla}_j \gamma_0\gamma_5 \gamma_j   b \right)  \right], \\
 V_i &=& \sqrt{Z_V^{(qq)} Z_V^{(bb)}} \rho_{V_i} \left[ \bar{q} \gamma_i b + 2a \left( c^R_{V_i}\, \bar{q} \gamma_i \gamma_j \overrightarrow{\nabla}_j  b + c^L_{V_i}\, \bar{q} \overleftarrow{\nabla}_j \gamma_i \gamma_j   b + d^R_{V_i}\, \bar{q} \overrightarrow{\nabla}_i  b + d^L_{V_i}\, \bar{q} \overleftarrow{\nabla}_i b \right)   \right], \\
 A_i &=& \sqrt{Z_V^{(qq)} Z_V^{(bb)}} \rho_{A_i} \left[ \bar{q} \gamma_i \gamma_5 b + 2a \left( c^R_{A_i}\, \bar{q} \gamma_i\gamma_5 \gamma_j \overrightarrow{\nabla}_j  b + c^L_{A_i}\, \bar{q} \overleftarrow{\nabla}_j \gamma_i\gamma_5 \gamma_j   b + d^R_{A_i}\, \bar{q} \gamma_5 \overrightarrow{\nabla}_i  b + d^L_{A_i}\, \bar{q} \label{eq:Ai} \overleftarrow{\nabla}_i\gamma_5 b \right)   \right],
\end{eqnarray}
where $Z_V^{(qq)}$ and $Z_V^{(bb)}$ are the matching factors of the flavor-conserving temporal vector currents $\bar{q} \gamma_0 q$ and $\bar{b} \gamma_0 b$, which
are computed nonperturbatively using charge conservation. These nonperturbative factors provide the bulk of the renormalization, resulting in a much improved
convergence of perturbation theory for the residual matching factors $\rho_{V_\mu,A_\mu}$. Above, $i$ denotes the spatial components ($i=1,2,3$), and the repeated index $j$ is summed from 1 to 3. The lattice currents containing the lattice derivatives
\begin{eqnarray}
 \overrightarrow{\nabla}_\mu  \,b  &=& \frac{1}{2a} \left[  U_\mu(x) b(x+\hat{\mu})- U_\mu^\dag(x-\hat{\mu}) b(x-\hat{\mu}) \right], \\
 \bar{q}\,\overleftarrow{\nabla}_\mu  &=& \frac{1}{2a} \left[  \bar{q}(x+\hat{\mu}) U^\dag_\mu(x) - \bar{q}(x-\hat{\mu}) U_\mu(x-\hat{\mu})  \right],
\end{eqnarray}
are needed to remove $\mathcal{O}(a)$-discretization errors from the currents. Time derivatives have been eliminated in Eqs.~(\ref{eq:V0}-\ref{eq:Ai}) using the equations of motion.
We have computed $\rho_{V_\mu,A_\mu}$ as well as all of the $\mathcal{O}(a)$-improvement coefficients $c^{R,L}_{V_\mu,A_\mu}$, $d^{R,L}_{V_\mu,A_\mu}$ to one loop in
mean-field improved lattice perturbation theory using the automated framework \texttt{PhySyHCAl} \cite{Lehner:2012bt, PhySyHCAl}. The results are given in
Table \ref{tab:Pmatchingfactors}. The central values are the average of plaquette and Landau-gauge mean-field improved results with perturbative
expansion in $\alpha_{\overline{\rm MS}}(\mu = a^{-1})$ \cite{Aoki:2012xaa}.
The uncertainties are the maximum of i) the difference of the respective mean-field improved results, ii) the numerical integration error, and iii) a power-counting estimate.
For consistency with earlier stages of this project a different power-counting estimate is used for the $b \to u$ and $b\to c$ cases.
For a perturbative quantity $h$ with tree-level result $h^{(0)}$ and full one-loop result $h^{(1)}$ we use $(h^{(0)} / h^{(1)} - 1)^2 h^{(1)}$ as an error estimate
for $b\to u$ and $(h^{(1)} - h^{(0)}) (\alpha_{\overline{\rm MS}}(\mu = a^{-1}) / \pi)$ as an estimate for $b\to c$.
For the $\rho$ factors (but not for the $\mathcal{O}(a)$-improvement coefficients) this estimate tends to be less conservative than estimates that are used in similar work
\cite{Flynn:2015mha, Lattice:2015tia, Lattice:2015rga}. The estimates of the combined uncertainty from the $\rho$ factors and $\mathcal{O}(a)$-improvement coefficients, see Figs.~\ref{fig:systLbp},
\ref{fig:systLbLc} and Table \ref{tab:errorbreakdown}, agree well with similar work \cite{Flynn:2015mha, Lattice:2015tia, Lattice:2015rga}.

\begin{table}
\begin{center}
\small
\begin{tabular}{cllllllll}
\hline\hline
Parameter         & \hspace{2ex}         & $b\to c$, coarse    & \hspace{2ex} & \hspace{1ex} $b\to c$, fine     & \hspace{2ex} & $b\to u$, coarse    & \hspace{2ex} & \hspace{2ex} $b\to u$, fine       \\
\hline
 $\rho_{V^0}$      & & $\wm0.9798(20)$     & &  $\wm0.9848(15)$   & &  $\wm1.02658(69)$  & &   $\wm1.01661(52)$     \\[0.5ex]
 $\rho_{A^0}$      & & $\wm1.0193(15)$     & &  $\wm1.0112(29)$   & &  $\wm1.02658(69)$  & &   $\wm1.01661(52)$     \\[0.5ex]
 $\rho_{V^j}$      & & $\wm1.0184(38)$     & &  $\wm1.0162(41)$   & &  $\wm0.99723(25)$  & &   $\wm0.99398(12)$     \\[0.5ex]
 $\rho_{A^j}$      & & $\wm0.9866(33)$     & &  $\wm0.9896(26)$   & &  $\wm0.99723(25)$  & &   $\wm0.99398(12)$     \\[0.5ex]
 $c_{V^0}^R$       & & $\wm0.0258(13)$     & &  $\wm0.02873(99)$  & &  $\wm0.0558(63)$   & &   $\wm0.0547(64)$      \\[0.5ex]
 $c_{A^0}^R$       & & $\wm0.03500(93)$    & &  $\wm0.03285(87)$  & &  $\wm0.0558(63)$   & &   $\wm0.0547(64)$      \\[0.5ex]
 $c_{V^0}^L$       & & $-0.0183(32)$       & &  $-0.0135(22)$     & &  $-0.0099(99)$     & &   $-0.0095(95)$        \\[0.5ex]
 $c_{A^0}^L$       & & $-0.0205(17)$       & &  $-0.0155(27)$     & &  $-0.0099(99)$     & &   $-0.0095(95)$        \\[0.5ex]
 $c_{V^j}^R$       & & $\wm0.03192(70)$    & &  $\wm0.0305(10)$   & &  $\wm0.0485(27)$   & &   $\wm0.0480(30)$      \\[0.5ex]
 $c_{A^j}^R$       & & $\wm0.0221(31)$     & &  $\wm0.0237(20)$   & &  $\wm0.0485(27)$   & &   $\wm0.0480(30)$      \\[0.5ex]
 $c_{V^j}^L$       & & $\wm0.0088(22)$     & &  $\wm0.0027(35)$   & &  $-0.0033(33)$     & &   $-0.0020(20)$        \\[0.5ex]
 $c_{A^j}^L$       & & $-0.0002(31)$       & &  $-0.0020(20)$     & &  $-0.0033(33)$     & &   $-0.0020(20)$        \\[0.5ex]
 $d_{V^j}^R$       & & $-0.0055(12)$       & &  $-0.0067(17)$     & &  $-0.00079(79)$    & &   $-0.0012(12)$        \\[0.5ex]
 $d_{A^j}^R$       & & $\wm0.0060(69)$     & &  $\wm0.0044(40)$   & &  $\wm0.00079(79)$  & &   $\wm0.0012(12)$      \\[0.5ex]
 $d_{V^j}^L$       & & $-0.0176(44)$       & &  $-0.0043(69)$     & &  $\wm0.0018(18)$   & &   $\wm0.00047(48)$     \\[0.5ex]
 $d_{A^j}^L$       & & $\wm0.0134(69)$     & &  $\wm0.0106(40)$   & &  $-0.0018(18)$     & &   $-0.00047(48)$       \\[0.5ex]
\hline\hline
\end{tabular}\vspace{-2ex}
\end{center}
\caption{\label{tab:Pmatchingfactors} Perturbative renormalization and improvement coefficients. The uncertainties given here include estimates
of the missing higher-loop corrections as well as uncertainties from the numerical evaluation of the one-loop integrals.
They are explained in more detail in the main text.}
\end{table}

\begin{figure}
\hfill \includegraphics[height=4.5cm]{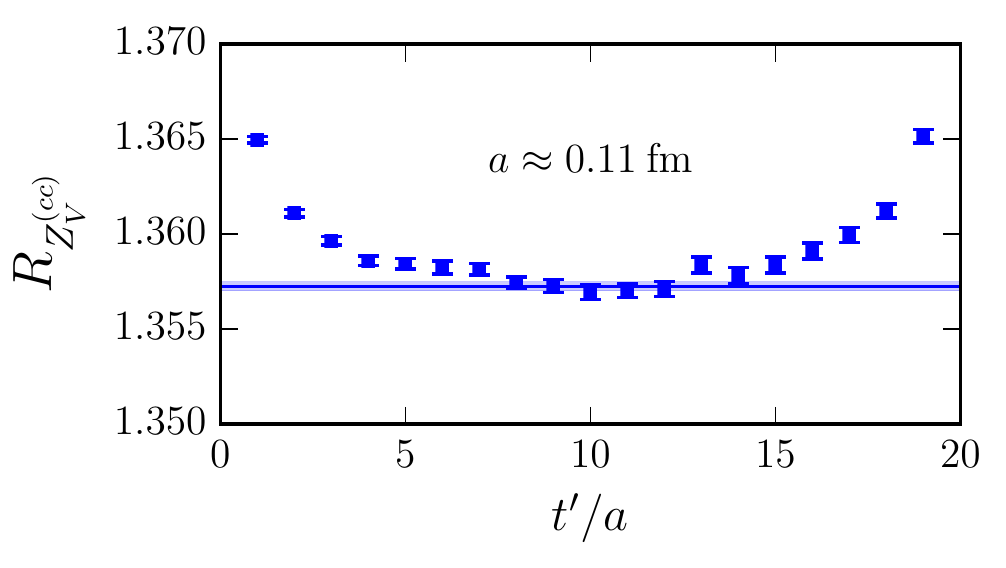} \hfill \includegraphics[height=4.5cm]{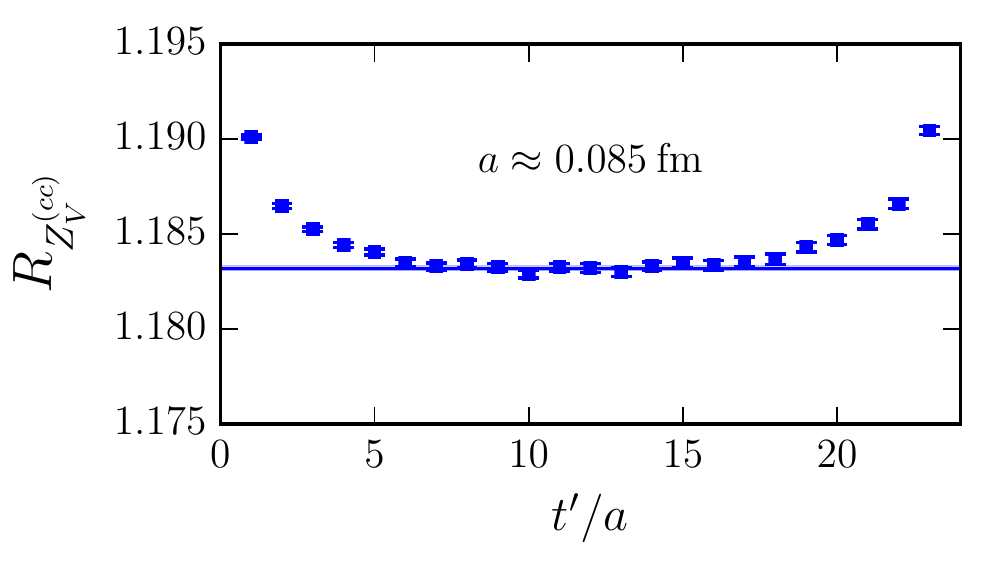} \hfill \null
\caption{\label{fig:ZVcc}Numerical results for the ratio $R_{Z_V^{(cc)}}(t,t^\prime)$ at $t/a=20$ for the \texttt{C54} data set (left)
and at $t/a=24$ for the \texttt{F43} data set (right). The horizontal lines indicate the extracted values of $Z_V^{(cc)}$. }
\end{figure}

The nonperturbative matching factors $Z_V^{(qq)}$ and $Z_V^{(bb)}$ are given in Table \ref{tab:NPmatchingfactors}. The light-quark and bottom-quark
$Z_V^{(uu)}$ and $Z_V^{(bb)}$ were computed by the RBC and UKQCD collaborations \cite{Aoki:2010dy, Christ:2014uea}. We determined the charm-quark $Z_V^{(cc)}$
using the method of Ref.~\cite{Christ:2014uea}, by computing the following ratio of $D_s$ meson correlation functions without and with insertion of the
current $J_0 = \bar{c}\gamma_0 c$:
\begin{equation}
 R_{Z_V^{(cc)}}(t,t^\prime) = \frac{\displaystyle\sum_{\mathbf{z}}\left\langle D_s(x_0+t,\mathbf{z})\:\:D_s^\dag(x_0,\mathbf{x}) \right\rangle}{\displaystyle\sum_{\mathbf{y},\mathbf{z}}\left\langle D_s(x_0+t,\mathbf{z})\:\: J_0(x_0+t^\prime,\mathbf{y})\:\:D_s^\dag(x_0,\mathbf{x}) \right\rangle}. \label{eq:ZVcc}
\end{equation}
Here, we used the following interpolating field with the quantum numbers of the $D_s$ meson,
\begin{equation}
 D_s = \tilde{\bar{s}}\gamma_5\tilde{c},
\end{equation}
where the tilde indicates gauge-covariant Gaussian smearing to suppress excited-state contamination.
For large Euclidean time separations $t$, $t^\prime$, and $|t-t^\prime|$, the ratio (\ref{eq:ZVcc}) becomes equal to $Z_V^{(cc)}$.
Our numerical results for $R_{Z_V^{(cc)}}(t,t^\prime)$, along with fits in the plateau region giving $Z_V^{(cc)}$, are shown in 
Fig.~\ref{fig:ZVcc} for the ensembles used in this calculation.

\begin{table}
\begin{center}
\small
\begin{tabular}{clll}
\hline\hline
Parameter                                & \hspace{1.2ex} coarse    & \hspace{2ex} & \hspace{1.2ex} fine       \\
\hline
 $Z_V^{(bb)}$                            & $10.037(34)$   & & $5.270(13)$    \\
 $Z_V^{(cc)}$                            & $1.35725(23)$  & & $1.18321(14)$  \\   
 $Z_V^{(uu)}$                            & $0.71651(46)$  & & $0.74475(12)$  \\
\hline\hline
\end{tabular}\vspace{-2ex}
\end{center}
\caption{\label{tab:NPmatchingfactors} Nonperturbative renormalization factors of the flavor-conserving temporal vector currents. For $Z_V^{(uu)}$,
we use the results in the chiral limit from Ref.~\cite{Aoki:2010dy}. For $Z_V^{(bb)}$, we use the results obtained in Ref.~\cite{Christ:2014uea} on
the coarse $am_{u,d}^{(\mathrm{sea})}=0.005$ and fine $am_{u,d}^{(\mathrm{sea})}=0.004$ ensembles. }
\end{table}

\FloatBarrier
\section{\label{sec:ratios}Extraction of the form factors from correlation functions}
\FloatBarrier

In this section we explain how we extract the form factors at the different lattice spacings and quark masses from nonperturbative Euclidean correlation
functions. The extrapolations of these results to the physical limit will be discussed in Sec.~\ref{sec:ccextrap}.

We use the following interpolating fields for the $\Lambda_b$, $\Lambda_c$, and the proton,
\begin{eqnarray}
 \Lambda_{b\alpha} &=& \epsilon^{abc}\:(C\gamma_5)_{\beta\gamma}\:\tilde{d}^a_\beta\:\tilde{u}^b_\gamma\: \tilde{b}^c_\alpha, \\
 \Lambda_{c\alpha} &=& \epsilon^{abc}\:(C\gamma_5)_{\beta\gamma}\:\tilde{d}^a_\beta\:\tilde{u}^b_\gamma\: \tilde{c}^c_\alpha, \\
 N_{\alpha} &=& \epsilon^{abc}\:(C\gamma_5)_{\beta\gamma}\:\tilde{u}^a_\beta\:\tilde{d}^b_\gamma\: \tilde{u}^c_\alpha, \label{eq:protoninterpol}
\end{eqnarray}
where $C$ is the charge-conjugation matrix, $a$, $b$, $c$ are color indices, and $\alpha$, $\beta$, $\gamma$ are spinor indices (the symbol $N$ is used for the proton
to avoid confusion with the $\Lambda_b$-momentum $p$). The tilde
on the quark fields indicates gauge-covariant Gaussian smearing. For the $u$ and $d$ quarks, the smearing parameters are the same as in Ref.~\cite{Detmold:2012ge}.
In the notation of Ref.~\cite{Detmold:2012ge}, for the charm quarks we used $(\sigma, n_S)=(3.0, 70)$ at the coarse lattice spacing and $(\sigma, n_S)=(4.0, 70)$
at the fine lattice spacing, and for the bottom quarks $(\sigma, n_S)=(2.0, 10)$ at the coarse lattice spacing and $(\sigma, n_S)=(2.67, 10)$ at the fine lattice spacing.
The smearing of both the charm and bottom quark fields was done using Stout-smeared gauge links \cite{Morningstar:2003gk} with ten iterations and staple weight $\rho=0.08$
in the spatial directions.

\begin{figure}
\includegraphics[width=0.5\linewidth]{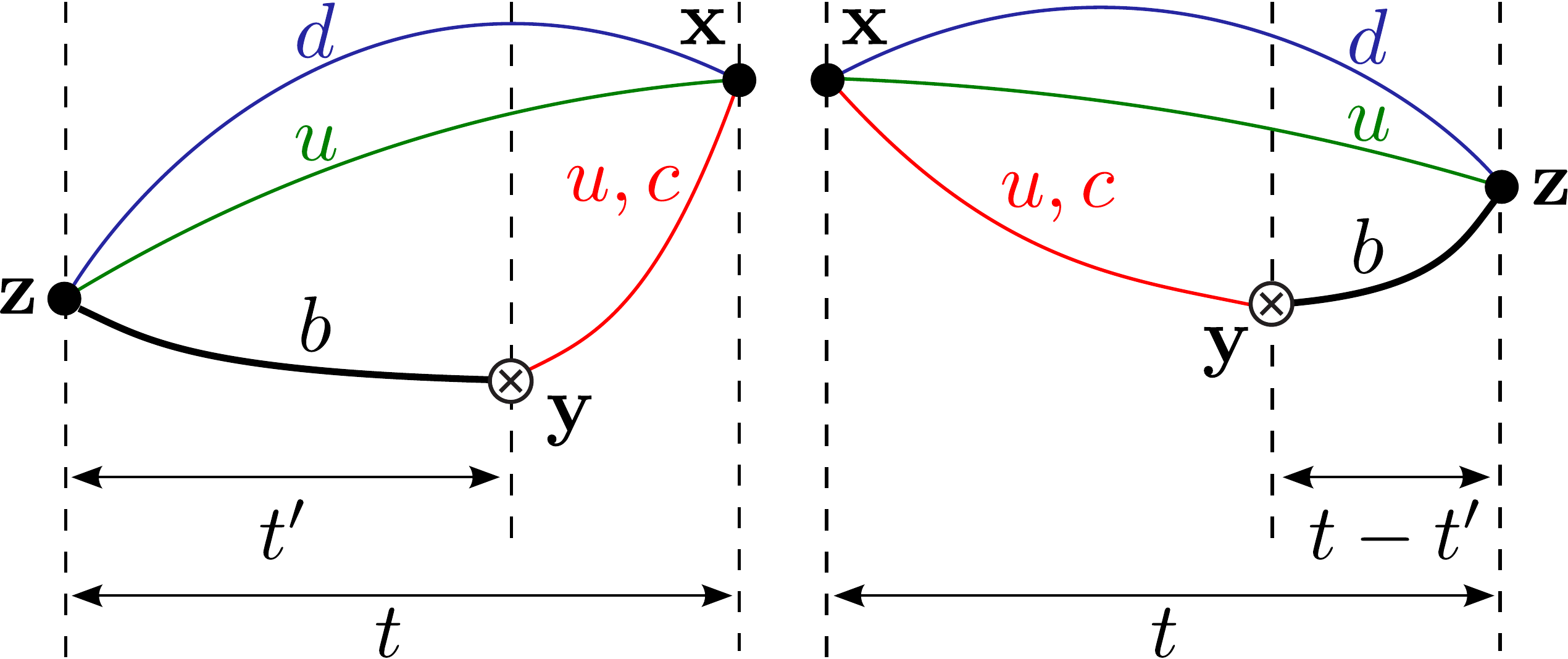}
\caption{\label{fig:threept}Illustration of the quark field contractions on a given background gauge field for the forward (left) and backward (right) three-point functions. The $u$, $d$, and $c$ quark
propagators are common to the forward and backward three-point functions and have a Gaussian-smeared source at $(x_0,\mathbf{x})$. We sum over the spatial
points $\mathbf{x}$ and $\mathbf{y}$ with the appropriate phases to project to definite momenta. The $b$-quark propagators
are computed using the sequential source method, with sequential sources on the time slices $x_0 \pm t$.}
\end{figure}

In the following, we denote the final-state interpolating field by $X_\alpha$ ($=N_{\alpha}, \Lambda_{c\alpha}$) and the renormalized currents as $J_\Gamma$, where
\begin{eqnarray}
 J_{\gamma_\mu} &=& V_\mu, \\
 J_{\gamma_\mu \gamma_5} &=& A_\mu,
\end{eqnarray}
with $V_\mu$ and $A_\mu$ given by Eqs.~(\ref{eq:V0}-\ref{eq:Ai}). We set the $\Lambda_b$ three-momentum $\mathbf{p}$ to zero, and compute ``forward'' and ``backward''
three-point functions (where $t \geq t^\prime \geq 0$),
\begin{eqnarray}
 C^{(3,\mathrm{fw})}_{\delta\alpha}(\Gamma,\:\mathbf{p^\prime}, t, t^\prime) &=& \sum_{\mathbf{y},\mathbf{z}} e^{-i\mathbf{p^\prime}\cdot(\mathbf{x}-\mathbf{y})}
 \Big\langle X_{\delta}(x_0,\mathbf{x})\:\:\:\: J_\Gamma^\dag(x_0-t+t^\prime,\mathbf{y})
 \:\:\:\: \bar{\Lambda}_{b\alpha} (x_0-t,\mathbf{z}) \Big\rangle, \hspace{4ex} \label{eq:threept} \\
C^{(3,\mathrm{bw})}_{\alpha\delta}(\Gamma,\:\mathbf{p^\prime}, t, t-t^\prime) &=& \sum_{\mathbf{y},\mathbf{z}}
e^{-i\mathbf{p^\prime}\cdot(\mathbf{y}-\mathbf{x})} \Big\langle \Lambda_{b\alpha}(x_0+t,\mathbf{z})\:\:\:\: J_\Gamma(x_0+t^\prime,\mathbf{y})
\:\:\:\: \bar{X}_{\delta} (x_0,\mathbf{x}) \Big\rangle, \label{eq:threeptbw}
\end{eqnarray}
as well as the two-point functions
\begin{eqnarray}
 C^{(2,X,\mathrm{fw})}_{\delta\alpha}(\mathbf{p^\prime}, t) &=&  \sum_{\mathbf{y}} e^{-i\mathbf{p^\prime}\cdot (\mathbf{y}-\mathbf{x})} \left\langle X_{\delta}(x_0+t,\mathbf{y})\: \overline{X}_{\alpha}(x_0,\mathbf{x})  \right\rangle, \label{eq:X2pt}\\
C^{(2,X,\mathrm{bw})}_{\delta\alpha}(\mathbf{p^\prime}, t) &=& \sum_{\mathbf{y}} e^{-i\mathbf{p^\prime}\cdot (\mathbf{x}-\mathbf{y})} \left\langle X_{\delta}(x_0,\mathbf{x})\: \overline{X}_{\alpha}(x_0-t,\mathbf{y}) \right\rangle,\\
 C^{(2,\Lambda_b,\mathrm{fw})}_{\delta\alpha}(t) &=&  \sum_{\mathbf{y}} \left\langle \Lambda_{b\delta}(x_0+t,\mathbf{y})\: \overline{\Lambda}_{b\alpha}(x_0,\mathbf{x})  \right\rangle, \label{eq:Lambdab2pt}\\
C^{(2,\Lambda_b,\mathrm{bw})}_{\delta\alpha}(t) &=& \sum_{\mathbf{y}}  \left\langle \Lambda_{b\delta}(x_0,\mathbf{x})\: \overline{\Lambda}_{b\alpha}(x_0-t,\mathbf{y}) \right\rangle.
\end{eqnarray}
These definitions are similar to those in the static $b$-quark case \cite{Detmold:2012vy, Detmold:2013nia}, but with the relativistic heavy-quark action used here,
the $b$ quark can propagate in all directions, and we included additional sums over the spatial coordinates for the momentum projections. The quark-field contractions
for the three-point functions are illustrated in Fig.~\ref{fig:threept}. Only the $b$-quark sequential propagators need to be recomputed for each source-sink separation,
$t$. For the proton final state, 16 times as many sequential propagators are needed as for the $\Lambda_c$ final state because of the different structure of diquark contractions.
The $b$-quark propagators decay extremely fast with distance, and care has to be taken to perform sufficiently many conjugate-gradient iterations to get an accurate solution up to the distance needed.

\begin{figure}
\begin{center}
\hfill \includegraphics[width=0.49\linewidth]{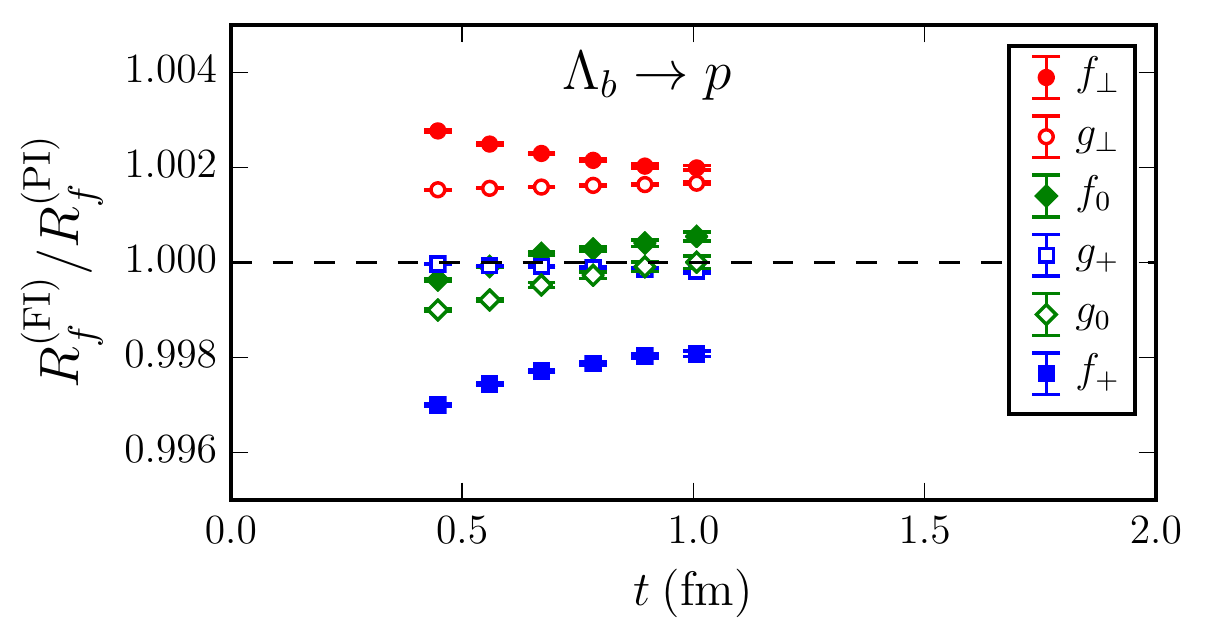} \hfill \includegraphics[width=0.49\linewidth]{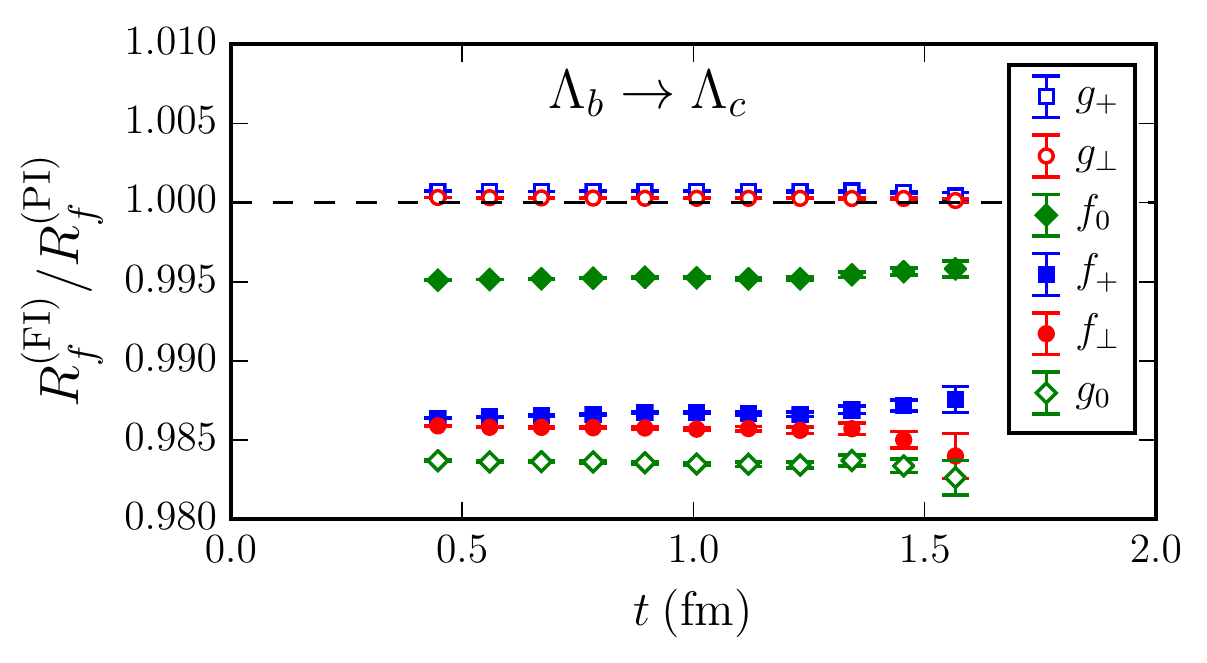} \hfill \null
 \caption{\label{fig:FIvsI} Ratios of the ``fully $\mathcal{O}(a)$-improved'' (``FI'') and ``partially $\mathcal{O}(a)$-improved'' (``PI'') data for $R_f(t)$
 for the six different helicity form factors, at $|\mathbf{p}^\prime|^2=3(2\pi/L)^2$, from the \texttt{C14} data set. The partially improved data only include
 the currents $\bar{q}\Gamma \gamma_j \protect\overrightarrow{\nabla}_j b$ for the $\mathcal{O}(a)$-improvement, where $\bar{q}=\bar{u},\bar{c}$.
 For $\Lambda_b\to p$, these are the only currents needed at tree level and the ratio is very close to 1. For $\Lambda_b\to\Lambda_c$, the ratio deviates from 1
 significantly more, because the currents $\bar{c} \protect\overleftarrow{\nabla}_j\gamma_j \Gamma b$ are missing  the partially improved data, but are needed
 already at tree level in this case. The range of source-sink separations shown for $\Lambda_b \to p$ is smaller because the statistical fluctuations in the
 correlators were too large to reliably compute the individual quantities $R_f(t)$ for $t>1.0\:{\rm fm}$ in this case.}
 \end{center}
\end{figure}

We computed the three-point functions for all final-state momenta $\mathbf{p}^\prime$ with $|\mathbf{p}^\prime|^2 \leq 12\:(2\pi/L)^2$, and for the ranges of source-sink separations shown in Table \ref{tab:tsep}.
In a first run we computed the three-point functions for all possible values of $t/a$ in the wide ranges shown in the left column of Table \ref{tab:tsep}, but only
for the lattice currents of the form $\bar{q}\Gamma b$ and $\bar{q}\Gamma\:\gamma_j\overrightarrow{\nabla}_j b$. In a second run, we then computed the three-point functions
for all of the remaining $\mathcal{O}(a)$-improvement currents shown in Eqs.~(\ref{eq:V0}-\ref{eq:Ai}), but only for the subsets of separations in the right column of Table \ref{tab:tsep}
to save computer time and disk space. For one of the data sets ($\texttt{C14}$), we performed the calculation of all the currents for the whole range of source-sink separations. As shown
in Fig.~\ref{fig:FIvsI}, the effects of the additional $\mathcal{O}(a)$ improvements are small. Our method for effectively including these corrections for all source-sink separations
will be explained further below.

\begin{table}
\begin{tabular}{ccccc}
\hline\hline
Set  & \hspace{1ex} & Partial $\mathcal{O}(a)$-improvement   & \hspace{1ex} & Full $\mathcal{O}(a)$-improvement \\
\hline
\texttt{C14}        && $t/a=4\, ...\, 15$   &&  $t/a=4\, ...\, 15$ \\
\texttt{C24}        && $t/a=4\, ...\, 15$   &&  $t/a=5,8,11$      \\
\texttt{C54}        && $t/a=4\, ...\, 15$   &&  $t/a=5,8,11$      \\
\texttt{F23}        && $t/a=5\, ...\, 15$   &&  $t/a=6,10,14$     \\
\texttt{F43}        && $t/a=5\, ...\, 15$   &&  $t/a=6,10,14$     \\ 
\texttt{F63}        && $t/a=5\, ...\, 17$   &&  $t/a=6,10,14$     \\ 
\hline\hline
\end{tabular}
\caption{\label{tab:tsep}Source-sink separations used for the three-point functions for each data set. For the separations in the column ``full $\mathcal{O}(a)$-improvement'', we computed the
three-point functions for all of the $\mathcal{O}(a)$-corrections in Eqs.~(\ref{eq:V0}-\ref{eq:Ai}). For the separations in the column ``partial $\mathcal{O}(a)$-improvement'', we computed only
the corrections with coefficients $c_{V_\mu}^R$ and $c_{A_\mu}^R$. As explained in the main text and illustrated in Fig.~\protect\ref{fig:FIvsI}, the effects of the missing terms
are very small and practically independent of the source-sink separation, and we achieve full $\mathcal{O}(a)$-improvement for all separations by applying $t$-independent correction factors
computed using the subsets of separations were all $\mathcal{O}(a)$-corrections are available.}
\end{table}

To discuss the spectral decomposition of the correlation functions, we introduce the overlap factors
\begin{eqnarray}
 \langle 0 | \Lambda_{b\alpha} (0) | \Lambda_b(p,s) \rangle &=& [(Z_{\Lambda_b}^{(1)}+Z_{\Lambda_b}^{(2)}\gamma^0) \: u(p, s)]_\alpha, \\
 \langle 0 | X_{\alpha} (0) | X(p^\prime,s^\prime) \rangle &=& [(Z_{X}^{(1)}+Z_{X}^{(2)}\gamma^0) \: u(p^\prime, s^\prime)]_\alpha.
\end{eqnarray}
The two separate $Z$ factors for each matrix element are needed because the spatial-only smearing of the quark fields in the interpolating field breaks hypercubic
symmetry \cite{Bowler:1997ej}.  Because we set $\mathbf{p}=0$, we can write
\begin{equation}
 \langle 0 | \Lambda_{b\alpha} (0) | \Lambda_p(p,s) \rangle = Z_{\Lambda_b} u(p, s)_\alpha,
\end{equation}
where $Z_{\Lambda_b}=Z_{\Lambda_b}^{(1)}+Z_{\Lambda_b}^{(2)}$. Further, we introduce the following short-hand notation for the form factor
decomposition of the matrix elements (cf.~Sec.~\ref{sec:FFdefinitions}):
\begin{equation}
\langle X(p^\prime, s^\prime) | J_\Gamma | \Lambda_b(p,s) \rangle = \overline{u}_X(p^\prime, s^\prime) \:\mathscr{G}[\Gamma]\: u_{\Lambda_b}(p,s)\,.
\end{equation}
The spectral decompositions of the correlation functions then read
\begin{eqnarray}
\nonumber C^{(3,{\rm fw})}(\mathbf{p^\prime},\:\Gamma, t, t^\prime) 
&=&  \: Z_{\Lambda_b} \: \frac{1}{2E_X} 
 \frac{1}{2 m_{\Lambda_b}} \: e^{-E_X(t-t^\prime)} \:e^{-m_{\Lambda_b} t^\prime}  \left[(Z_X^{(1)}+Z_X^{(2)}\gamma^0) (m_X+\slashed{p}^\prime) \:\mathscr{G}[\Gamma]\: m_{\Lambda_b}(1+\gamma^0)  \right] \\
 &&+({\rm excited\text{-}state\:\:contributions}), \\
\nonumber C^{(3,{\rm bw})}(\mathbf{p^\prime},\:\Gamma, t, t-t^\prime) 
&=&  \: Z_{\Lambda_b} \: \frac{1}{2E_X} 
 \frac{1}{2 m_{\Lambda_b}} \: e^{-m_{\Lambda_b}(t-t^\prime)} \:e^{-E_X t^\prime}  \left[m_{\Lambda_b}(1+\gamma^0) \:\overline{\mathscr{G}[\Gamma]}\: (m_X+\slashed{p}^\prime) (Z_X^{(1)}+Z_X^{(2)}\gamma^0) \right] \\
 &&+({\rm excited\text{-}state\:\:contributions}), \\
\nonumber C^{(2,X,{\rm fw})}(\mathbf{p^\prime}, t) = C^{(2,X,\mathrm{bw})}(\mathbf{p^\prime}, t) &=&  \frac{1}{2E_X} \: e^{-E_X t}  \left[(Z_X^{(1)}+Z_X^{(2)}\gamma^0)(m_X+\slashed{p}^\prime) (Z_X^{(1)}+Z_X^{(2)}\gamma^0) \right] \\
 &&+({\rm excited\text{-}state\:\:contributions}), \\
\nonumber C^{(2,\Lambda_b,{\rm fw})}(t) = C^{(2,\Lambda_b,\mathrm{bw})}(t) &=& \frac{1}{2m_{\Lambda_b}} e^{-m_{\Lambda_b} t} \left[ Z_{\Lambda_b}^2 m_{\Lambda_b}(1+\gamma^0 )  \right] \\
 &&+({\rm excited\text{-}state\:\:contributions}),
\end{eqnarray}
where $\overline{\mathscr{G}[\Gamma]}=\gamma_0 \mathscr{G}[\Gamma]^\dag \gamma_0$, and all correlators are $4\times 4$ matrices in spinor space.
In the above expressions, we have explicitly shown only the ground-state contributions, which correspond to the positive-parity baryons of interest. The excited-state contributions
decay exponentially faster with the time separations $t^\prime, t$.

To extract individual form factors, we contract the currents in the three-point functions with suitable polarization vectors and form certain double ratios that eliminate
all the time-dependence and overlap factors for the ground-state contributions. For an arbitrary four-vector $n$, we define
\begin{equation}
r[n]=n-\frac{(q\cdot n)}{q^2}q,
\end{equation}
where $q=p-p^\prime$ is the four-momentum transfer. By construction, $r[n]$ is orthogonal to $q$. For the vector current, we define the three ratios
\begin{eqnarray}
\mathscr{R}_{+}^V(\mathbf{p}^\prime,t,t^\prime) &=& \frac{ r_\mu[(1,\mathbf{0})] \: r_\nu[(1,\mathbf{0})] \:
\mathrm{Tr}\Big[   C^{(3,{\rm fw})}(\mathbf{p^\prime},\:\gamma^\mu, t, t^\prime) \:    C^{(3,{\rm bw})}(\mathbf{p^\prime},\:\gamma^\nu, t, t-t^\prime)  \Big] }
{\mathrm{Tr}\Big[C^{(2,X,{\rm av})}(\mathbf{p^\prime}, t)\Big] \mathrm{Tr}\Big[C^{(2,\Lambda_b,{\rm av})}(t)\Big] }, \label{eq:RVplus} \\
\mathscr{R}_{\perp}^V(\mathbf{p}^\prime,t,t^\prime) &=& \frac{ r_\mu[(0,\mathbf{e}_j\times \mathbf{p}^\prime)] \:   r_\nu[(0,\mathbf{e}_k\times \mathbf{p}^\prime)] \:
\mathrm{Tr}\Big[  C^{(3,{\rm fw})}(\mathbf{p^\prime},\:\gamma^\mu, t, t^\prime) \gamma_5 \gamma^j \:    C^{(3,{\rm bw})}(\mathbf{p^\prime},\:\gamma^\nu, t, t-t^\prime) \gamma_5 \gamma^k  \Big] }
{\mathrm{Tr}\Big[C^{(2,X,{\rm av})}(\mathbf{p^\prime}, t)\Big] \mathrm{Tr}\Big[C^{(2,\Lambda_b,{\rm av})}(t)\Big] }, \label{eq:RVperp} \\
\mathscr{R}_{0}^V(\mathbf{p}^\prime,t,t^\prime) &=& \frac{ q_\mu \: q_\nu \:
\mathrm{Tr}\Big[   C^{(3,{\rm fw})}(\mathbf{p^\prime},\:\gamma^\mu, t, t^\prime) \:    C^{(3,{\rm bw})}(\mathbf{p^\prime},\:\gamma^\nu, t, t-t^\prime)  \Big] }
{\mathrm{Tr}\Big[C^{(2,X,{\rm av})}(\mathbf{p^\prime}, t)\Big] \mathrm{Tr}\Big[C^{(2,\Lambda_b,{\rm av})}(t)\Big] }, \label{eq:RV0}
\end{eqnarray}
where $\mathbf{e}_j$ is the three-dimensional unit vector in $j$-direction, and $\times$ is the three-dimensional vector cross product. We sum over repeated indices $\mu$, $\nu$ from 0 to
3 and over repeated indices $j,k$ from 1 to 3. The quantities $C^{(2,\Lambda_b,{\rm av})}$ and $C^{(2,X,{\rm av})}$ in the denominators are the averages of the forward- and backward two-point functions.

\begin{table}
\begin{tabular}{ccccccccccc}
\hline\hline
Set  & \hspace{1ex} & $a m_{\Lambda_b}$   & \hspace{1ex} & $a m_{\Lambda_c}$  & \hspace{1ex} & $am_N$  & \hspace{1ex} & $am_{B_c}$ & \hspace{1ex} & $am_{B}$    \\
\hline
\texttt{C14}        && $3.305(11)\nb$   &&  $1.3499(51)\nb$  &&  $0.6184(76)$  && $3.60327(42)$   &&  $3.0649(27)$   \\
\texttt{C24}        && $3.299(10)\nb$   &&  $1.3526(57)\nb$  &&  $0.6259(57)$  && $3.60312(45)$   &&  $3.0628(29)$   \\
\texttt{C54}        && $3.3161(71)$     &&  $1.3706(40)\nb$  &&  $0.6580(39)$  && $3.60326(44)$   &&  $3.0638(33)$   \\ 
\texttt{F23}        && $2.469(16)\nb$   &&  $1.008(12)\nb$   &&  $0.4510(86)$  && $2.73156(44)$   &&  $2.3198(32)$   \\ 
\texttt{F43}        && $2.492(11)\nb$   &&  $1.0185(67)\nb$  &&  $0.4705(42)$  && $2.73169(44)$   &&  $2.3230(26)$   \\ 
\texttt{F63}        && $2.5089(70)$     &&  $1.0314(40)\nb$  &&  $0.5004(25)$  && $2.73257(33)$   &&  $2.3221(22)$   \\ 
\hline\hline
\end{tabular}
\caption{\label{tab:hadronmasses}Hadron masses in lattice units.}
\end{table}

\begin{figure}
\begin{center}
 \includegraphics[width=0.8\linewidth]{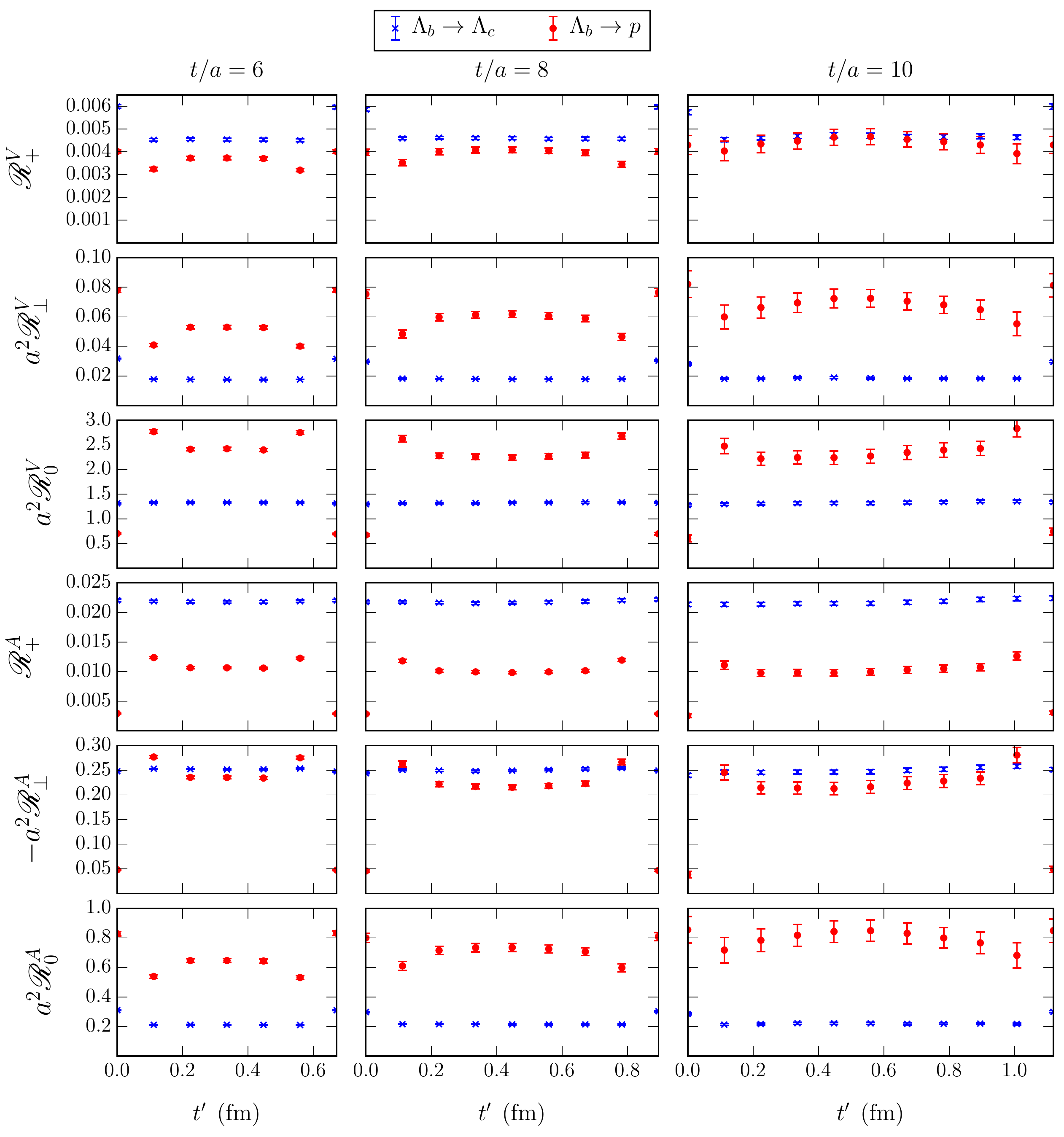}
 \caption{\label{fig:ratios}Numerical results for the vector-current ratios (\protect\ref{eq:RVplus}), (\protect\ref{eq:RVperp}), (\protect\ref{eq:RV0})
 and their axial-vector counterparts,  at $|\mathbf{p}^\prime|^2=3(2\pi/L)^2$, plotted for three different source-sink separations $t$. The data shown here
 are from the $\mathtt{C24}$ data set.}
\end{center}
\end{figure}

\begin{figure}
\begin{center}
   \includegraphics[width=0.28\linewidth]{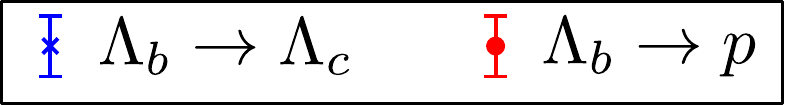}
   
   \vspace{2ex}
   
 \hfill \includegraphics[width=0.48\linewidth]{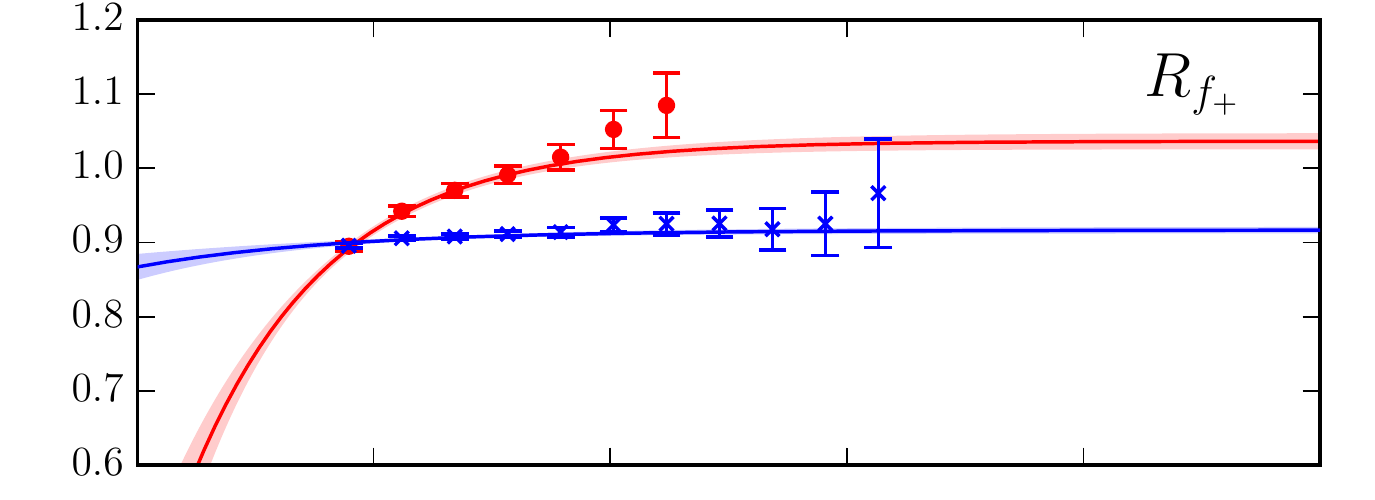} \hfill \includegraphics[width=0.48\linewidth]{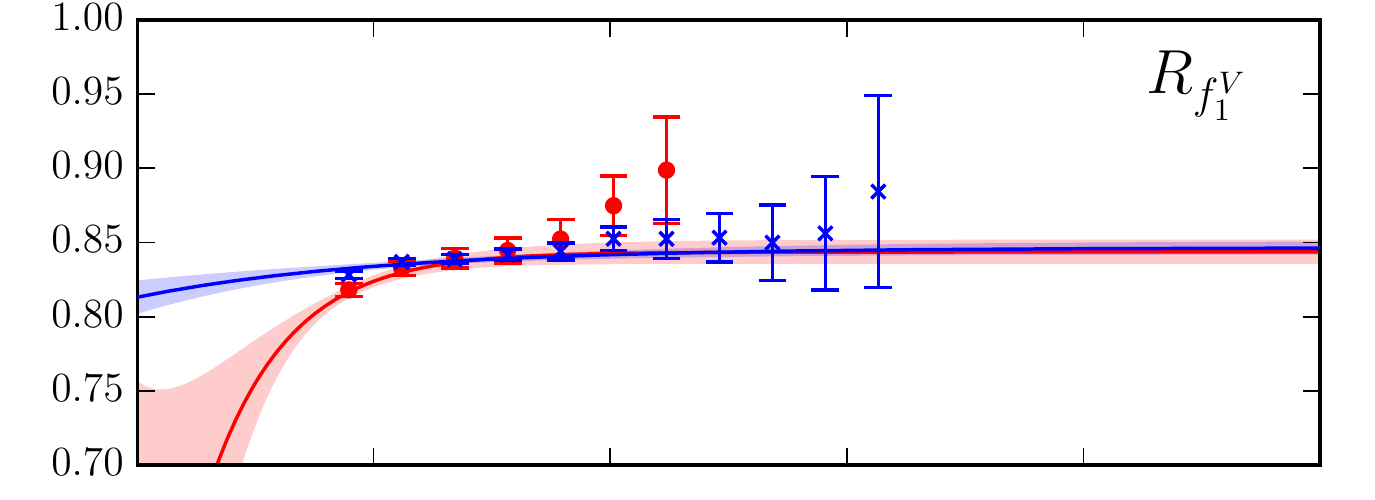} \hfill \null \\
 \hfill \includegraphics[width=0.48\linewidth]{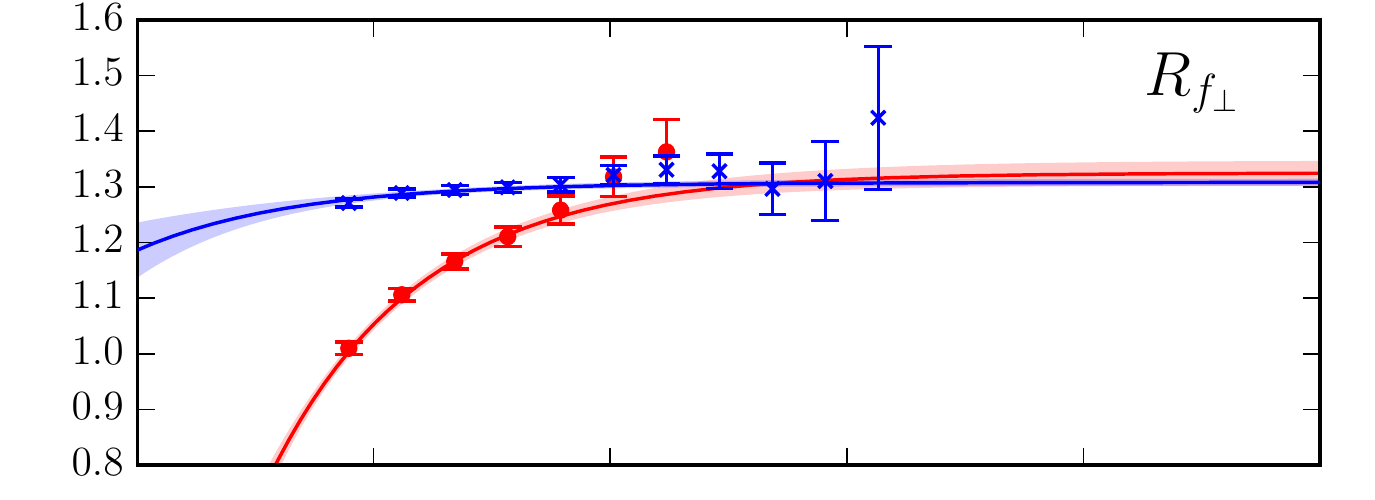} \hfill \includegraphics[width=0.48\linewidth]{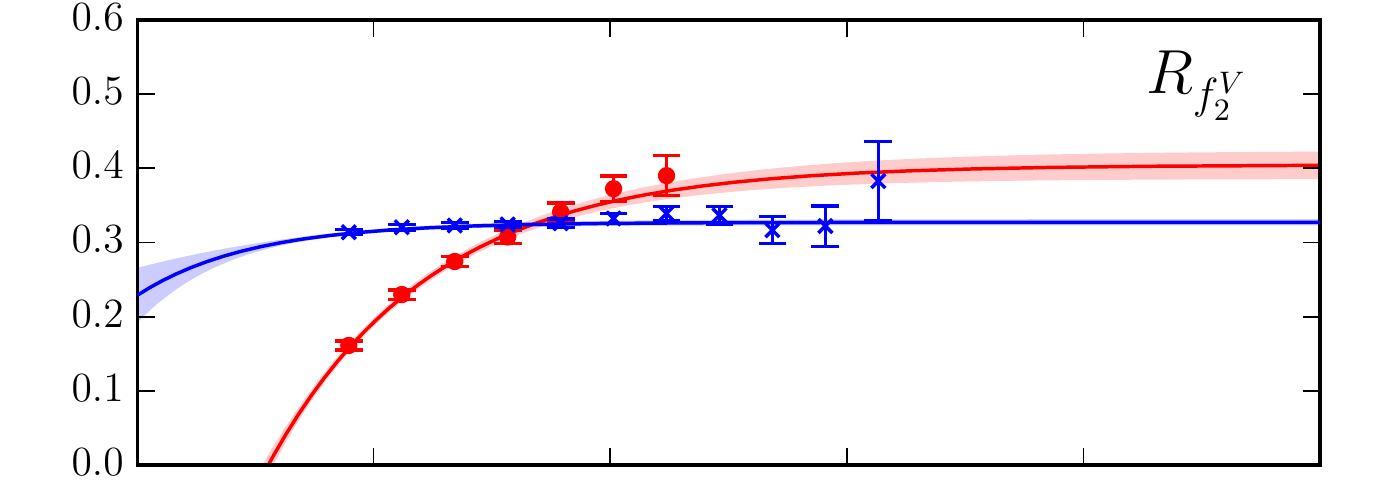} \hfill \null \\
 \hfill \includegraphics[width=0.48\linewidth]{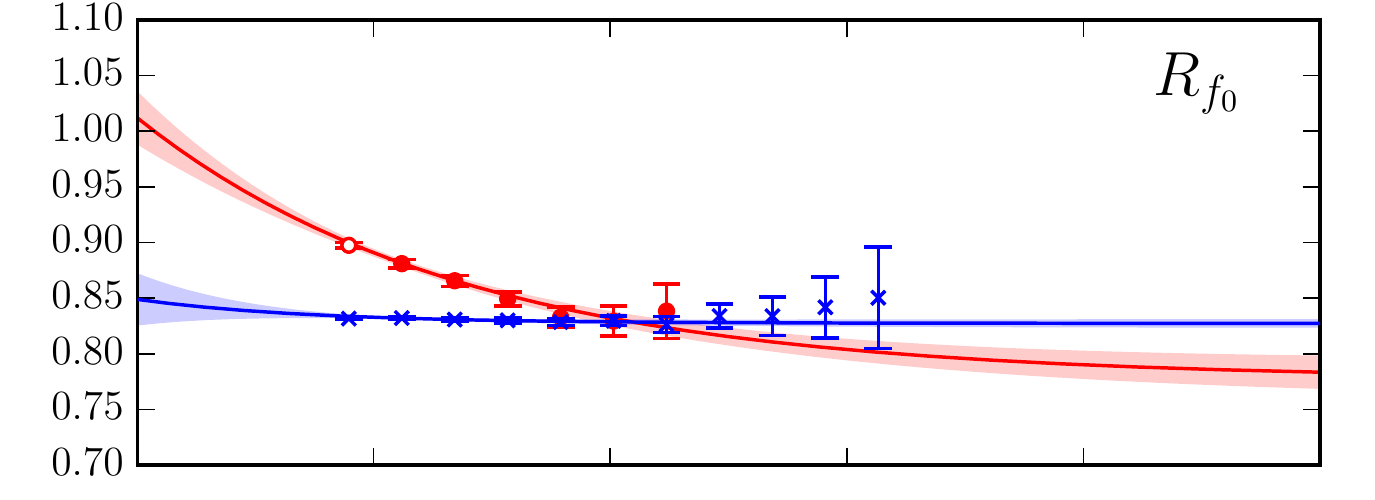}    \hfill \includegraphics[width=0.48\linewidth]{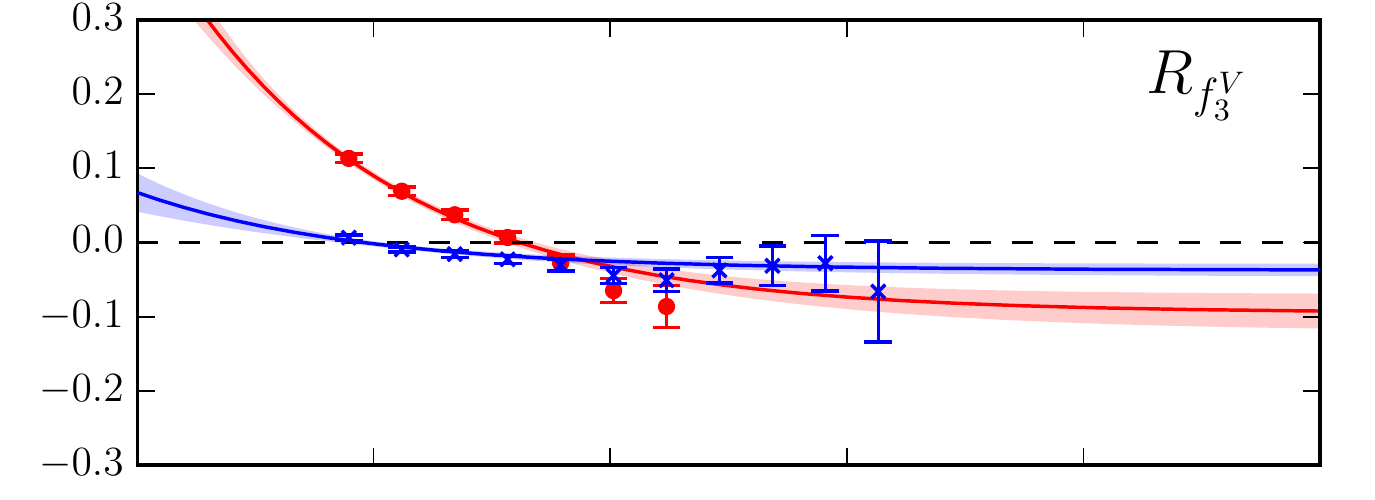} \hfill \null \\
 \hfill \includegraphics[width=0.48\linewidth]{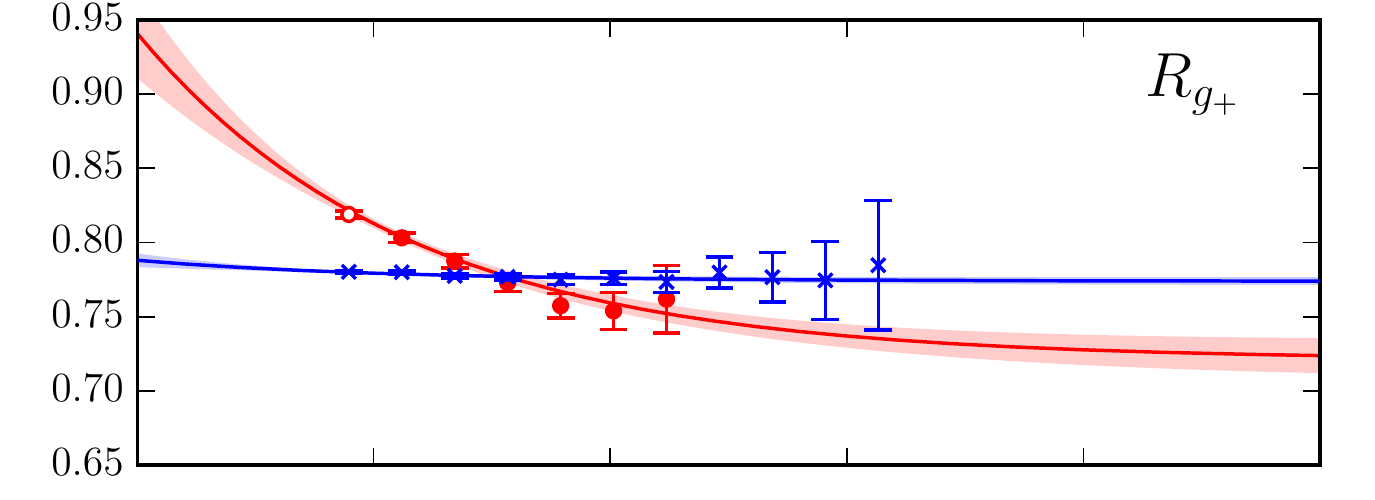} \hfill \includegraphics[width=0.48\linewidth]{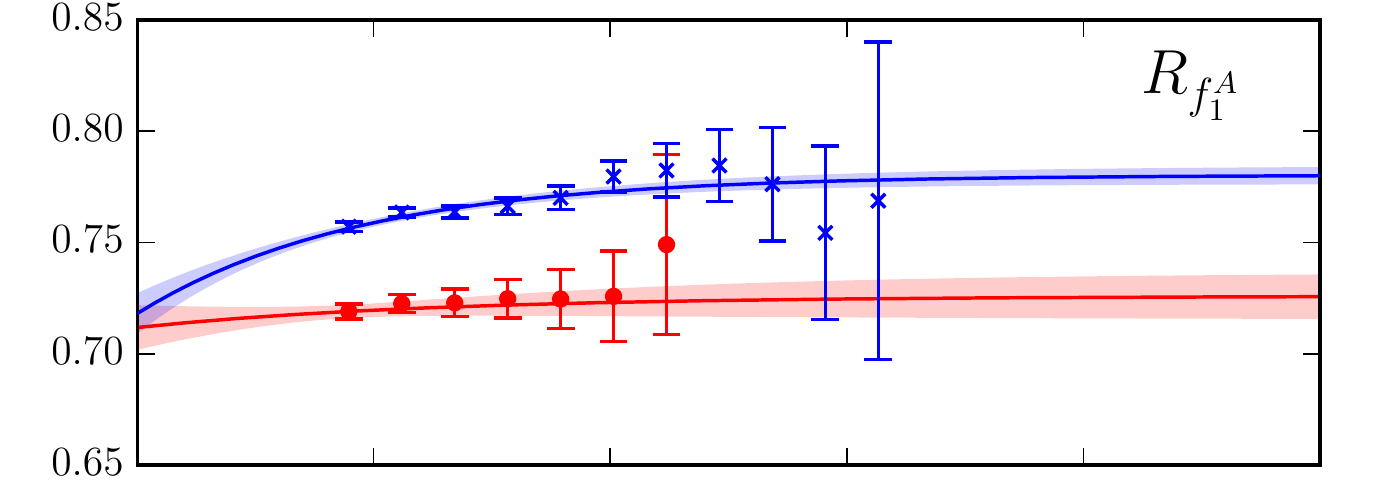} \hfill \null \\
 \hfill \includegraphics[width=0.48\linewidth]{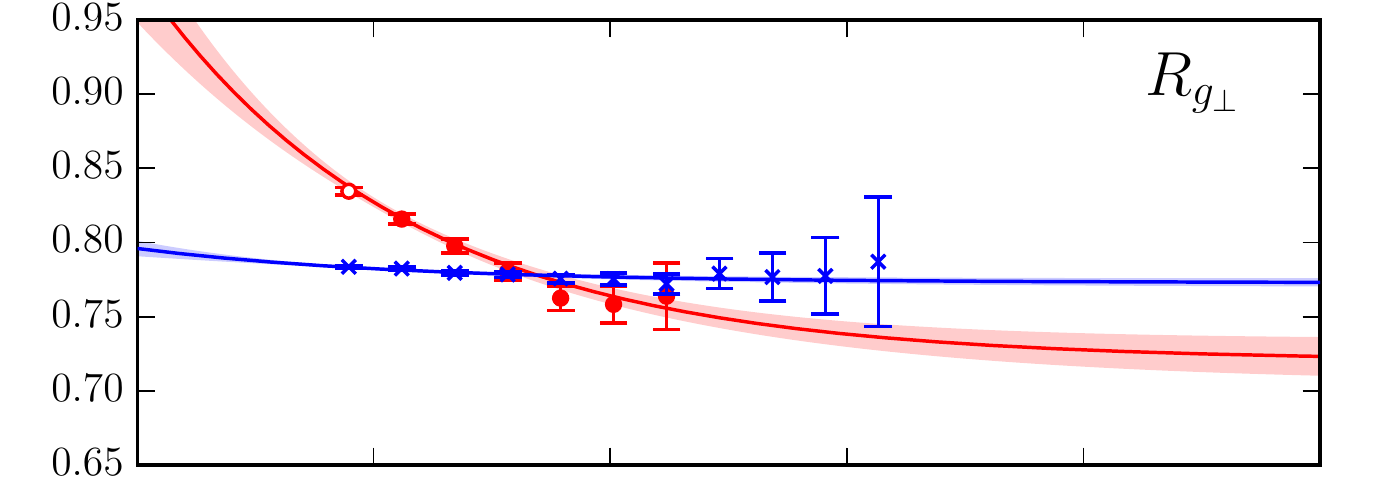} \hfill \includegraphics[width=0.48\linewidth]{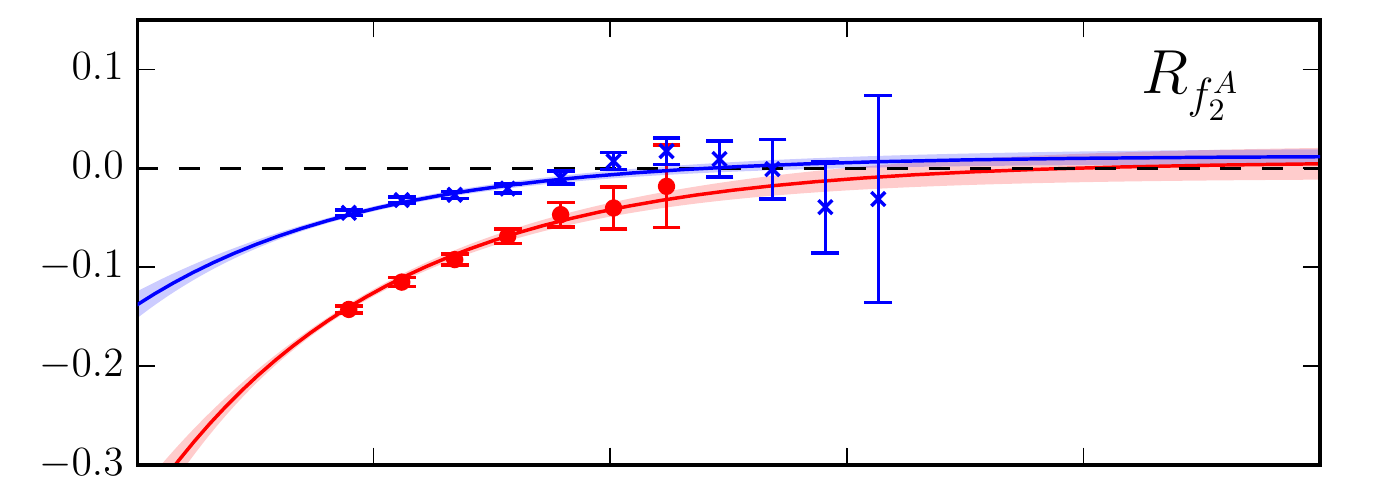} \hfill \null \\
 \hfill \includegraphics[width=0.48\linewidth]{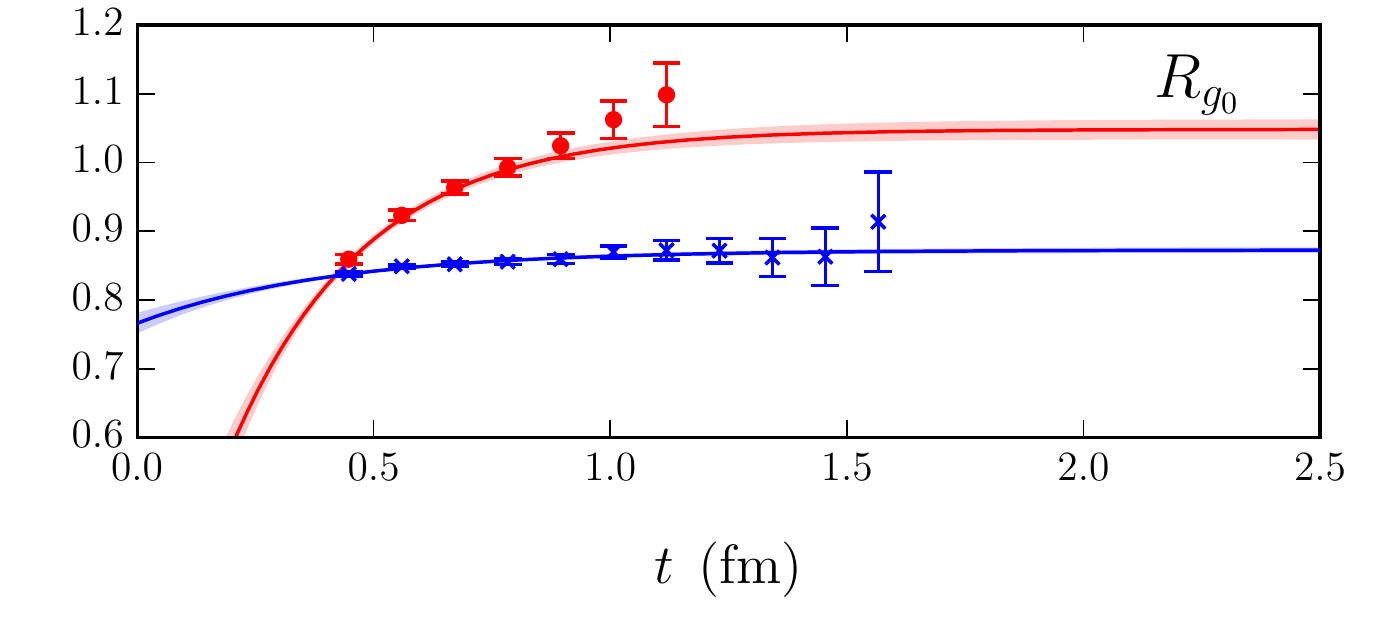}    \hfill \includegraphics[width=0.48\linewidth]{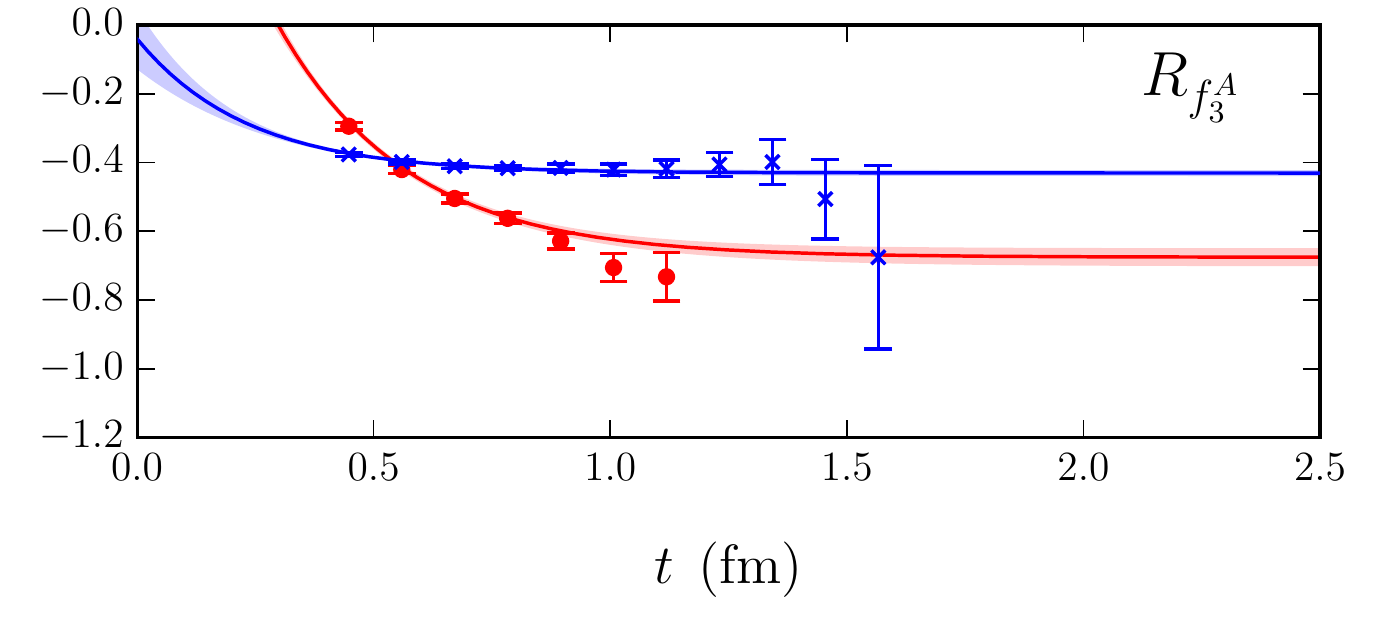} \hfill \null \\
 \caption{\label{fig:tsepextrap} Extrapolations of $R_{f}(|\mathbf{p}^\prime|, t)$ to infinite source-sink separation. The data shown here
 are at momentum $|\mathbf{p}^\prime|^2=3(2\pi/L)^2$,  and are from the $\mathtt{C24}$ data set. For each momentum, all vector (or axial vector)
 form factors from all data sets are fitted simultaneously as explained in the main text.}
\end{center}
\end{figure}

These ratios are designed to isolate particular helicity form factors and are equal to
\begin{eqnarray}
\mathscr{R}_{+}^V(\mathbf{p}^\prime,t,t^\prime) &=& \frac{(E_X-m_X)^2 (E_X+m_X) \Big[m_{\Lambda_b} (m_{\Lambda_b}+m_X)  f_+   \Big]^2}
{4\, m_{\Lambda_b}^2 E_X \:  q^4}
+({\rm excited\text{-}state\:\:contributions}), \\
\mathscr{R}_{\perp}^V(\mathbf{p}^\prime,t,t^\prime) &=& \frac{(E_X-m_X)^2 (E_X+m_X) \Big[ m_{\Lambda_b} f_\perp  \Big]^2}{m_{\Lambda_b}^2 E_X}
+({\rm excited\text{-}state\:\:contributions}), \\
\mathscr{R}_{0}^V(\mathbf{p}^\prime,t,t^\prime) & = & \frac{(E_X+m_X) \Big[ m_{\Lambda_b} (m_{\Lambda_b}-m_X)f_0 \Big]^2}{4 E_X m_{\Lambda_b}^2}
+({\rm excited\text{-}state\:\:contributions}).
\end{eqnarray}
Sample numerical results for $\mathscr{R}_{+}^V$, $\mathscr{R}_{\perp}^V$, and $\mathscr{R}_{0}^V$ are shown in Fig.~\ref{fig:ratios}. We further define the quantities
\begin{eqnarray}
 R_{f_+}(|\mathbf{p}^\prime|, t)      &=& \frac{2\, q^2 } {(E_X-m_X)(m_{\Lambda_b}+m_X)} \sqrt{\frac{ E_X  }{ E_X+m_X } \mathscr{R}_{+}^V(|\mathbf{p}^\prime|, t, t/2)} \label{eq:Rfplus} , \\
 R_{f_\perp}(|\mathbf{p}^\prime|, t)  &=& \frac{1}{E_X-m_X} \sqrt{\frac{ E_X}{ E_X+m_X } \mathscr{R}_{\perp}^V(|\mathbf{p}^\prime|, t, t/2)}, \\
 R_{f_0}(|\mathbf{p}^\prime|, t)      &=& \frac{2}{ m_{\Lambda_b}-m_X}  \sqrt{\frac{ E_X}{E_X+m_X } \mathscr{R}_{0}^V(|\mathbf{p}^\prime|, t, t/2)}, \label{eq:Rf0}
\end{eqnarray}
where we evaluate the ratios $\mathscr{R}$ at $t^\prime=t/2$ to minimize excited-state contamination at a given value of the source-sink separation $t$
(if $t/a$ is odd, we average $\mathscr{R}$ over $t^\prime=(t+a)/2$ and $t^\prime=(t-a)/2$ instead). The notation with the absolute value indicates that we
average over the directions of $\mathbf{p}^\prime$. Equations (\ref{eq:Rfplus}-\ref{eq:Rf0}) yield
\begin{eqnarray}
 R_{f_+}(|\mathbf{p}^\prime|, t)      &=& f_+ +({\rm excited\text{-}state\:\:contributions}), \\
 R_{f_\perp}(|\mathbf{p}^\prime|, t)  &=& f_\perp +({\rm excited\text{-}state\:\:contributions}), \\
 R_{f_0}(|\mathbf{p}^\prime|, t)      &=& f_0 +({\rm excited\text{-}state\:\:contributions}),
\end{eqnarray}
where the excited-state contributions decay exponentially with $t$. We checked that the helicity form factors (plus the corresponding excited-state contributions,
for the separations we utilize) are all positive by analyzing individual three-point functions, so that the square roots in Eqs.~(\ref{eq:Rfplus}-\ref{eq:Rf0}) give
the correct signs. Although not explicitly annotated, the form factors in all of the above expressions depend on $|\mathbf{p}^\prime|$ and on the lattice parameters.
For the axial-vector current, we define $\mathscr{R}_{+,\perp,0}^A$ as in Eq.~(\ref{eq:RVplus}-\ref{eq:RV0}) but with $\Gamma=\gamma^\mu\gamma_5$ in the three-point functions.
The axial-vector helicity form factors are then extracted as
\begin{eqnarray}
 R_{g_+}(|\mathbf{p}^\prime|, t)      &=& \frac{2\,  q^2 }{(E_X+m_X)(m_{\Lambda_b}-m_X)}\sqrt{\frac{E_X}{E_X-m_X} \mathscr{R}_{+}^A(|\mathbf{p}^\prime|, t, t/2)} , \\
 R_{g_\perp}(|\mathbf{p}^\prime|, t)  &=& \frac{1}{E_X+m_X}\sqrt{-\frac{E_X}{E_X-m_X} \mathscr{R}_{\perp}^A(|\mathbf{p}^\prime|, t, t/2)}, \\
 R_{g_0}(|\mathbf{p}^\prime|, t)      &=& \frac{2}{m_{\Lambda_b}+m_X} \sqrt{\frac{E_X}{E_X-m_X}\mathscr{R}_{0}^A(|\mathbf{p}^\prime|, t, t/2)}.
\end{eqnarray}
When evaluating the ratios, we take the baryon masses in lattice units, $a m_{\Lambda_b}$, $a m_{\Lambda_c}$, and $am_N$, from exponential fits to the zero-momentum
two-point functions for each data set; see Table \ref{tab:hadronmasses}. We then compute the energies $a E_{\Lambda_c}(\mathbf{p}^\prime)$, and $a E_N(\mathbf{p}^\prime)$
from these masses using the relativistic continuum dispersion relation, and we also compute $a^2 q^2$ from these masses and energies. Because the form factors are dimensionless,
the values of the lattice spacing are not needed at this stage. The ratios are evaluated using statistical bootstrap, and we use corresponding bootstrap samples for the
masses to take into account all correlations.

As mentioned earlier, except in the case of the \texttt{C14} data set, we have ``full-$\mathcal{O}(a)$ improvement'' (``FI'') data only for three source-sink separations
in each data set, but we have data with ``partial $\mathcal{O}(a)$-improvement'' (``PI'') for all source-sink separations in the ranges shown in Table \ref{tab:tsep}.
To account for this, we computed the ratios
\begin{equation}
 \frac{R_{f}^{(\rm FI)}(|\mathbf{p}^\prime|, t)}{R_{f}^{(\rm PI)}(|\mathbf{p}^\prime|, t)}, \label{eq:RFIPI}
\end{equation}
where $f=f_{+},f_\perp,f_0,g_+,g_\perp,g_0$, for those source-sink separations where both $R_{f}^{(\rm FI)}$ and $R_{f}^{(\rm PI)}$ are available.
Numerical results for Eq.~(\ref{eq:RFIPI}) from the \texttt{C14} data set (where we have FI data for all values of $t$) are shown in Fig.~\ref{fig:FIvsI}.
In the case of $\Lambda_b \to \Lambda_c$ (Fig.~\ref{fig:FIvsI} right), the correction is as large as 2\% for some of the form factors, but is
independent of the source-sink separation to a high degree (even though $R_{f}^{(\rm FI)}$ and $R_{f}^{(\rm PI)}$ individually have a strong $t$-dependence). The same
behavior is found at other values of the momentum. In the case of $\Lambda_b \to p$ (Fig.~\ref{fig:FIvsI} left), the correction shows a more significant dependence
on the source-sink separation, but is smaller than 0.3\% for all form factors. For the data sets other than \texttt{C14} we therefore
performed constant fits to the ratios (\ref{eq:RFIPI}) as a function of $t$, individually for each form factor $f$, each momentum $|\mathbf{p}^\prime|$,
and each data set. If these fits had a poor $\chi^2/{\rm dof}$, we excluded the shortest or the two shortest separations. In this way, we obtained
correction factors, which we then applied to $R_{f}^{(\rm PI)}(|\mathbf{p}^\prime|, t)$ at all separations, to effectively obtain $R_{f}^{(\rm FI)}(|\mathbf{p}^\prime|, t)$
at all separations. This procedure is accurate to better than permille level. In the following, all ratios $R_{f}(|\mathbf{p}^\prime|, t)$ are understood
to be corrected using this procedure for all source-sink separations.

For the further data analysis, we then also formed the linear combinations
\begin{eqnarray}
 R_{f_1^V}      &=& \frac{(m_{\Lambda_b}+m_X)^2 R_{f_+} - q^2 R_{f_\perp} }{s_+}, \label{eq:FFRR1} \\
 R_{f_2^V}      &=& \frac{m_{\Lambda_b}(m_{\Lambda_b}+m_X)( R_{f_\perp}-R_{f_+} ) }{s_+}, \\
 R_{f_3^V}      &=& \frac{m_{\Lambda_b}(m_{\Lambda_b}-m_X) \left[ (m_{\Lambda_b}+m_X)^2(R_{f_0}-R_{f_+})+q^2(R_{f_\perp}-R_{f_0}) \right]  }{q^2\, s_+}, \label{eq:FFRR3} \\
 R_{f_1^A}      &=& \frac{(m_{\Lambda_b}-m_X)^2 R_{g_+} - q^2 R_{g_\perp} }{s_-}, \\
 R_{f_2^A}      &=& \frac{m_{\Lambda_b}(m_{\Lambda_b}-m_X)( R_{g_+}-R_{g_\perp} ) }{s_-}, \\
 R_{f_3^A}      &=& \frac{m_{\Lambda_b}(m_{\Lambda_b}+m_X) \left[ (m_{\Lambda_b}-m_X)^2(R_{g_+}-R_{g_0})+q^2(R_{g_0}-R_{g_\perp}) \right]  }{q^2\, s_-}, \label{eq:FFRR2}
\end{eqnarray}
which, according to the relations in Eqs.~(\ref{eq:FFR1})-(\ref{eq:FFR6}), become equal to the Weinberg form factors at large $t$. To extract the ground-state form factors
from $R_{f}(|\mathbf{p}^\prime|, t)$ (for both the helicity and Weinberg form factors), we performed correlated fits of the $t$-dependence including exponential correction
terms to account for the leading excited-state contributions, thereby extrapolating $R_{f}(|\mathbf{p}^\prime|, t)$ to $t=\infty$. To discuss these extrapolations in more detail, it
is convenient to denote the data for $R_{f}(|\mathbf{p}^\prime|, t)$ by 
\begin{equation}
 R_{f,i,n}(t),
\end{equation}
where $f=f_+,f_\perp,f_0,g_+,g_\perp,g_0,f_1^V,f_2^V,f_3^V,f_1^A,f_2^A,f_3^A$ labels the form factors, $i=\texttt{C14},\texttt{C24},\texttt{C54},\texttt{F23},\texttt{F43},\texttt{F63}$
labels the data set (cf.~Table \ref{tab:params}), and $n$ labels the final-state momentum via $|\mathbf{p^\prime}|^2=n\, (2\pi)^2/L^2$. We performed the fits using the functions
\begin{equation}
 R_{f,i,n}(t) = f_{i,n} + A_{f,i,n} \: e^{-\delta_{f,i,n}\:t},\hspace{2ex}\delta_{f,i,n}=\delta_{\rm min} + e^{\,l_{f,i,n}}\:\:{\rm GeV},
\end{equation}
with parameters $f_{i,n}$, $A_{f,i,n}$, and $l_{f,i,n}$, where $f_{i,n}$ are the form factors we aim to extract. By writing the energy-gaps $\delta_{f,i,n}$ in the above form,
we impose the constraint $\delta_{f,i,n}>\delta_{\rm min}$. We chose $\delta_{\rm min}=170\:\:{\rm MeV}$, which is smaller than any expected energy gap
(given our prior knowledge of the hadron spectrum at our values of the pion masses). This constraint has negligible effect in most cases, but prevents numerical instabilities
for some form factors at certain momenta where the data show no discernible excited-state contamination.

At each momentum $n$, we perform one coupled fit to all the data for the vector-current form factors ($f_+$, $f_\perp$, $f_0$, $f_1^V$, $f_2^V$, $f_3^V$) and another coupled fit
to all the data for the axial-vector-current form factors ($g_+$, $g_\perp$, $g_0$, $f_1^A$, $f_2^A$, $f_3^A$). This allows us to implement two additional constraints
to stabilize the fits, based on the following knowledge:
\begin{itemize}
 \item Because the lattice size, $L$ (in physical units), is equal within uncertainties for all data sets ($L\approx 2.7\:{\rm fm}$), the squared momentum
 $|\mathbf{p'}|^2= n\, (2\pi/L)^2$ for a given $n$ is also equal within uncertainties for all data sets. This means that the energy levels, and hence the parameters
 $l_{f,i,n}$, are expected to be approximately equal across all data sets $i$, up to some dependence on the pion mass and the lattice spacing.
 \item By construction, the data $R_{f,i,n}(t)$ for the helicity and Weinberg form factors exactly satisfy the defining relations (\ref{eq:FFRR1})-(\ref{eq:FFRR1})
 at each value of the source-sink separation. The extracted ground-state form factors $f_{i,n}$ should also satisfy these relations.
\end{itemize}
For the coupled fit to all vector form factor data at a given momentum $n$, we therefore add the following terms, corresponding to Gaussian priors, to the $\chi^2$ function:
\begin{eqnarray}
\nonumber \chi^2_{V,n} &\rightarrow \chi^2_{V,n} & + \sum_f \left[ \frac{(l_{f,\mathtt{C14},n}-l_{f,\mathtt{C24},n})^2}{[\sigma_m^{\mathtt{C14},\mathtt{C24}}]^2} + \frac{(l_{f,\mathtt{C24},n}-l_{f,\mathtt{C54},n})^2}{[\sigma_m^{\mathtt{C24},\mathtt{C54}}]^2}  + \frac{(l_{f,\mathtt{F23},n}-l_{f,\mathtt{F43},n})^2}{[\sigma_m^{\mathtt{F23},\mathtt{F43}}]^2} \right. \\
&& \left. \hspace{5ex} + \frac{(l_{f,\mathtt{F43},n}-l_{f,\mathtt{F63},n})^2}{[\sigma_m^{\mathtt{F43},\mathtt{F63}}]^2}
\nonumber + \frac{(l_{f,\mathtt{C54},n}-l_{f,\mathtt{F63},n})^2}{[\sigma_m^{\mathtt{C54},\mathtt{F63}}]^2+\sigma_a^2} \right] \\
\nonumber && + \sum_i \left( f_{+,i,n} -  f_{1,i,n}^V - \frac{q^2_{i,n}}{m_{\Lambda_b,i}(m_{\Lambda_b,i}+m_{X,i})} f_{2,i,n}^V \right)^2 / \sigma_f^2 \\
\nonumber && + \sum_i \left( f_{\perp,i,n} -  f_{1,i,n}^V - \frac{m_{\Lambda_b,i}+m_{X,i}}{m_{\Lambda_b,i}} f_{2,i,n}^V \right)^2 / \sigma_f^2 \\
&& + \sum_i \left( f_{0,i,n} -  f_{1,i,n}^V - \frac{q^2_{i,n}}{m_{\Lambda_b,i}(m_{\Lambda_b,i}-m_{X,i})} f_{3,i,n}^V \right)^2 / \sigma_f^2, \label{eq:chisqr}
\end{eqnarray}
where $\sigma_a=0.1$ and
\begin{equation}
 [\sigma_m^{i,j}]^2 = w_m^2 [ (m_\pi^i)^2-(m_\pi^j)^2 ]^2,
\end{equation}
with $w_m = 4\:{\rm GeV}^{-2}$. With these widths, the first two lines in Eq.~(\ref{eq:chisqr}) implement the constraint that the energy gaps $(\delta_{f,i,n}-\delta_{\rm min})$
at given momentum $n$ should not change by more than 10\% when going from the fine to the coarse lattice spacing and not more than 400\% times the change in $m_\pi^2$ (in ${\rm GeV}^2$); both
are reasonable assumptions given the prior experience with hadron spectroscopy in lattice QCD.
Note that absolute variations of $l_{f,i,n}$ translate to relative variations of $(\delta_{f,i,n}-\delta_{\rm min})$ because $\mathrm{d}[\exp(l_{f,i,n})]/\exp(l_{f,i,n}) = \mathrm{d} l_{f,i,n}$.
The last three lines in Eq.~(\ref{eq:chisqr}) enforce the relations (\ref{eq:FFRR1})-(\ref{eq:FFRR3}) between the ground-state vector form factors in the helicity and Weinberg definitions (we set $\sigma_f=10^{-4}$).
For the fit to the axial vector form factor data, analogous terms are added to $\chi^2_{A,n}$.

We initially included all available values of $t$ in the fits, and then removed data points for each form factor at the smallest $t$ until the fits had good quality as determined by the correlated $\chi^2/{\rm dof}$.
To estimate the remaining systematic uncertainties associated with higher excited states, we then further removed the next-lowest values of $t$ simultaneously for all $R_{f,i,n}$ and computed the resulting shifts in $f_{i,n}$.
We then took the larger of the following two as our estimate of the excited-state systematic uncertainty: i) the shift in $f_{i,n}$ at the given momentum $n$, and ii) the average of the shifts $f_{i,n}$ over all momenta $n$.
We added these excited-state uncertainties in quadrature to the statistical uncertainties in $f_{i,n}$. All result for $f_{i,n}$ are listed in Appendix \ref{sec:fftables}. As can be seen in Tables
\ref{LbpWeinberg} and \ref{LbLcWeinberg}, the results for the second-class form factor $f_2^A$ are very close to or consistent with zero for both $\Lambda_b \to \Lambda_c$ and $\Lambda_b \to \Lambda_c$,
despite the rather large mass differences $m_b-m_u$ and $m_b-m_c$. The results for the other second-class form factor $f_3^A$ are significantly nonzero, but are still noticeably smaller than the results for the first-class form factors.

\FloatBarrier
\section{\label{sec:ccextrap}Chiral/continuum/kinematic extrapolation of the form factors}
\FloatBarrier

The last step in the data analysis is to perform fits of form factor results $\{ f_{i,n} \}$ using suitable functions
describing the dependence on the momentum transfer, the dependence on the up and down quark masses (or equivalently the pion mass), and the dependence on the lattice spacing. We perform
global fits of the helicity form factors based on the simplified $z$-expansion \cite{Bourrely:2008za}, modified to account for pion-mass and lattice-spacing dependence. The expansion
parameter $z^f$ for a form factor $f$ is defined as
\begin{equation}
z^f(q^2) = \frac{\sqrt{t_+^f-q^2}-\sqrt{t_+^f-t_0}}{\sqrt{t_+^f-q^2}+\sqrt{t_+^f-t_0}},
\end{equation}
where we choose
\begin{equation}
t_0 = (m_{\Lambda_b} - m_{X})^2,
\end{equation}
so that the point $z=0$ corresponds to $q^2=q^2_{\rm max}$ (i.e. $\mathbf{p^\prime}=0$ in the $\Lambda_b$ rest frame). The values of 
$t_+^f$ are discussed further below. After factoring out the leading pole contribution, we expand the form factors in a power series in $z^f$. We find that our lattice data can be described well by keeping only the zeroth and first order
in $z^f$. As explained further below, we also perform higher-order fits to estimate systematic uncertainties. Our nominal (as opposed to higher-order) fits are of the form
\begin{eqnarray}
\nonumber f(q^2) &=& \frac{1}{1-q^2/(m_{\rm pole}^f)^2} \bigg[ a_0^f\bigg(1+c_0^f \frac{m_\pi^2-m_{\pi,{\rm phys}}^2}{\Lambda_\chi^2}\bigg) + a_1^f\:z^f(q^2)  \bigg] \\
 & & \times \bigg[1  + b^f\, \frac{|\mathbf{p^\prime}|^2}{(\pi/a)^2} + d^f\, \frac{\Lambda_{\rm QCD}^2}{(\pi/a)^2} \bigg],  \label{eq:FFccfit}
\end{eqnarray}
with fit parameters $a_0^f$, $a_1^f$, $c_0^f$, $b^f$, and $d^f$. Here, $m_\pi$ are the valence pion masses of each data set (see Table \ref{tab:params}), and
$m_{\pi,{\rm phys}}=134.8\:{\rm MeV}$ is the physical pion mass in the isospin limit \cite{Colangelo:2010et}. As discussed in Ref.~\cite{Detmold:2013nia}, chiral-perturbation-theory predictions
for the pion-mass dependence of the form factors considered here are unavailable and would be of limited use because of the large momentum scales in these matrix elements,
and because of the large number of low-energy constants. In Eq.~(\ref{eq:FFccfit}) we describe the pion-mass dependence
through the factor $1+c_0^f (m_\pi^2-m_{\pi,{\rm phys}}^2)/\Lambda_\chi^2$ multiplying
$a_0$. Here, we introduced the scale $\Lambda_\chi=4\pi f$ with $f=132\:{\rm MeV}$ so that $c_0$ becomes dimensionless. Because our lattice actions and currents are $\mathcal{O}(a)$-improved,
we allow for a quadratic dependence on the lattice spacing via the factor in the second line of Eq.~(\ref{eq:FFccfit}), where $\Lambda_{\rm QCD}=0.5\:\:{\rm GeV}$.
The parameters $b^f$ and $d^f$ describe the momentum-dependent and momentum-independent parts of the lattice discretization errors. We use the individual lattice QCD results
for the baryon masses from each dataset (see Table \ref{tab:hadronmasses}) to evaluate $a^2 q^2$ and $z$, and we take into account the uncertainties and correlations of these masses.
We set the pole masses equal to
\begin{equation}
 a m_{\rm pole}^f = a m_{\rm PS} + a \Delta^f,
\end{equation}
where $a m_{\rm PS}$ is the pseudoscalar $B_u$ or $B_c$ mass (in lattice units) computed individually for each data set (and also listed in Table \ref{tab:hadronmasses}),
and $\Delta^f$ is the mass splitting (in GeV) between the meson with the relevant quantum numbers and the pseudoscalar $B_u$ (for $\Lambda_b\to p$) or $B_c$
(for $\Lambda_b \to \Lambda_c$). We use fixed values of $\Delta^f$ for all data sets, based on experimental data (where available) \cite{Agashe:2014kda}
and averages of our lattice QCD results over the different data sets. These values are given in Table \ref{tab:polemasses}. The pole factor is then written as
\begin{equation}
 \frac{1}{1-(a^2 q^2)/(a m_{\rm PS} + a \Delta^f)^2}, \label{eq:polefactorlattice}
\end{equation}
so that the explicit value of the lattice spacing is needed only for the term $a \Delta^f$. Note that when the input values of $\Delta^f$ are varied, the shape
parameters $a_0^f$ and $a_1^f$ returned from the fit change in such a way as to largely cancel the effect of this variation on the form factors (varying $\Delta^f$ by
10\% changes the form factors themselves by less than 1\%).

The parameter $t_+^f$ should be set equal to or below the location of any singularities remaining after factoring out the leading pole contribution. The $\Lambda_b \to p$
and $\Lambda_b \to \Lambda_c$ form factors (in infinite volume) have branch cuts starting at $q^2=(m_B + m_\pi)^2$ and $q^2=(m_B + m_D)^2$, respectively.
In the case of $\Lambda_b \to p$, the form factors $f_0$, $g_+$, $g_\perp$ have no poles below this branch cut, and the form factors $f_+$, $f_\perp$, $g_0$
only have a single pole below $q^2=(m_B + m_\pi)^2$, which gets removed by the explicit factor of $1/[1-q^2/(m_{\rm pole}^f)^2]$.
We therefore set $t_+^f=(m_B + m_\pi)^2$ for all $\Lambda_b \to p$ form factors. More precisely, to evaluate the dimensionless quantity $a^2 t_+^f$ in the fit,
we use
\begin{equation}
 a^2 t_+^f = (a m_{\rm PS} + a m_{\pi,{\rm phys}})^2\hspace{4ex}(\text{for}\:\:\Lambda_b \to p),
\end{equation}
where $a m_{\rm PS}$ is the pseudoscalar $B_u$ mass (in lattice units) computed individually for each data set, and $m_{\pi,{\rm phys}}=134.8\:{\rm MeV}$. This means that the value of the lattice spacing
is needed only for the term $a m_{\pi,{\rm phys}}$. We checked that using the individual lattice pion masses of each data set instead of the physical pion mass has a
negligible effect on the extrapolated form factors.
In the case of $\Lambda_b \to \Lambda_c$, the onset of the branch cut, $m_B + m_D$, is several hundred MeV above the lowest pole for all form factors,
and there may be additional poles below $m_B + m_D$. We therefore set $t_+^f=(m_{\rm pole}^f)^2$ for the $\Lambda_b \to \Lambda_c$ form factors; more precisely,
\begin{equation}
 a^2 t_+^f = (a m_{\rm PS} + a \Delta^f)^2\hspace{4ex}(\text{for}\:\:\Lambda_b \to \Lambda_c),
\end{equation}
where $a m_{\rm PS}$ is the pseudoscalar $B_c$ mass (in lattice units) computed individually for each data set, as discussed above.
With this choice of $t_+^f$, the factors of $1/[1-q^2/(m_{\rm pole}^f)^2]$ are not strictly necessary, but we find that they improve
the quality of the fit at first order in the $z$ expansion.

We implement the constraint $g_\perp(q^2_{\rm max}) = g_+(q^2_{\rm max})$
[Eq.~(\ref{eq:FFC3})] at $z^{g_\perp,g_+}=0$ and $a=0$ by using shared parameters $a_0^{g_\perp,g_+}$ and $c_0^{g_\perp,g_+}$ for these two form factors. We impose the constraints
$f_0(0) = f_+(0)$ and $g_0(0) = g_+(0)$ [Eqs.~(\ref{eq:FFC1}) and (\ref{eq:FFC2})] using Gaussian priors with widths equal to $\mathrm{max}[z^{f_0}(0),z^{f_+}(0)]^2$
and $\mathrm{max}[z^{g_0}(0),z^{g_+}(0)]^2$, respectively, to allow for missing higher-order terms in $z^f$.
For $\Lambda_b \to p$, we performed one global fit to all helicity form factors, taking into account the correlations between different form factors, different momenta,
and different data sets. For $\Lambda_b \to \Lambda_c$, such a global fit showed indications of problems associated with a poorly conditioned data covariance matrix,
and we additionally performed fits of the subsets $\{ f_+, f_0 \}$, $\{ f_\perp \}$, $\{ g_+, g_\perp, g_0 \}$ to reduce the sizes of the data covariance matrices.
We then took the central values and covariances of the form factor parameters within each subset from these subset fits, and only used the global fit to estimate
the cross-covariances between the parameters in different subsets.

\begin{figure}
 \includegraphics[width=0.8\linewidth]{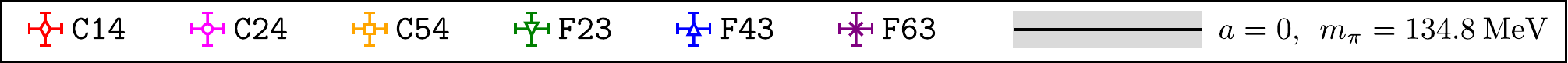} \\
 \includegraphics[width=\linewidth]{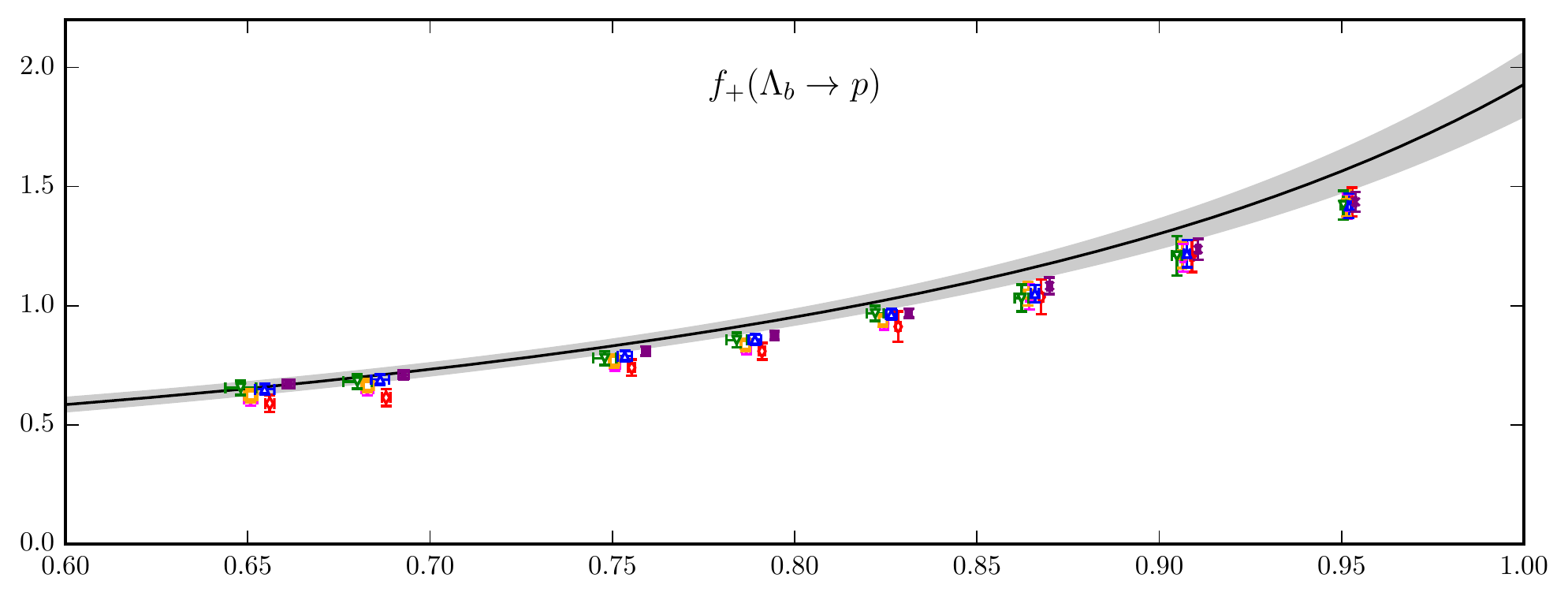} \\
 \includegraphics[width=\linewidth]{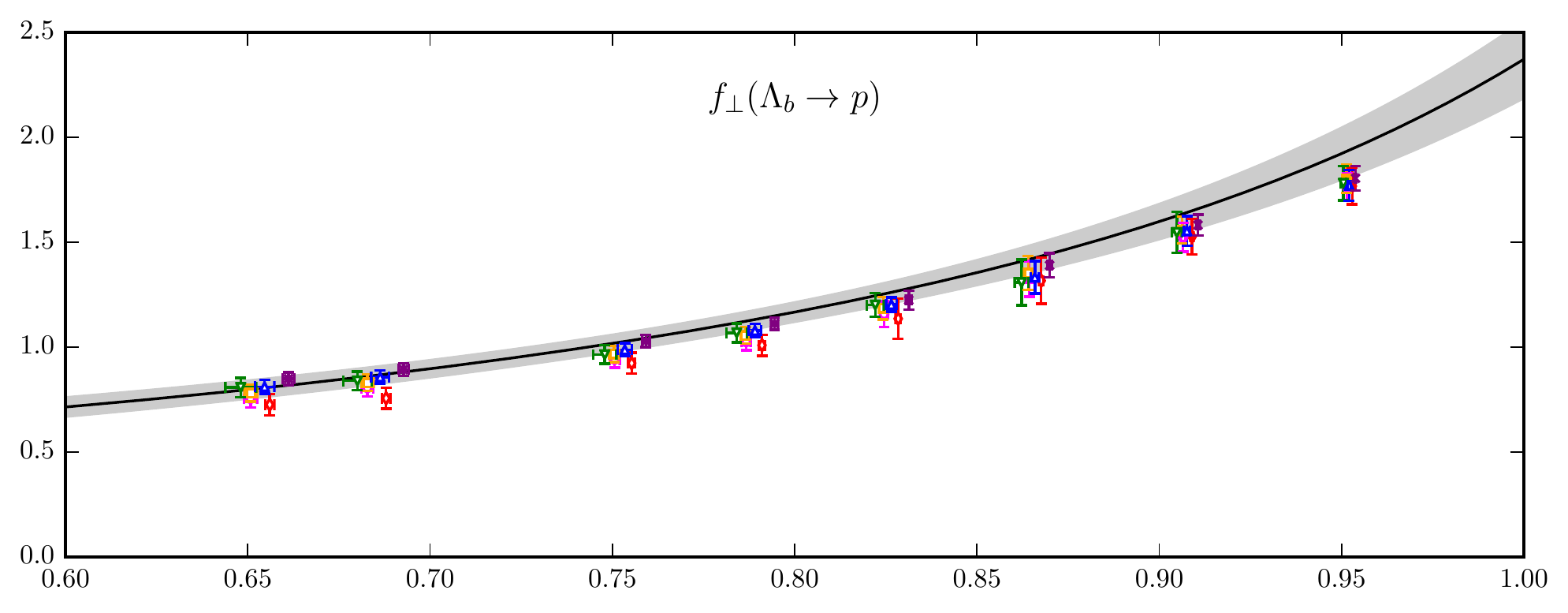} \\
 \includegraphics[width=\linewidth]{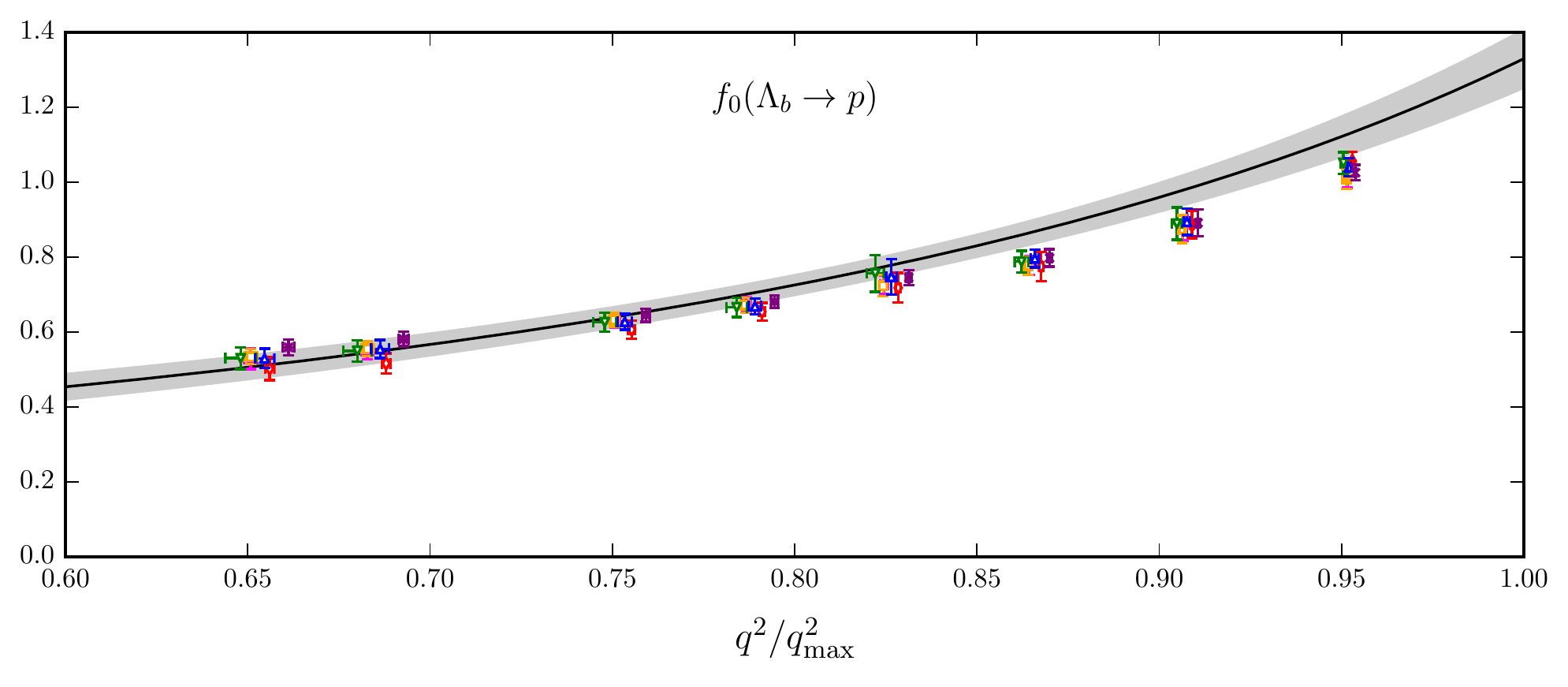}
 \caption{\label{fig:LbpVfit}$\Lambda_b\to p$ vector form factors: lattice results and extrapolation to the physical limit (nominal fit).
 The bands indicate the $1\sigma$ statistical uncertainty.}
\end{figure}

\begin{figure}
 \includegraphics[width=0.8\linewidth]{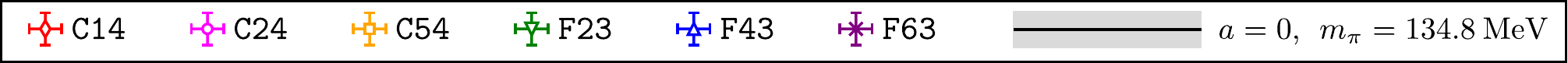} \\
 \includegraphics[width=\linewidth]{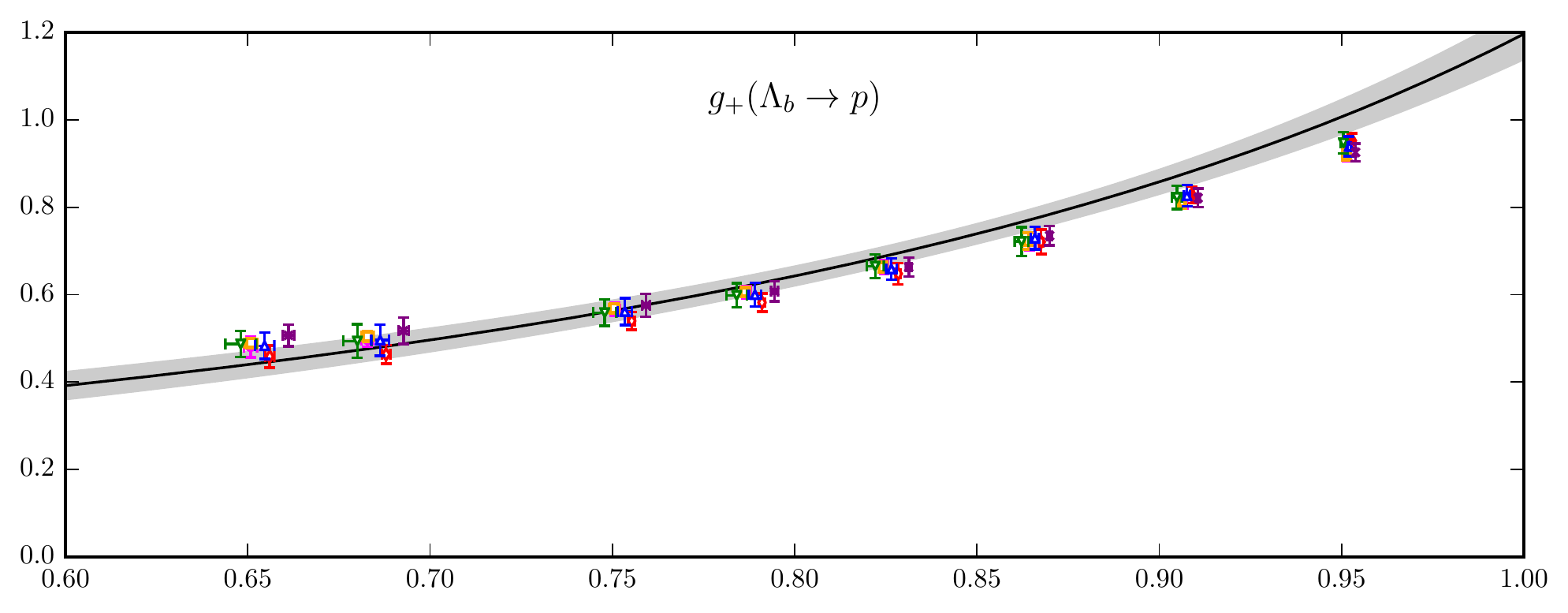} \\
 \includegraphics[width=\linewidth]{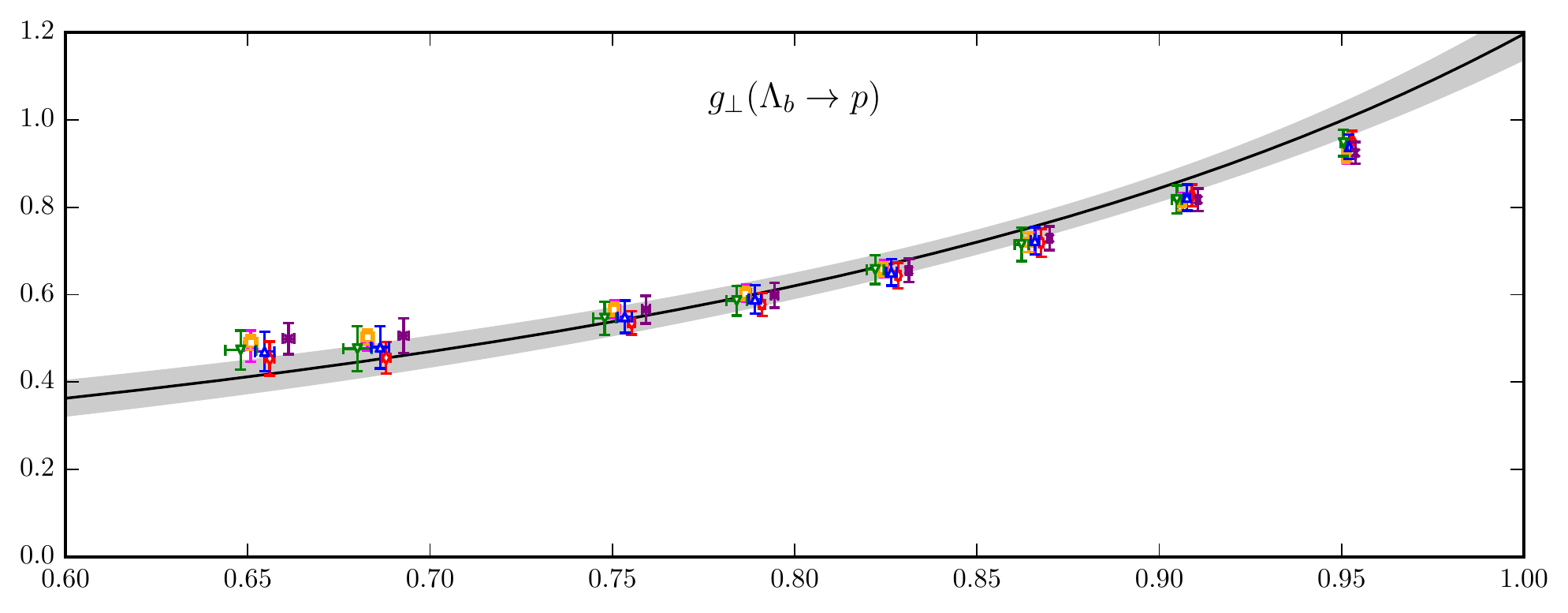} \\
 \includegraphics[width=\linewidth]{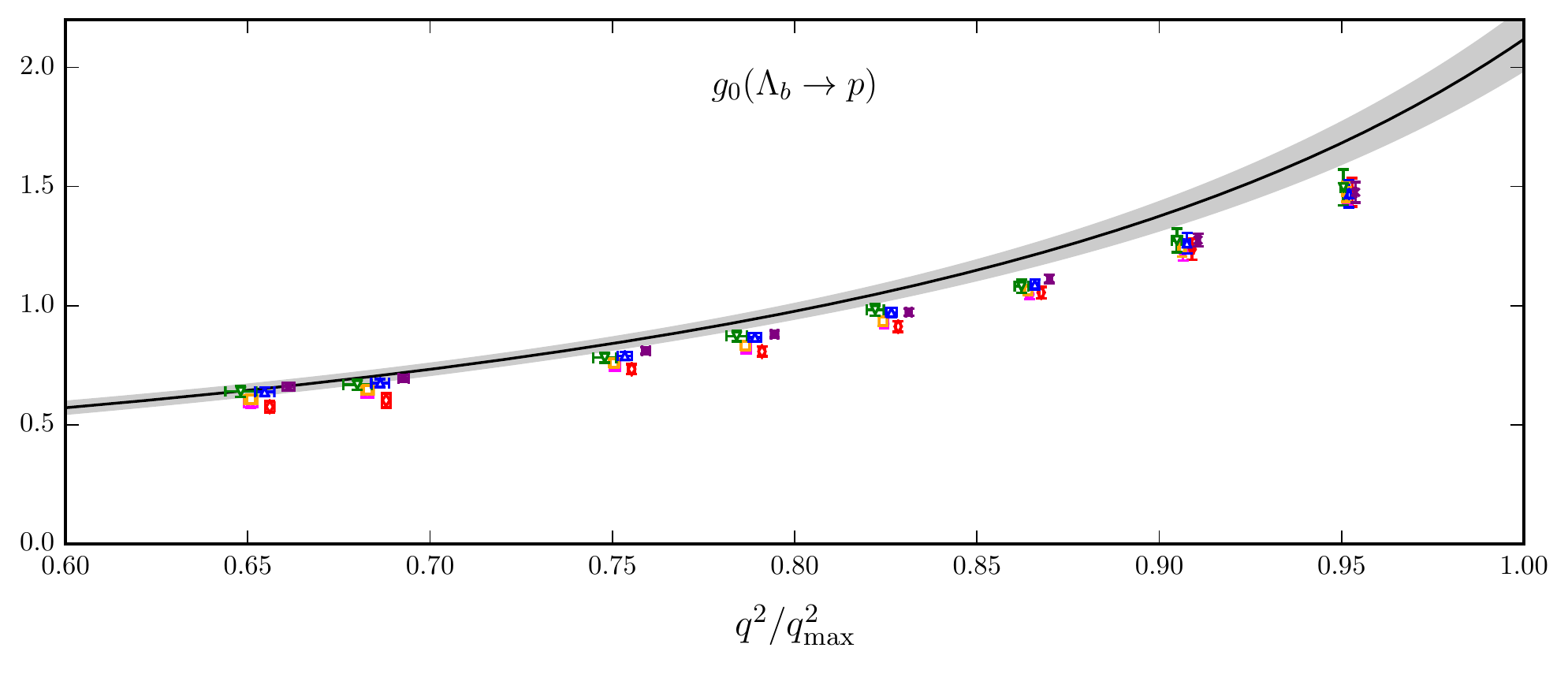}
 \caption{\label{fig:LbpAfit}$\Lambda_b\to p$ axial-vector form factors: lattice results and extrapolation to the physical limit (nominal fit).
 The bands indicate the $1\sigma$ statistical uncertainty.}
\end{figure}

\begin{figure}
 \includegraphics[width=0.8\linewidth]{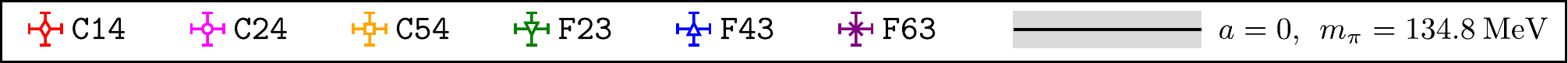} \\
 \includegraphics[width=\linewidth]{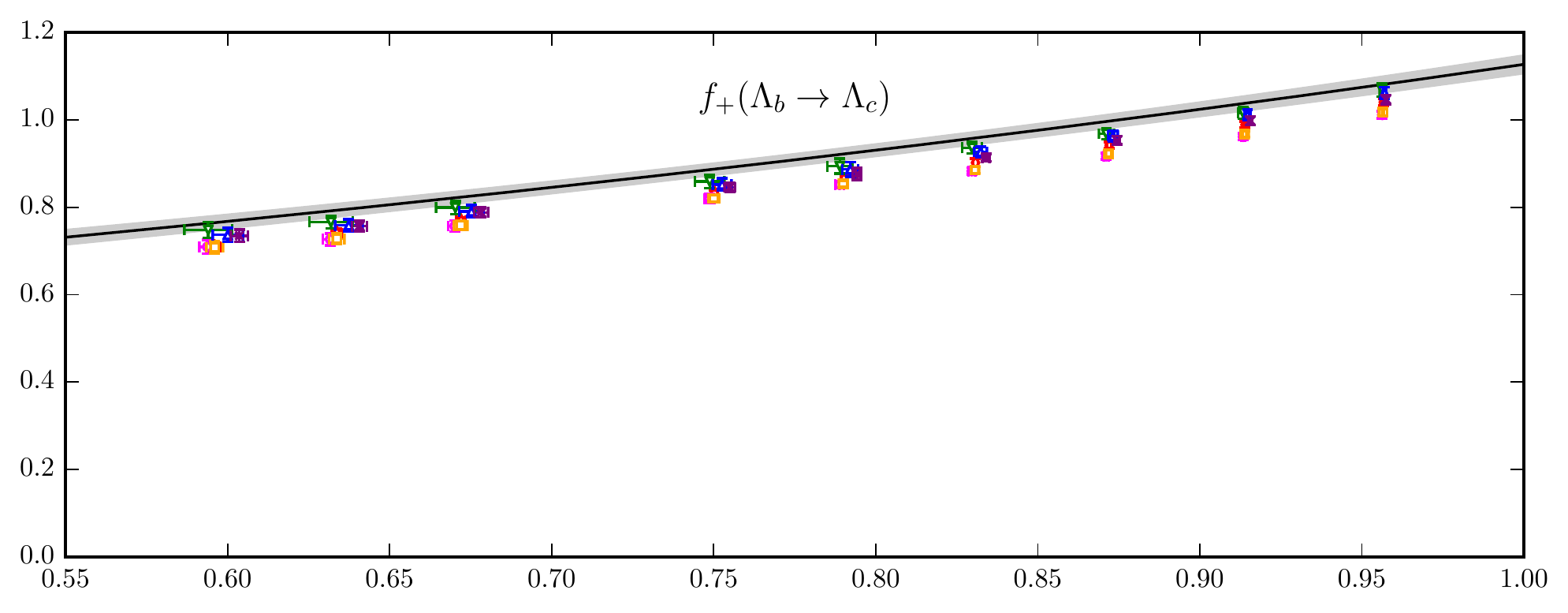} \\
 \includegraphics[width=\linewidth]{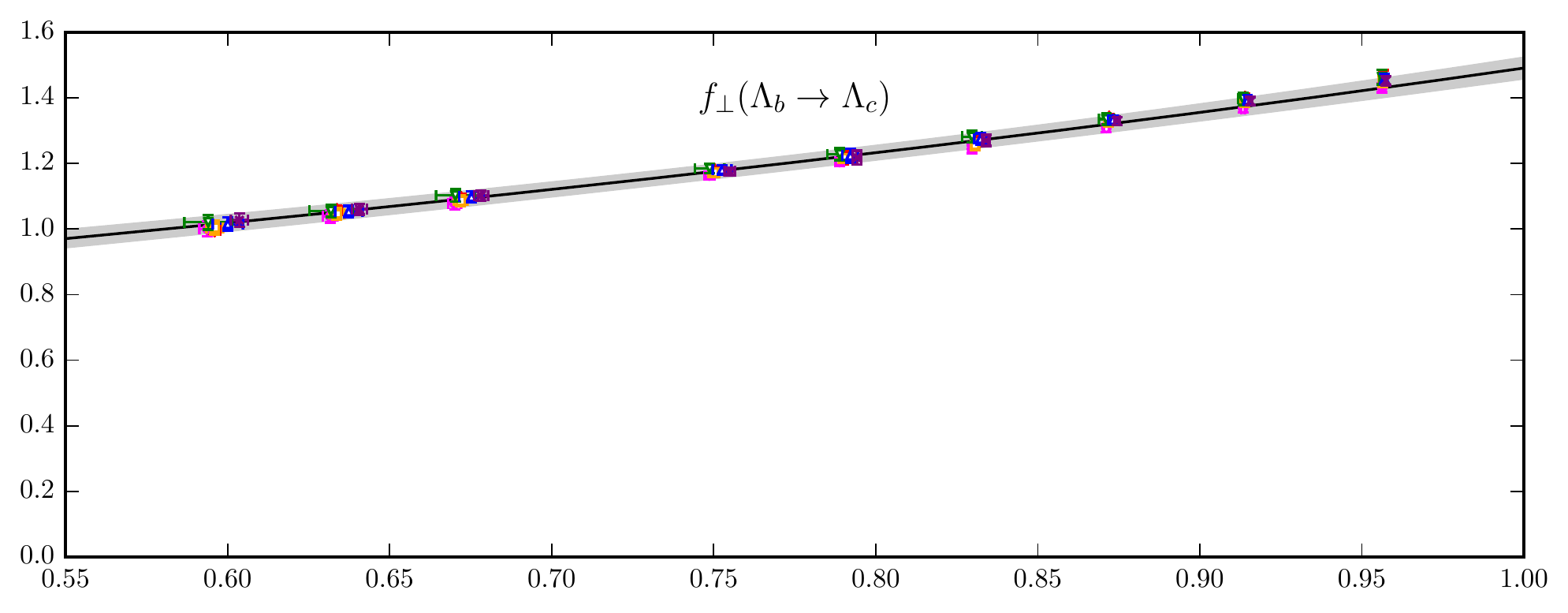} \\
 \includegraphics[width=\linewidth]{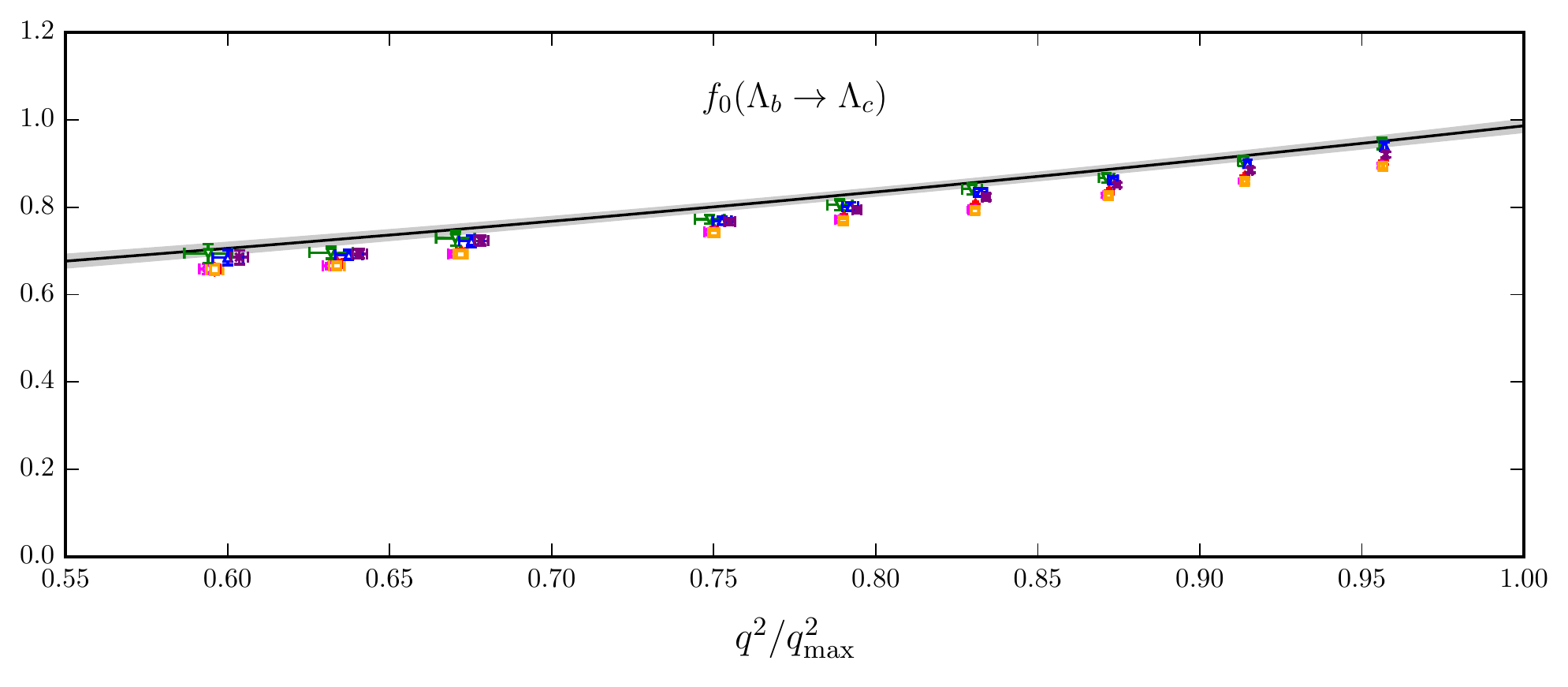}
 \caption{\label{fig:LbLcVfit}$\Lambda_b\to \Lambda_c$ vector form factors: lattice results and extrapolation to the physical limit (nominal fit).
 The bands indicate the $1\sigma$ statistical uncertainty.}
\end{figure}

\begin{figure}
 \includegraphics[width=0.8\linewidth]{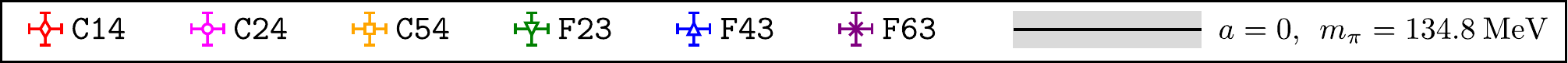} \\
 \includegraphics[width=\linewidth]{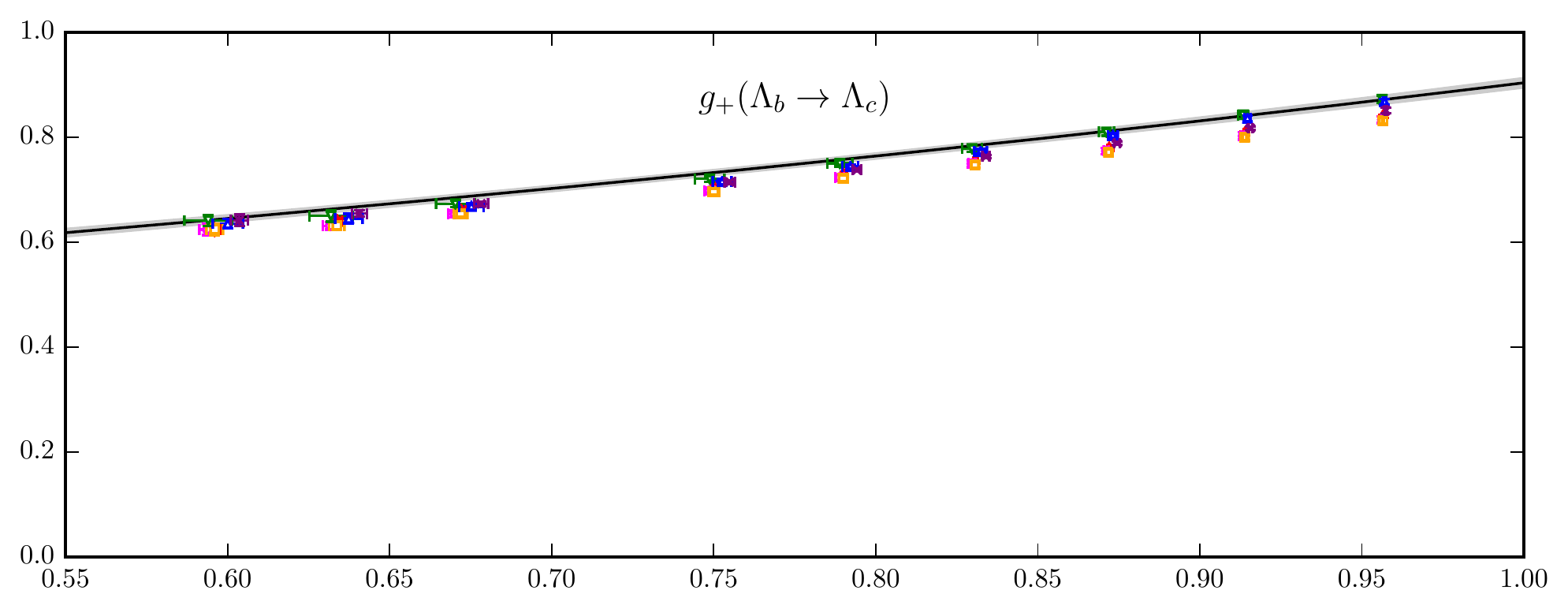} \\
 \includegraphics[width=\linewidth]{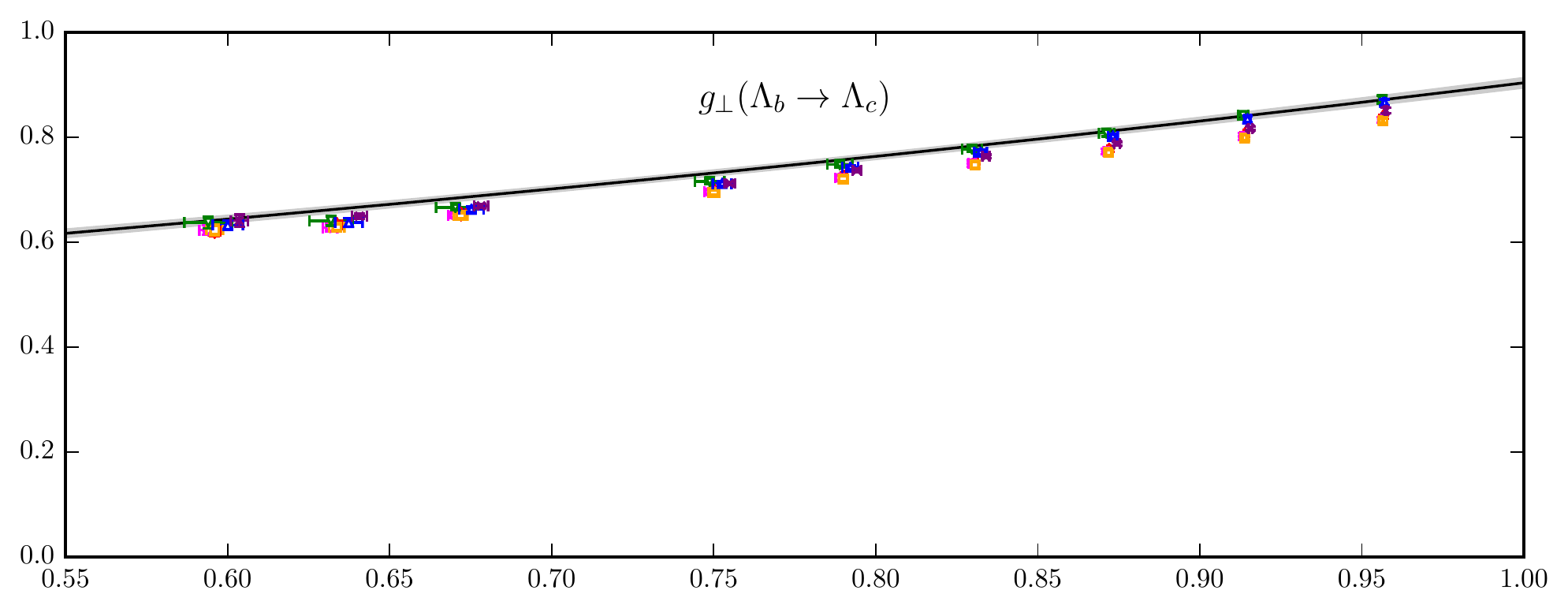} \\
 \includegraphics[width=\linewidth]{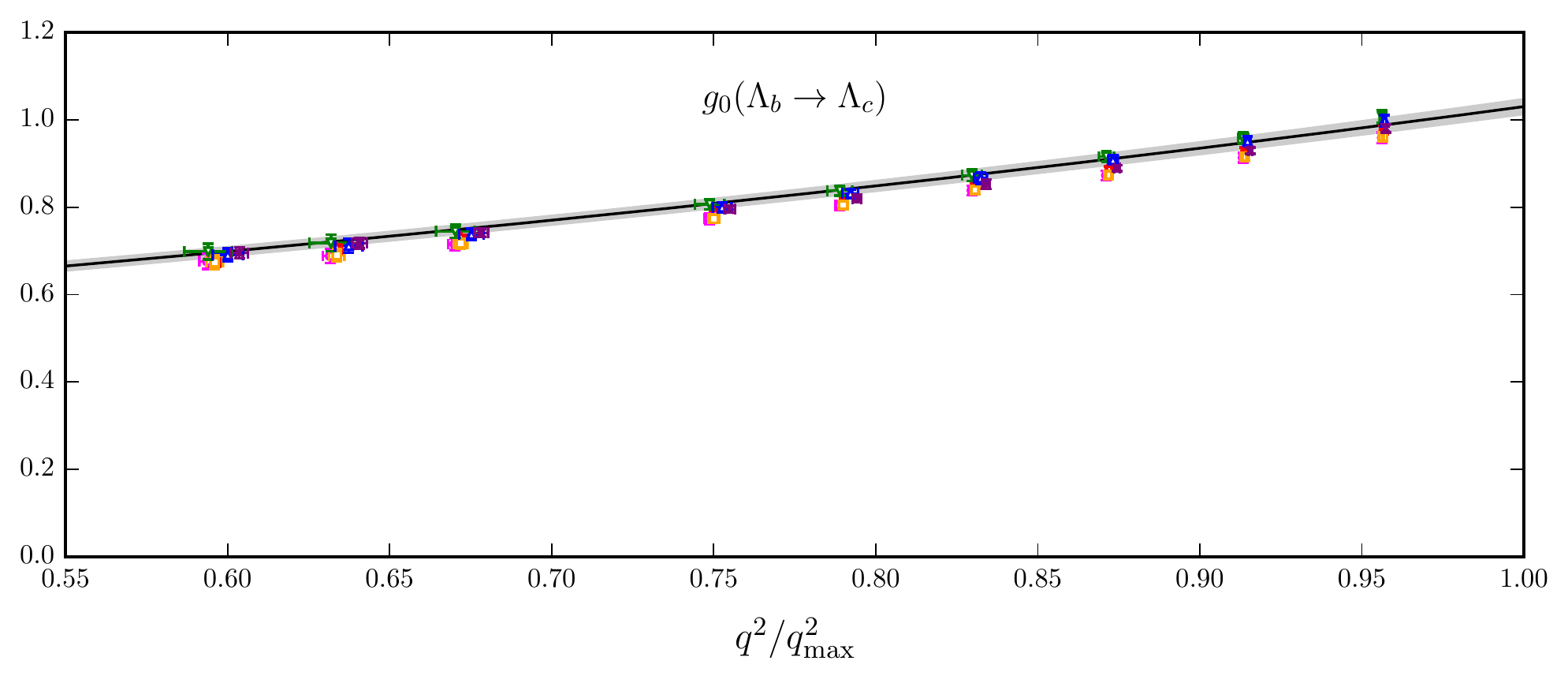}
 \caption{\label{fig:LbLcAfit}$\Lambda_b\to \Lambda_c$ axial-vector form factors: lattice results and extrapolation to the physical limit (nominal fit).
 The bands indicate the $1\sigma$ statistical uncertainty.}
\end{figure}

The physical limit is given by $a\to 0$ and $m_\pi \to m_{\pi,{\rm phys}}$, and correspondingly Eq.~(\ref{eq:FFccfit}) reduces to the simple form
\begin{eqnarray}
f(q^2) &=& \frac{1}{1-q^2/(m_{\rm pole}^f)^2} \big[ a_0^f + a_1^f\:z^f(q^2)  \big], \label{eq:nominalfitphys}
\end{eqnarray}
where $q^2$ should be evaluated using the experimental values of the baryon masses, and $m_{\rm pole}^f$, $t_+^f$ should be set to the values given in Table \ref{tab:polemasses},
with $m_B=5.279\:{\rm GeV}$, $m_\pi=134.8\:{\rm MeV}$, $m_{B_c}=6.276\:{\rm GeV}$.
The central values and uncertainties of the parameters $\{ a_0^f$, $a_1^f \}$ from the nominal fit are given in Table \ref{tab:nominal}, and the
correlation matrices are given in Table \ref{tab:nominalcorr}. The parameter \emph{covariances} ${\rm cov}(p,\:q)$ can be obtained from the correlations
${\rm corr}(p,\:q)$ and uncertainties $\sigma_p$, $\sigma_q$ using ${\rm cov}(p,\:q)=\sigma_p\:\sigma_q\:{\rm corr}(p,\:q)$; the central values and
covariance matrices of the fit parameters are also provided as ancillary files with the \texttt{arXiv} submission of this article. Plots of the lattice data along
with the physical-limit fit curves are shown in Figs.~\ref{fig:LbpVfit}, \ref{fig:LbpAfit},
\ref{fig:LbLcVfit}, and \ref{fig:LbLcAfit}.


\begin{table}
\begin{tabular}{ccccccccccccccc}
\hline\hline
 $f$     & \hspace{1ex} & $J^P$ & \hspace{1ex} & $t_+^f(\Lambda_b \to p)$  & \hspace{1ex} & $m_{\rm pole}^f(\Lambda_b \to p)$   & \hspace{1ex} &  $\Delta^f(\Lambda_b \to p)$ & \hspace{4ex} & $t_+^f(\Lambda_b \to \Lambda_c)$  & \hspace{1ex} & $m_{\rm pole}^f(\Lambda_b \to \Lambda_c)$ & \hspace{1ex} & $\Delta^f(\Lambda_b \to \Lambda_c)$ \\
\hline
$f_+$, $f_\perp$ && $1^-$ && $(m_B + m_\pi)^2$      && $m_{B} + \Delta^f$ && $\nb$46 MeV && $(m_{\rm pole}^f)^2$ && $m_{B_c} + \Delta^f$ && $\nb$56 MeV \\
$f_0$            && $0^+$ && $(m_B + m_\pi)^2$      && $m_{B} + \Delta^f$ && 377 MeV     && $(m_{\rm pole}^f)^2$ && $m_{B_c} + \Delta^f$ && 449 MeV     \\
$g_+$, $g_\perp$ && $1^+$ && $(m_B + m_\pi)^2$      && $m_{B} + \Delta^f$ && 427 MeV     && $(m_{\rm pole}^f)^2$ && $m_{B_c} + \Delta^f$ && 492 MeV     \\
$g_0$            && $0^-$ && $(m_B + m_\pi)^2$      && $m_{B} + \Delta^f$ && 0           && $(m_{\rm pole}^f)^2$ && $m_{B_c} + \Delta^f$ && 0           \\
\hline\hline
\end{tabular}
\caption{\label{tab:polemasses} Values of $t_+^f$ and $m_{\rm pole}^f$. To evaluate the form factors in the physical limit, $m_B=5.279\:{\rm GeV}$, $m_\pi=134.8\:{\rm MeV}$, and $m_{B_c}=6.276\:{\rm GeV}$ should be used.}
\end{table}

\begin{table}
\begin{tabular}{lllll}
\hline \hline
 Parameter         &  & \hspace{5ex} $\Lambda_b \to p$  & \hspace{1ex}  & \hspace{5ex}  $\Lambda_b \to \Lambda_c$    \\
\hline
$a_0^{f_+}$ && $\wm 0.4382\pm 0.0315$  && $\wm 0.8146\pm 0.0167$ \\ 
$a_1^{f_+}$ && $-0.6452\pm 0.2093$  && $-4.8990\pm 0.5425$ \\
$a_0^{f_0}$ && $\wm 0.4189\pm 0.0256$  && $\wm 0.7439\pm 0.0125$ \\ 
$a_1^{f_0}$ && $-0.7862\pm 0.2038$  && $-4.6480\pm 0.6084$ \\ 
$a_0^{f_\perp}$ && $\wm 0.5389\pm 0.0435$  && $\wm 1.0780\pm 0.0256$ \\ 
$a_1^{f_\perp}$ && $-0.8069\pm 0.3039$  && $-6.4170\pm 0.8480$ \\
$a_0^{g_\perp,g_+}$ && $\wm 0.3912\pm 0.0198$  && $\wm 0.6847\pm 0.0086$ \\ 
$a_1^{g_+}$ && $-0.8167\pm 0.1749$  && $-4.4310\pm 0.3572$ \\ 
$a_0^{g_0}$ && $\wm 0.4526\pm 0.0292$  && $\wm 0.7396\pm 0.0143$ \\ 
$a_1^{g_0}$ && $-0.7817\pm 0.1886$  && $-4.3660\pm 0.3314$ \\ 
$a_1^{g_\perp}$ && $-0.9061\pm 0.1956$  && $-4.4630\pm 0.3613$ \\
\hline\hline
\end{tabular}
\caption{\label{tab:nominal}Central values and uncertainties of the nominal form factor parameters for $\Lambda_b \to p$ and $\Lambda_b \to \Lambda_c$.
See Table \protect\ref{tab:nominalcorr} for the correlation matrices.}
\end{table}

To estimate the systematic uncertainties caused by our assumptions on the lattice-spacing, quark-mass, and $q^2$-dependence, we also perform fits that include additional
higher-order terms, employing the form
\begin{eqnarray}
\nonumber f_{\rm HO}(q^2) &=& \frac{1}{1-q^2/(m_{\rm pole}^f)^2} \bigg[ a_0^f\bigg(1+c_0^f \frac{m_\pi^2-m_{\pi,{\rm phys}}^2}{\Lambda_\chi^2}+\tilde{c}_0^f \frac{m_\pi^3-m_{\pi,{\rm phys}}^3}{\Lambda_\chi^3}\bigg) + a_1^f\bigg(1+c_1^f\frac{m_\pi^2-m_{\pi,{\rm phys}}^2}{\Lambda_\chi^2}\bigg)\:z^f(q^2)  \\
 & &  + a_2^f\:[z^f(q^2)]^2 \bigg] \bigg[1  + b^f\, \frac{|\mathbf{p^\prime}|^2}{(\pi/a)^2} + d^f\, \frac{\Lambda_{\rm QCD}^2}{(\pi/a)^2}
                     + \tilde{b}^f\, \frac{|\mathbf{p^\prime}|^3}{(\pi/a)^3}
                     + \tilde{d}^f\, \frac{\Lambda_{\rm QCD}^3}{(\pi/a)^3}
                     + j^f   \frac{|\mathbf{p^\prime}|^2\Lambda_{\rm QCD}}{(\pi/a)^3}
                     + k^f \frac{|\mathbf{p^\prime}|\Lambda_{\rm QCD}^2}{(\pi/a)^3} \bigg]. \hspace{5ex}  \label{eq:FFccfitHO}
\end{eqnarray}
This allows for higher-order variation in the lattice spacing, quark masses, and momentum dependence. The data themselves do not determine this more complex form sufficiently
well, so we constrain the higher-order coefficients $\tilde{c}_0^f$, $c_1^f$, $\tilde{b}^f$, $\tilde{d}^f$, $j^f$, $k^f$ to be natural-sized using Gaussian priors with central value
0 and width 10. We constrain the second-order $z$-expansion coefficients $a_2^f$ using Gaussian priors with central values 0 and widths given by approximately twice the magnitude of the previous (nominal) fit results for $a_1^f$.
Given that this fit is quadratic in $z^f$, we now impose the kinematic constraints (\ref{eq:FFC1}) and (\ref{eq:FFC2}) at $q^2=0$ up to widths of $\mathrm{max}[z^{f_0}(0),z^{f_+}(0)]^3$ and $\mathrm{max}[z^{g_0}(0),z^{g_+}(0)]^3$, respectively.

In the higher-order fit, we use bootstrap data for the correlator ratios in which the matching- and $\mathcal{O}(a)$-improvement coefficients were drawn from Gaussian random distributions
with central values and widths according to Table \ref{tab:Pmatchingfactors}. Thus, the higher-order fit results also include the perturbation-theory systematic uncertainty.
To take into account the uncertainties of the lattice spacings, we promote the lattice spacings of the different ensembles to fit parameters, constrained with Gaussian priors
according to the central values and uncertainties given in Table \ref{tab:params}.
The systematic uncertainties caused by the finite lattice volume cannot easily be estimated from the data, because all of our data sets have approximately
the same lattice size, $L\approx 2.7$ fm.
Finite-volume effects have been calculated using chiral perturbation theory for the nucleon magnetic moment \cite{Beane:2004tw}
and axial charge \cite{Beane:2004rf}, and, specifically for the ensembles used herein, for the heavy-baryon axial couplings \cite{Detmold:2012ge, Detmold:2011rb}. Based on this experience, we estimate that
the finite-volume systematic uncertainties in our results are 3\% for the $\Lambda_b \to p$ form factors, and 1.5\% for the $\Lambda_b \to \Lambda_c$ form factors.
The neglected isospin breaking effects in the form factors are estimated to be of order $\mathcal{O}((m_d-m_u)/\Lambda_{\rm QCD})\approx 0.5\%$ and $\mathcal{O}(\alpha_{\rm e.m.})\approx 0.7\%$.
Finally, there is an uncertainty resulting from the tuning of the relativistic heavy quark (RHQ) parameters, which was performed in Ref.~\cite{Aoki:2012xaa} for the $b$ quark and in Ref.~\cite{Brown:2014ena}
for the $c$ quark. In Ref.~\cite{Flynn:2015mha}, the same $b$-quark parameters were used to compute the $B\to\pi$ form factors on the same gauge field configurations as in the present work,
and the uncertainties of $a m_Q^{(b)}$, $\xi^{(b)}$, and $c_{E,B}^{(b)}$ were propagated to the form factors by repeating the calculation for multiple values of $a m_Q^{(b)}$, $\xi^{(b)}$, and $c_{E,B}^{(b)}$.
The resulting uncertainties in the $B\to\pi$ form factors were found to be 1\%. We could not afford to repeat the present calculation for multiple values of the RHQ parameters, and
therefore adopt the 1\% estimate also for the $b$-quark parameter uncertainty in the $\Lambda_b \to p$ and $\Lambda_b \to \Lambda_c$ form factors.
We are unable to estimate the $c$-quark parameter uncertainty in the $\Lambda_b \to \Lambda_c$ form factors at this time, but we note that our choice of parameters
precisely reproduces the experimental values of the charmonium masses and hyperfine splittings, as well as the relativistic continuum dispersion relation,
on both the coarse and the fine lattices \cite{Brown:2014ena}. To estimate the effect of the light-quark-mass uncertainties, we promote the pion masses (in lattice units)
to fit parameters, constrained with Gaussian priors according to the central values and widths given in Ref.~\cite{Aoki:2010dy}. We find that this results in a smaller than $0.1\%$
uncertainty in the $\Lambda_b \to p$ and $\Lambda_b \to \Lambda_c$ form factors.
To incorporate the finite-volume, isospin breaking, and RHQ parameter tuning uncertainties in the higher-order fit, we add these uncertainties to the data covariance matrix before performing
the fit. We assume that these uncertainties are 100\% correlated between the different data sets and different final-state momenta, and between the three different form factors
corresponding to the same type of current (vector or axial vector).

In the physical limit, the higher-order fit functions reduce to
\begin{eqnarray}
f_{\rm HO}(q^2) &=& \frac{1}{1-q^2/(m_{\rm pole}^f)^2} \big[ a_0^f + a_1^f\:z^f(q^2) + a_2^f\:[z^f(q^2)]^2  \big]. \label{eq:HOfitphys}
\end{eqnarray}
The values of the parameters $a_0^f$, $a_1^f$, $a_2^f$, their total uncertainties, and their correlation matrices are given in Tables \ref{tab:HO} and \ref{tab:HOcorr},
and are also included as ancillary files with the \texttt{arXiv} submission.
The recommended procedure for computing the central value, statistical uncertainty, and total systematic uncertainty of a general observable depending on the form factor parameters
(for example, a differential decay rate at a particular value of $q^2$, or an integrated decay rate, or a ratio of decay rates) is the following:
\begin{enumerate}
 \item Compute the observable and its uncertainty using the nominal form factors given by Eq.~(\ref{eq:nominalfitphys}), with the parameter values and correlation matrices from Tables
 \ref{tab:nominal} and \ref{tab:nominalcorr}. Denote the so-obtained central value and uncertainty as \vspace{-2ex}
 \begin{equation}
  O, \:\:\sigma_O. \label{eq:O}
 \end{equation}
  \item Compute the same observable and its uncertainty using the higher-order form factors given by Eq.~(\ref{eq:HOfitphys}), with the parameter values and correlation
  matrices from Tables \ref{tab:HO} and \ref{tab:HOcorr}. Denote the so-obtained central value and uncertainty as \vspace{-2ex}
 \begin{equation}
  O_{\rm HO}, \:\:\sigma_{O,{\rm HO}}.  \label{eq:OHO}
 \end{equation}
  \item The final result for the observable is then given by
  \begin{equation}
  O \:\pm\: \underbrace{\sigma_O}_{\rm stat.} \:\pm \:\underbrace{{\rm max}\left( |O_{\rm HO}-O|,\: \sqrt{|\sigma_{O,{\rm HO}}^2-\sigma_O^2|}  \right)}_{\rm syst.}. \label{eq:finalres}
 \end{equation}
\end{enumerate}
In other words, the central value and statistical uncertainty are obtained from the nominal fit, and the systematic uncertainty is given by the larger of the following two quantities: i) the shift in
the central value between the nominal fit and the higher-order fit, and ii) the increase in the uncertainty (computed in quadrature as shown above) from the nominal fit to the higher-order fit.
The statistical and systematic uncertainties in Eq.~(\ref{eq:finalres}) should be added in quadrature. By construction, the above procedure gives the combined systematic uncertainty associated
with the continuum extrapolation, chiral extrapolation, $z$ expansion, perturbative matching, scale setting, $b$-quark parameter tuning, finite volume, and missing isospin symmetry breaking/QED.

\begin{figure}
 \includegraphics[height=5.3cm]{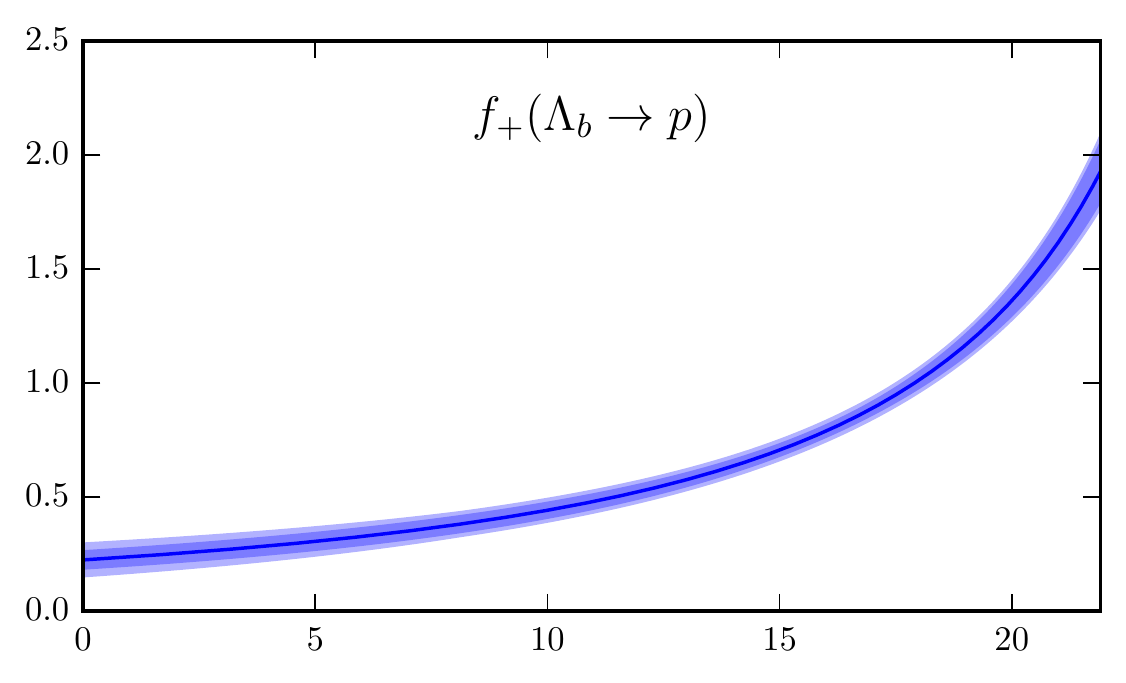} \hfill
 \includegraphics[height=5.3cm]{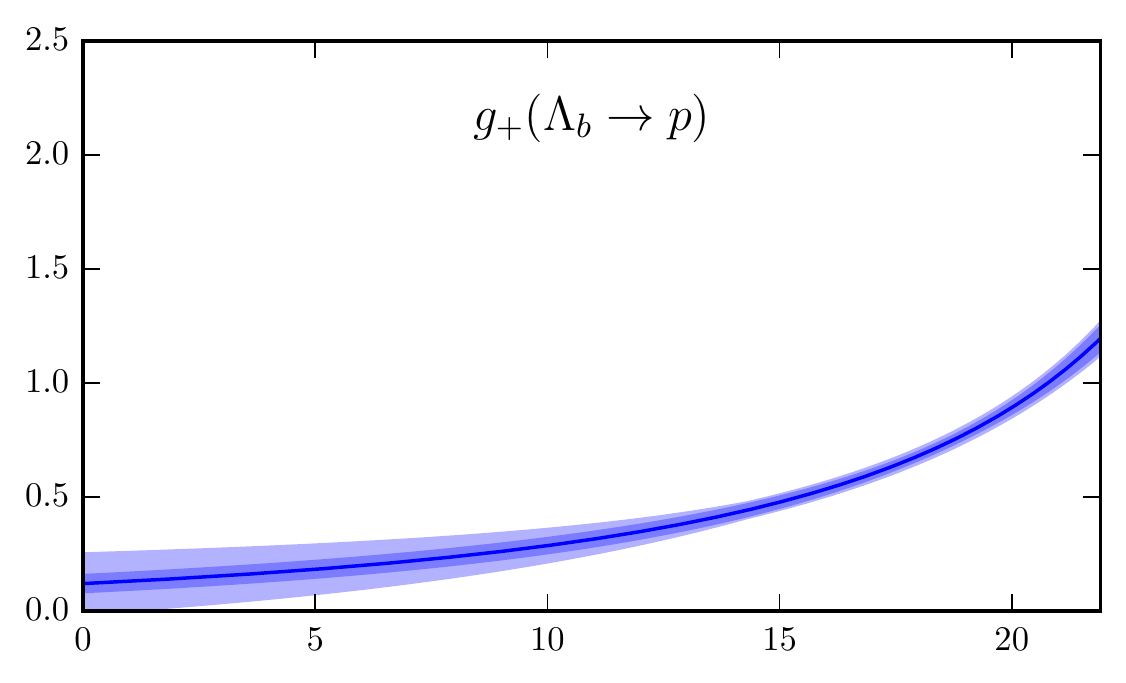} \\
 \includegraphics[height=5.3cm]{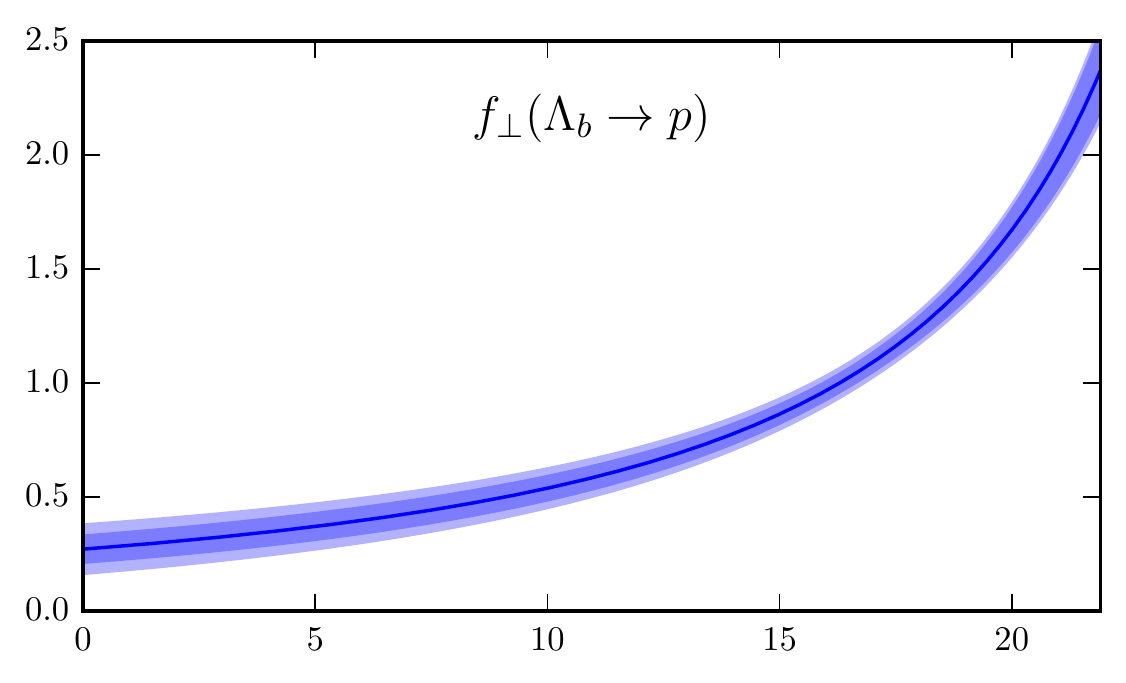} \hfill
 \includegraphics[height=5.3cm]{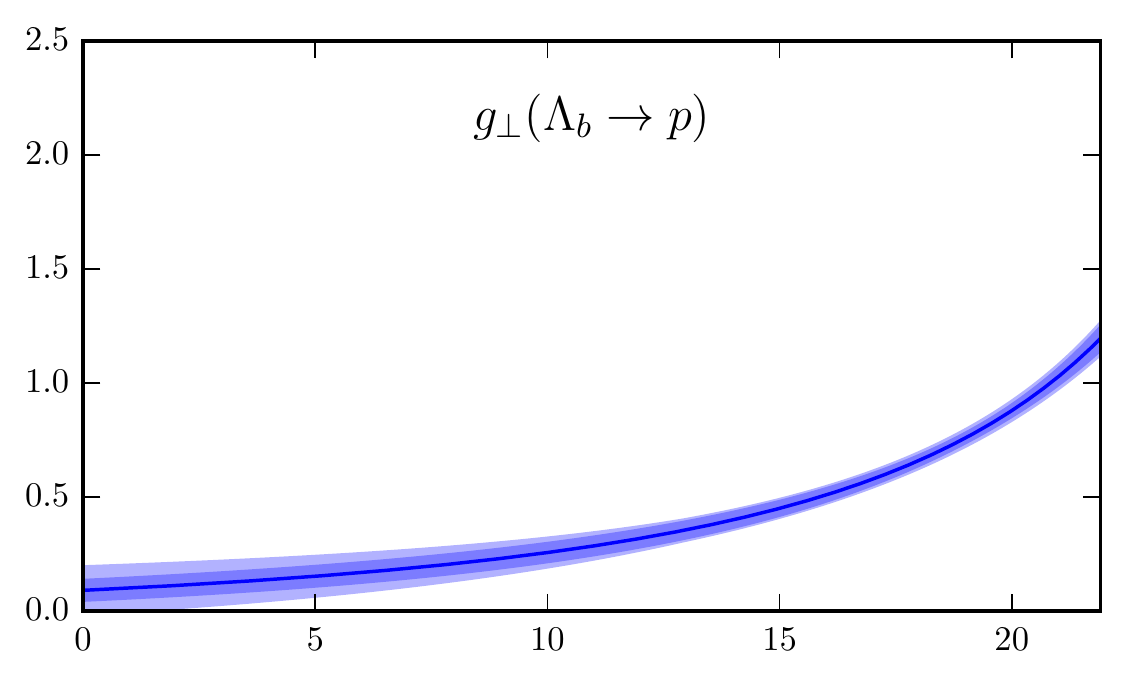} \\
 \includegraphics[height=6.1cm]{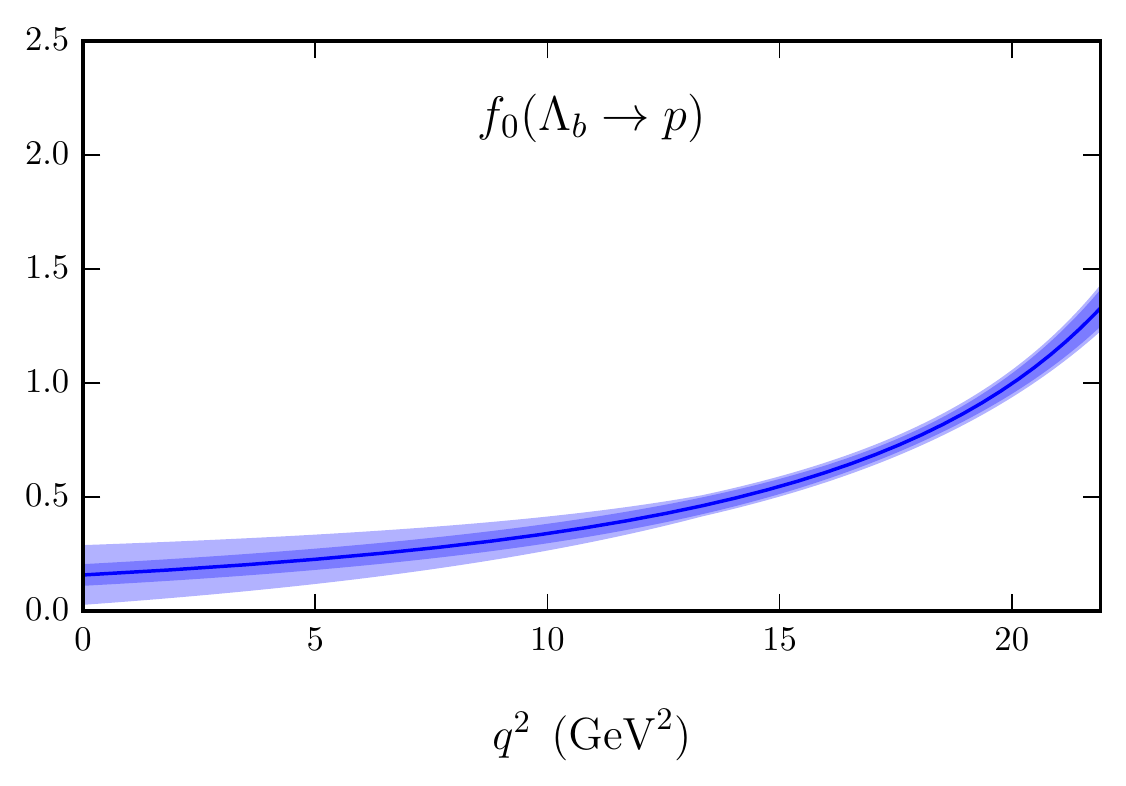} \hfill
 \includegraphics[height=6.1cm]{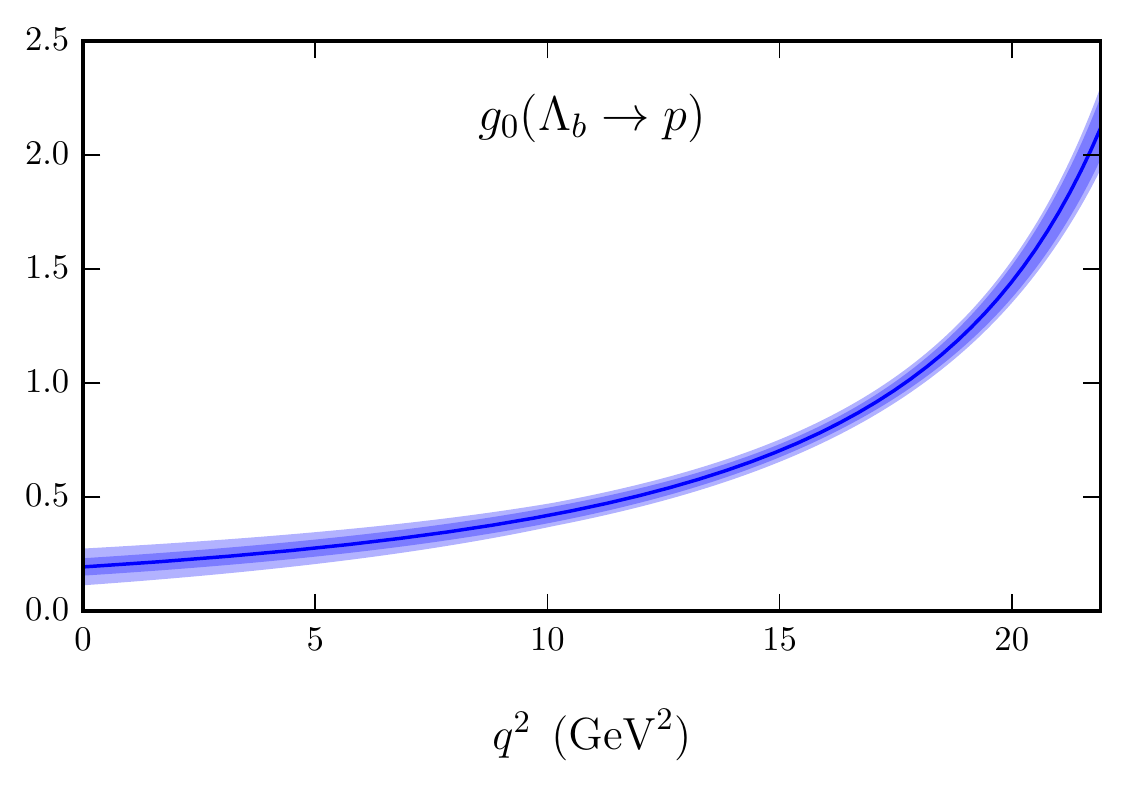} \\
 \caption{\label{fig:finalFFsLbp}Final results for the $\Lambda_b\to p$ form factors. The inner bands show the statistical uncertainty and the outer
 bands show the total uncertainty.}
\end{figure}

\begin{figure}
 \hspace{1cm}\includegraphics[width=\linewidth]{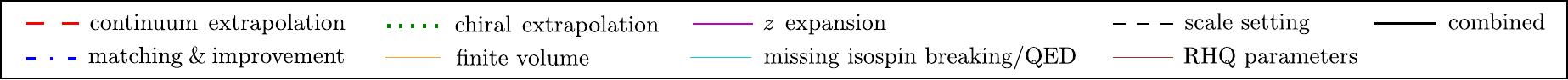} \\
 \includegraphics[height=5.4cm]{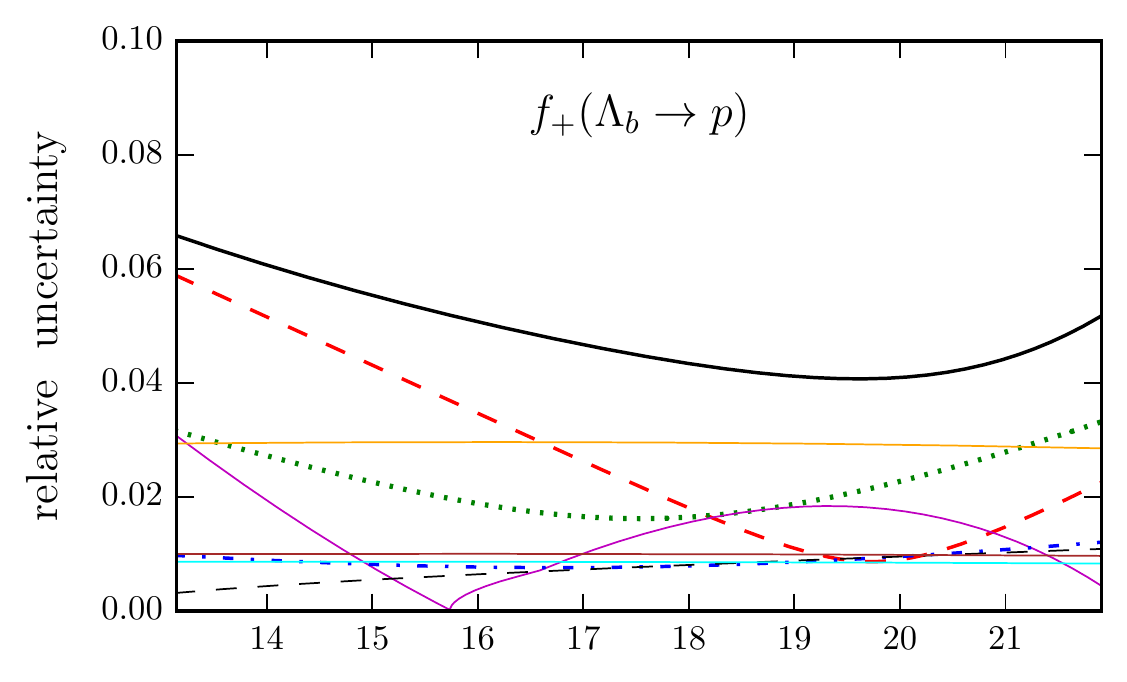} \hfill
 \includegraphics[height=5.4cm]{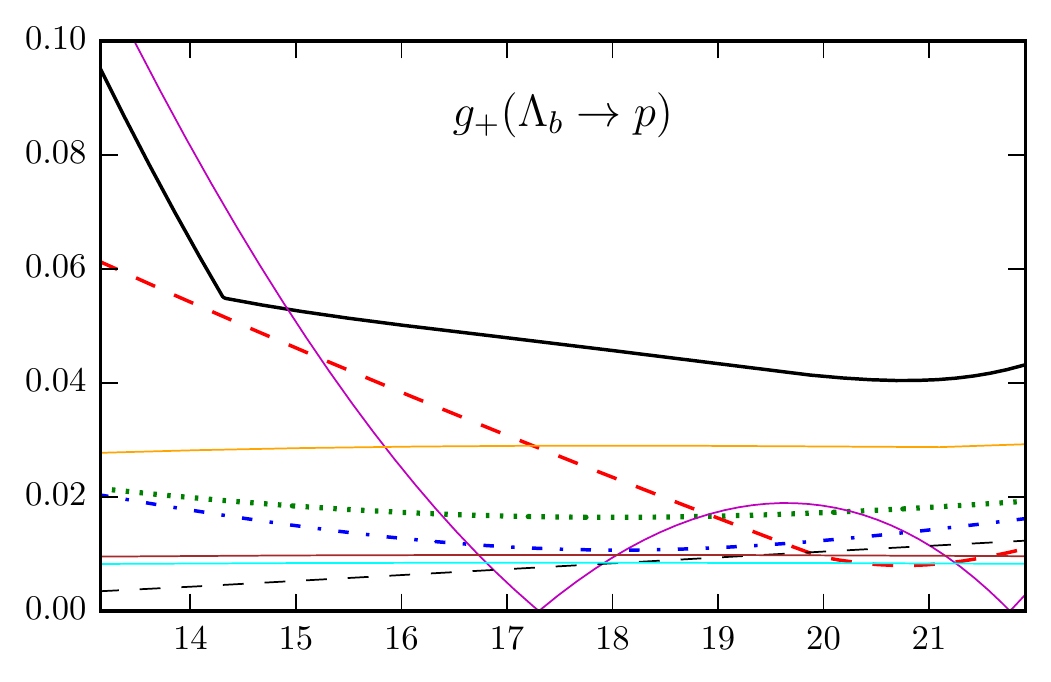} \\
 \includegraphics[height=5.4cm]{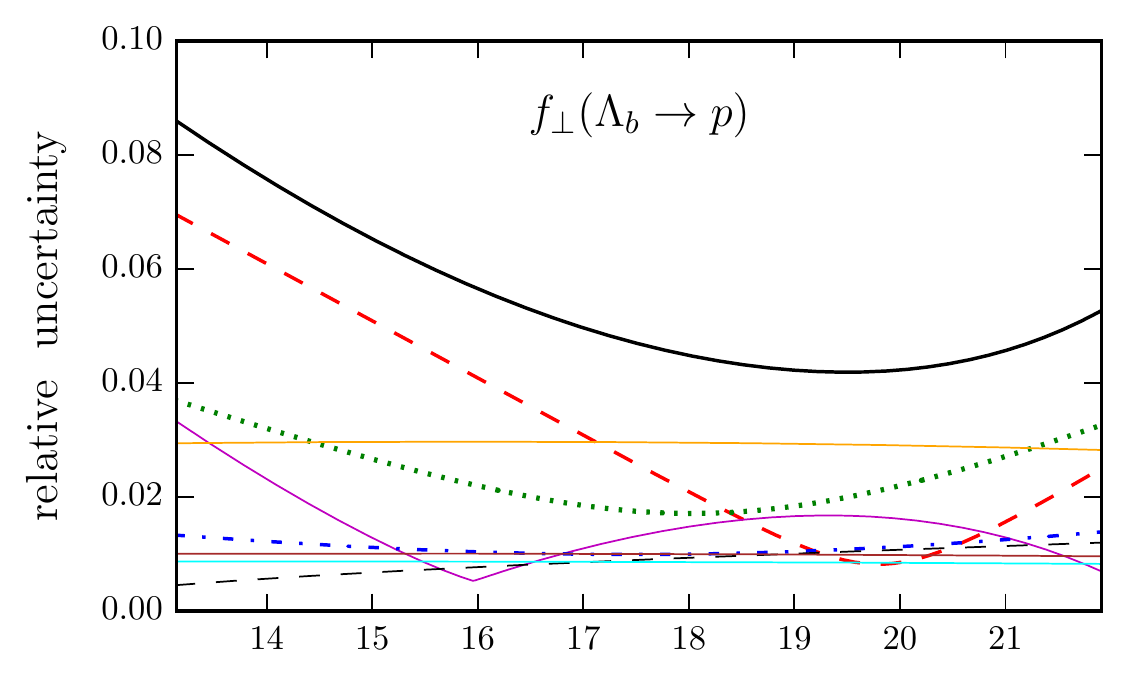} \hfill
 \includegraphics[height=5.4cm]{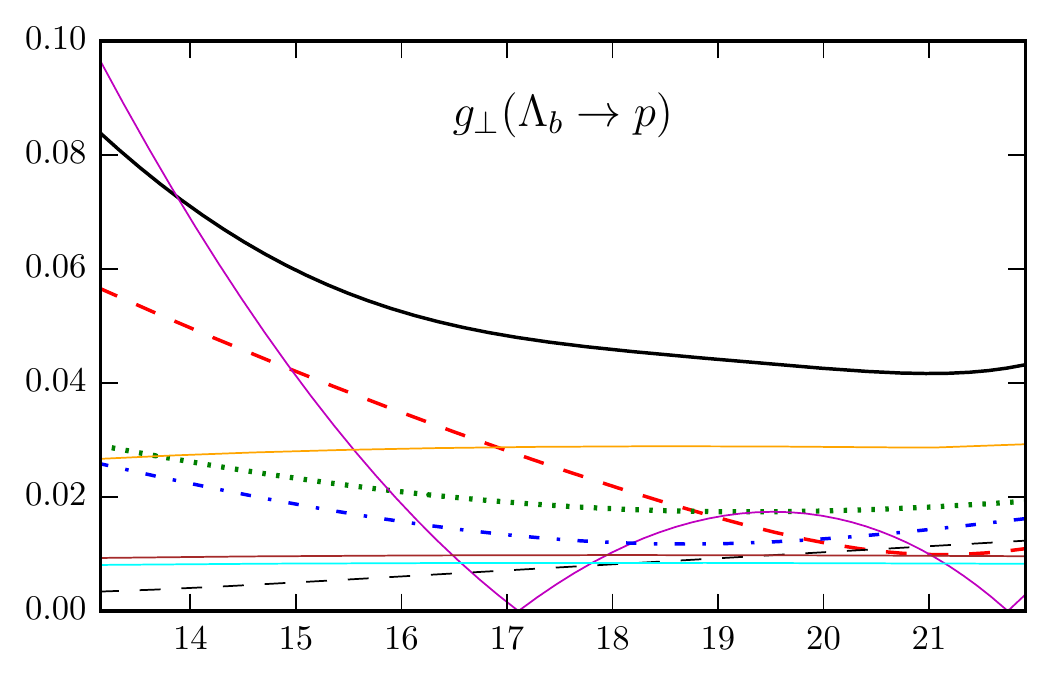} \\
 \includegraphics[height=6.2cm]{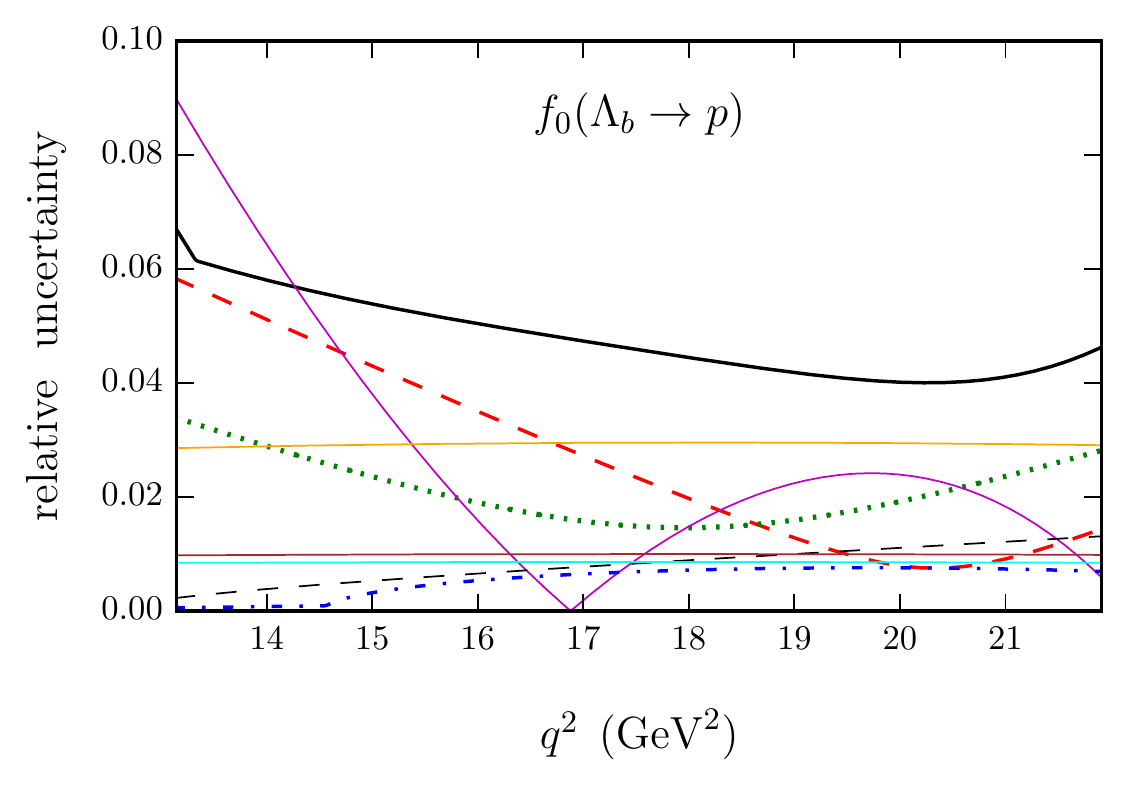} \hfill
 \includegraphics[height=6.2cm]{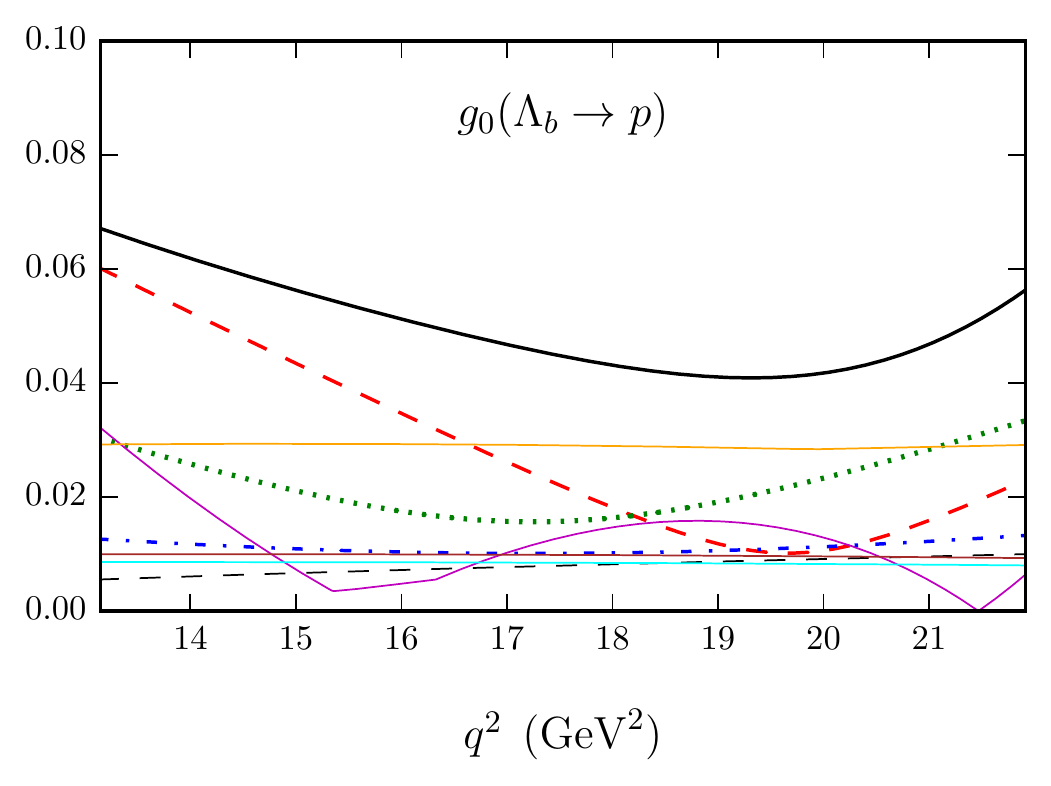}
 \caption{\label{fig:systLbp}Systematic uncertainties in the $\Lambda_b\to p$ form factors in the high-$q^2$ region. As explained in the main text,
 the combined uncertainty is not simply the quadratic sum of the individual uncertainties.}
\end{figure}

\begin{figure}
 \includegraphics[height=5.3cm]{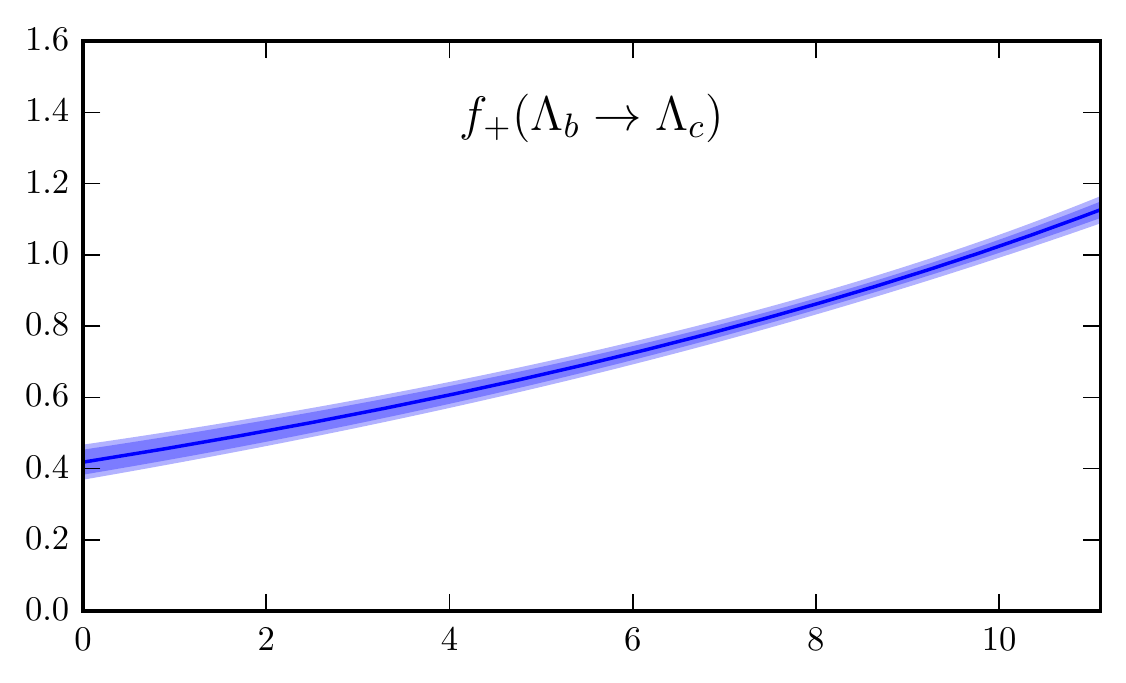} \hfill
 \includegraphics[height=5.3cm]{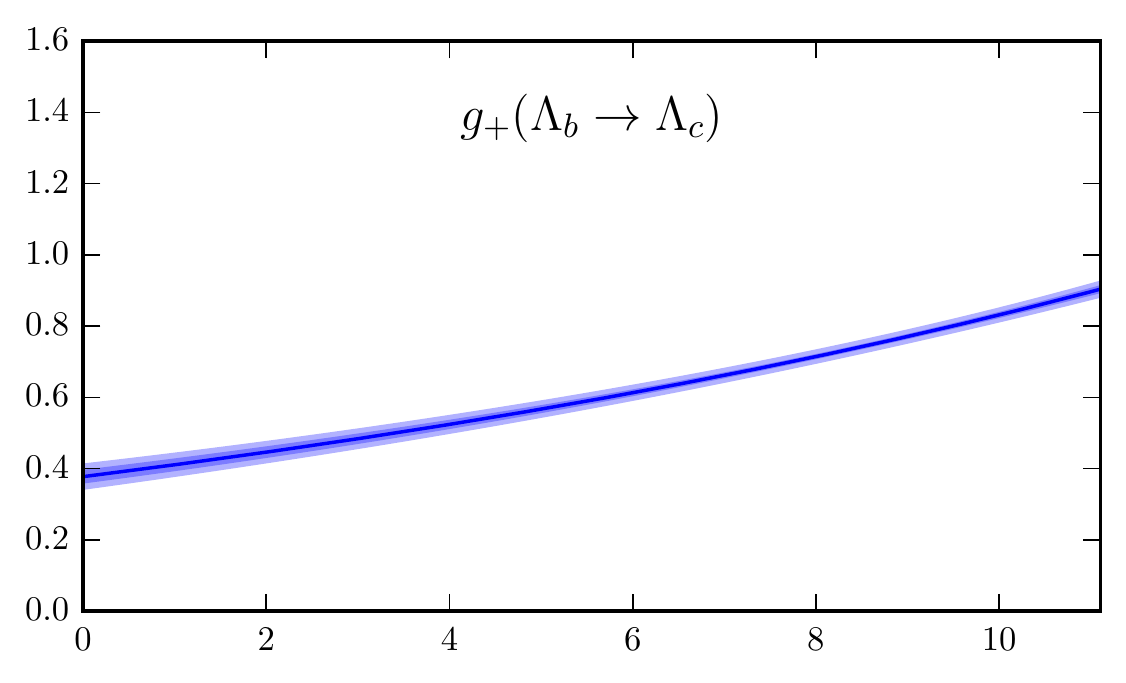} \\
 \includegraphics[height=5.3cm]{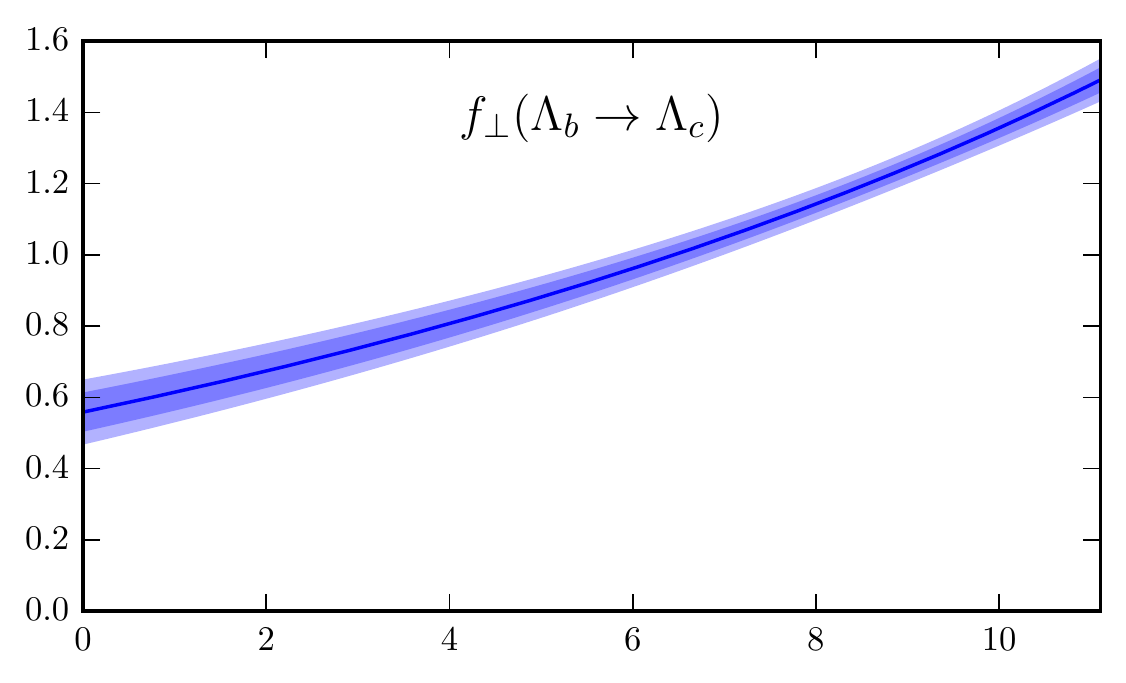} \hfill
 \includegraphics[height=5.3cm]{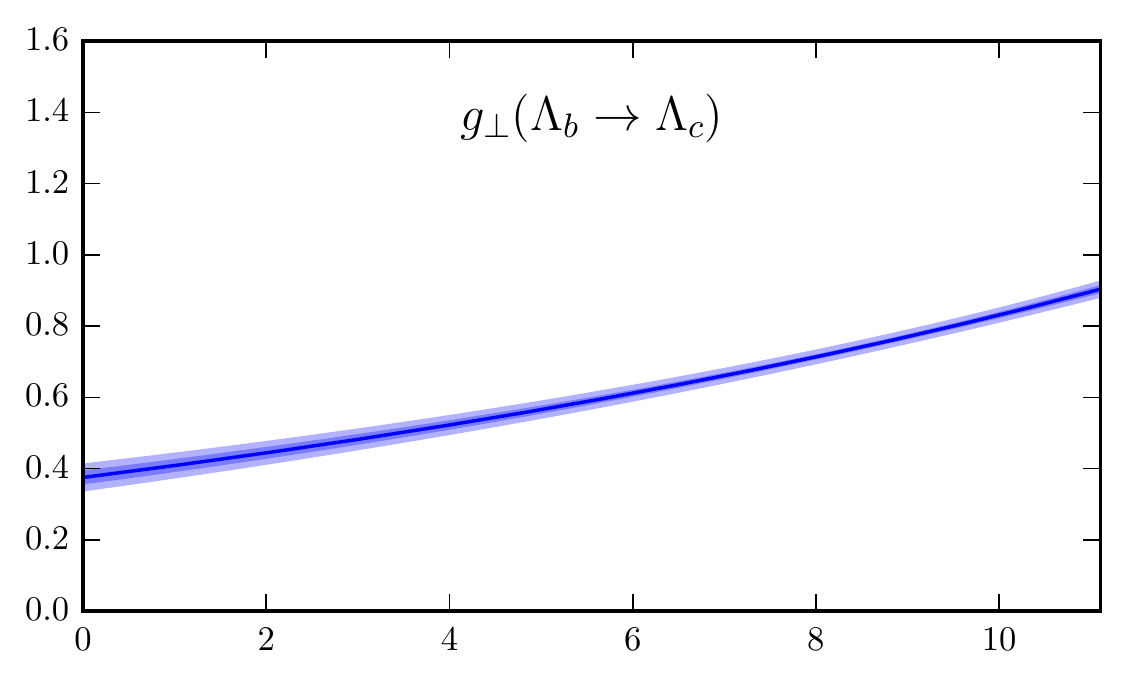} \\
 \includegraphics[height=6.1cm]{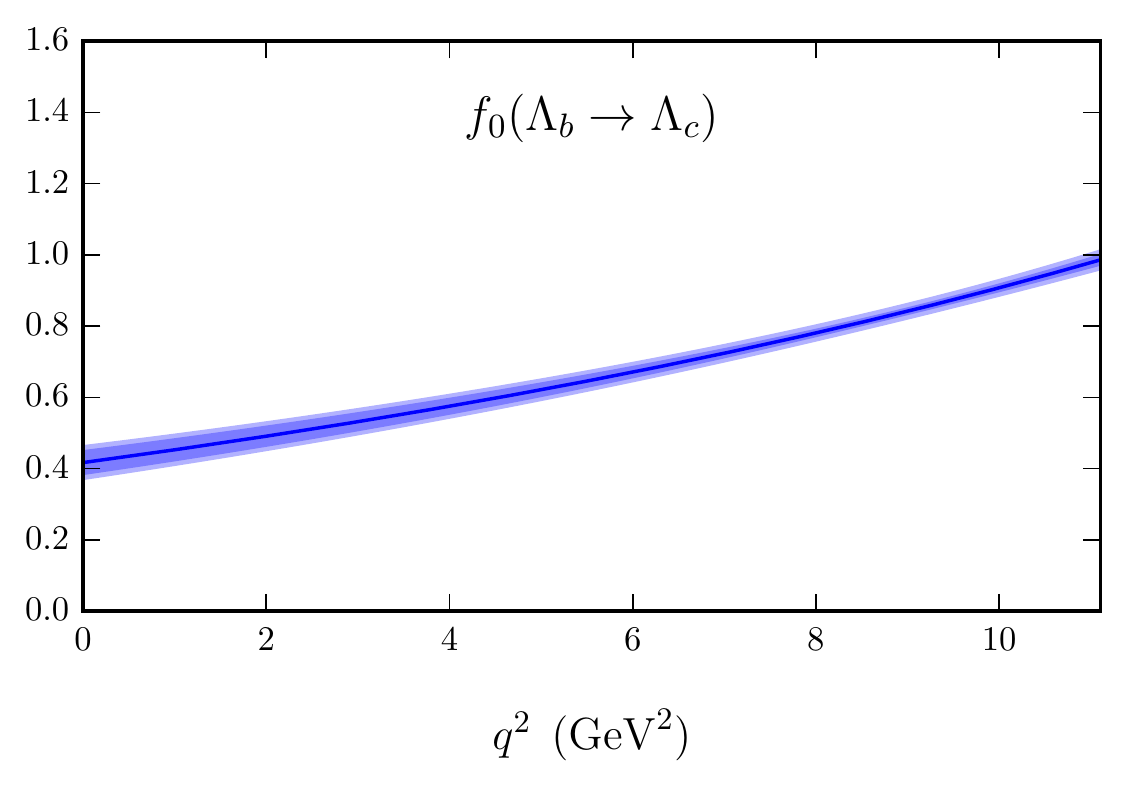} \hfill
 \includegraphics[height=6.1cm]{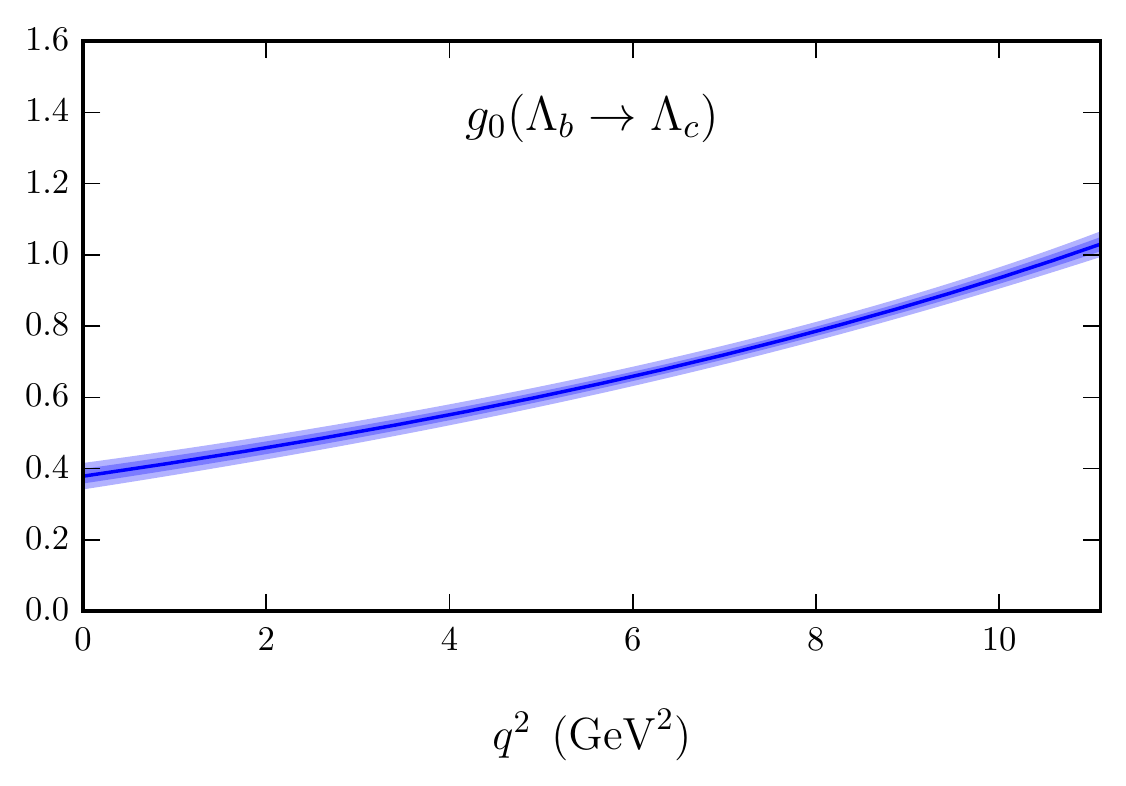} \\
 \caption{\label{fig:finalFFsLbLc}Final results for the $\Lambda_b\to \Lambda_c$ form factors. The inner bands show the statistical uncertainty and the outer
 bands show the total uncertainty.}
\end{figure}

\begin{figure}
 \hspace{1cm}\includegraphics[width=\linewidth]{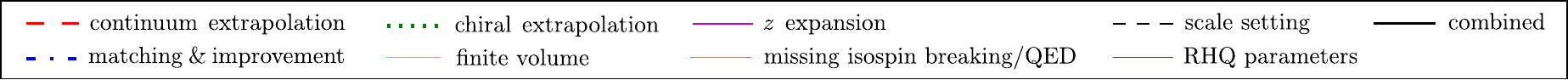} \\
 \includegraphics[height=5.4cm]{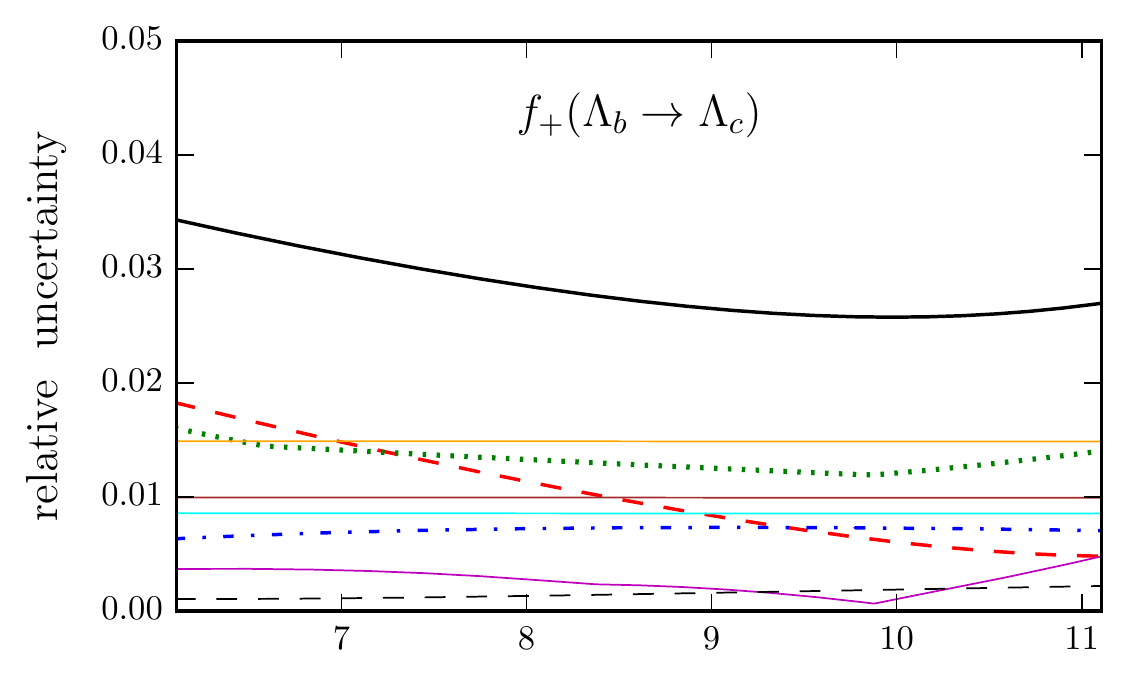} \hfill
 \includegraphics[height=5.4cm]{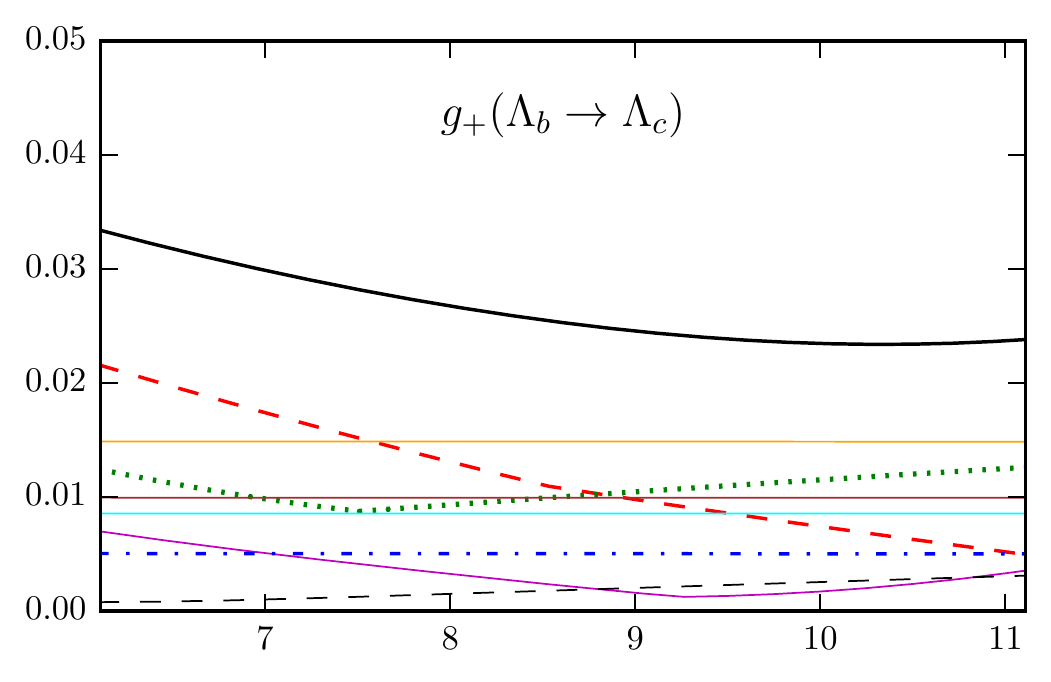} \\
 \includegraphics[height=5.4cm]{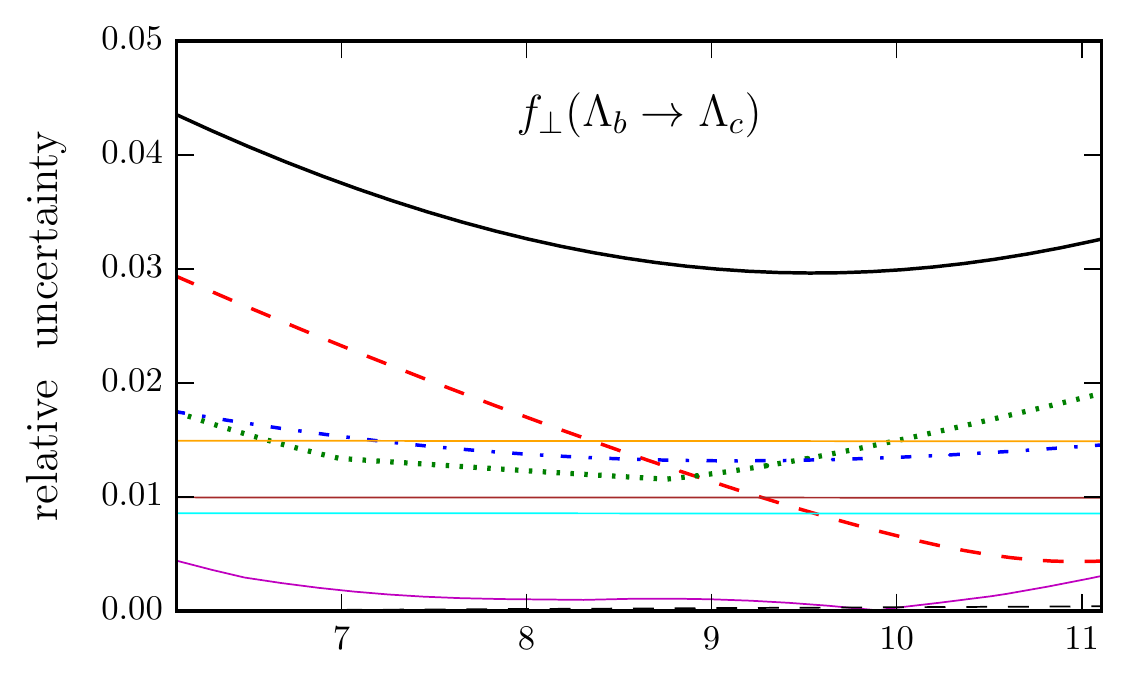} \hfill
 \includegraphics[height=5.4cm]{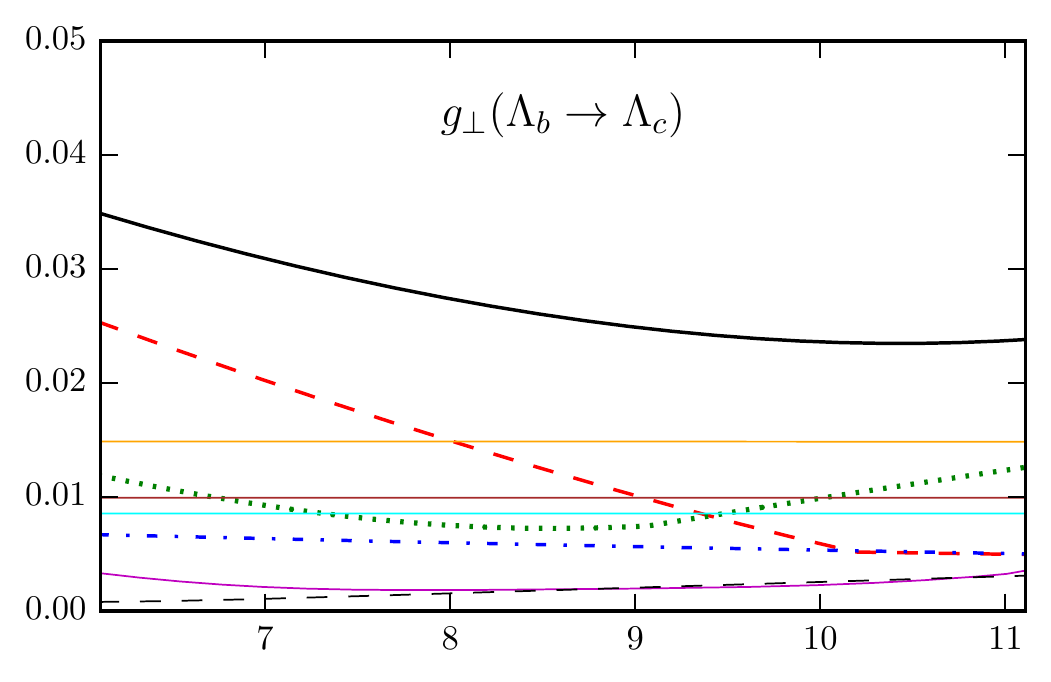} \\
 \includegraphics[height=6.2cm]{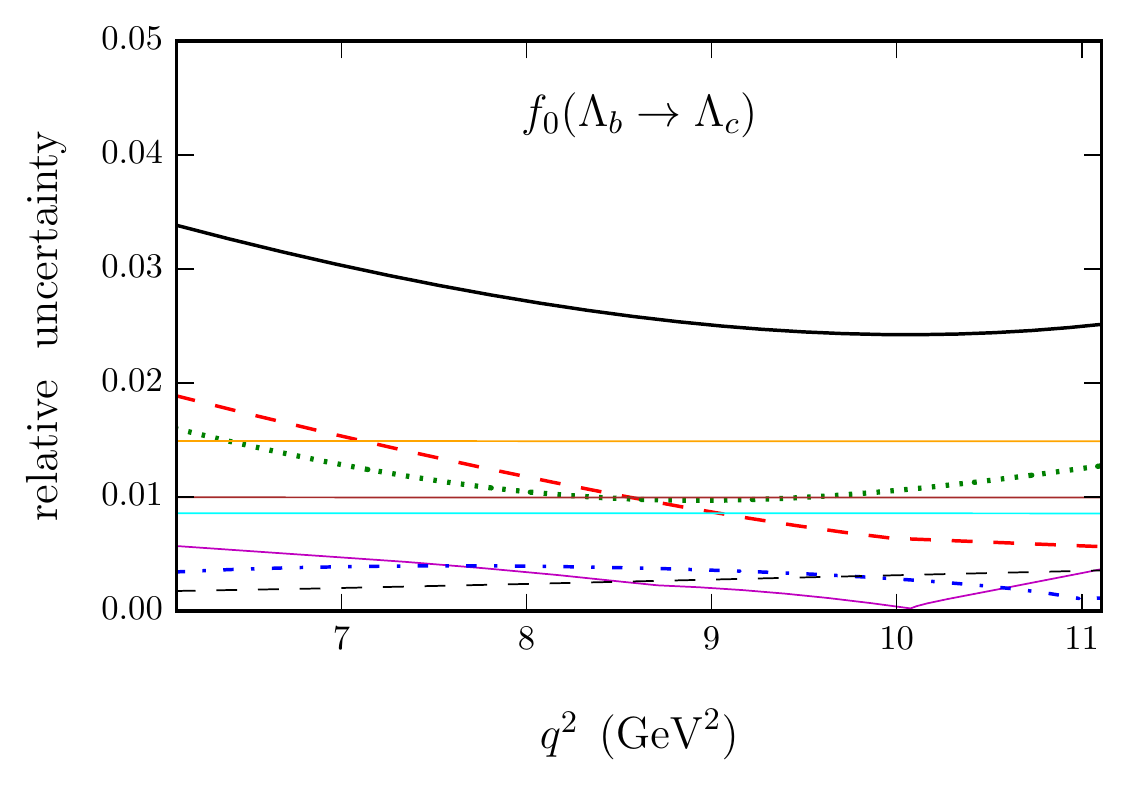} \hfill
 \includegraphics[height=6.2cm]{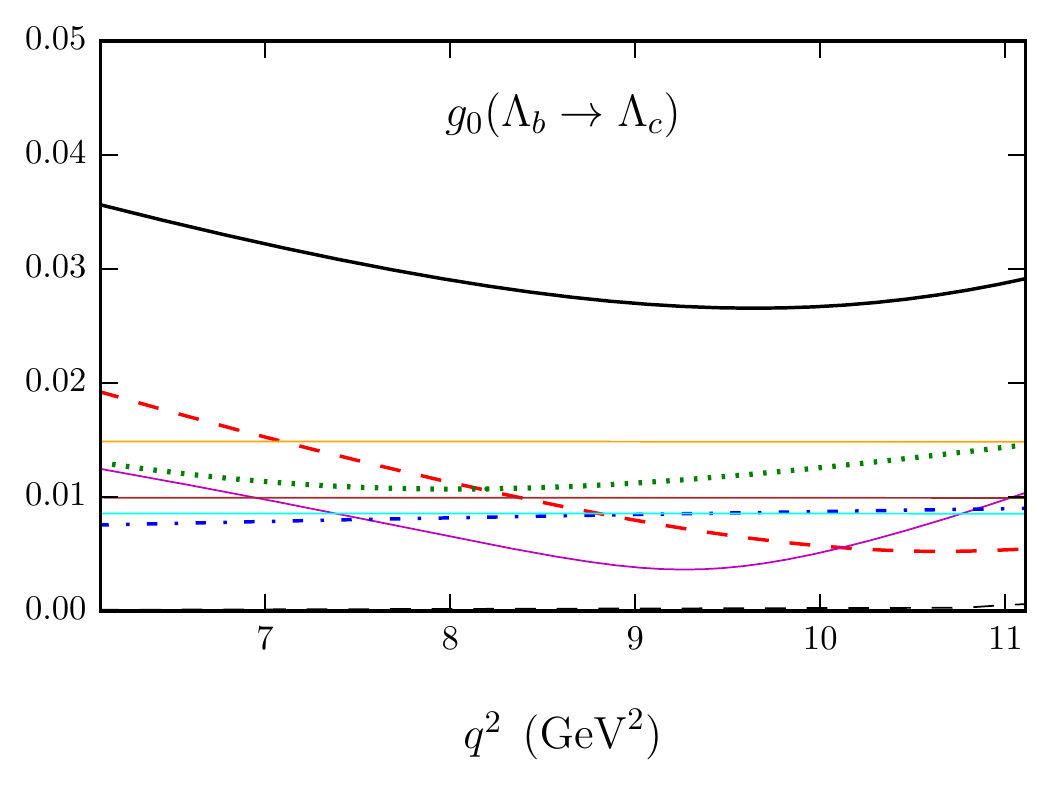}
 \caption{\label{fig:systLbLc}Systematic uncertainties in the $\Lambda_b\to \Lambda_c$ form factors in the high-$q^2$ region. As explained in the main text,
 the combined uncertainty is not simply the quadratic sum of the individual uncertainties.}
\end{figure}

Plots of the form factors including the systematic uncertainties, computed as explained above, are shown in Figs.~\ref{fig:finalFFsLbp} and \ref{fig:finalFFsLbLc}. The relative
systematic uncertainties in the form factors are shown in Figs.~\ref{fig:systLbp} and \ref{fig:systLbLc}. In addition to the combined systematic uncertainty (thick black curves),
these figures also show the individual sources of uncertainty. The individual systematic uncertainties were estimated using additional fits as follows:
\begin{itemize}
 \item Continuum extrapolation uncertainty: only the higher-order terms with coefficients $\tilde{b}^f$, $\tilde{d}^f$, $j^f$, $k^f$ were added to Eq.~(\ref{eq:FFccfit}).
 \item Chiral extrapolation uncertainty: only the higher-order terms with coefficients $\tilde{c}_0^f$, $c_1^f$ were added to Eq.~(\ref{eq:FFccfit}).
 \item $z$ expansion uncertainty: only the higher-order term $a_2^f\:[z^f(q^2)]^2$ was added to Eq.~(\ref{eq:FFccfit}).
 \item Matching \& improvement uncertainty: no higher-order terms were added to Eq.~(\ref{eq:FFccfit}), but the the matching- and $\mathcal{O}(a)$-improvement coefficients
 were drawn from Gaussian random distributions with central values and widths according to Table \ref{tab:Pmatchingfactors} when computing the correlator ratios using bootstrap.
 \item Scale setting (i.e., lattice spacing) uncertainty: no higher-order terms were added to Eq.~(\ref{eq:FFccfit}), but the lattice spacings were promoted to fit parameters
 constrained with Gaussian priors according to the central values and uncertainties given in Table \ref{tab:params}.
 \item Finite-volume effects, missing isospin breaking/QED, and RHQ parameter tuning uncertainties: no higher-order terms were added to Eq.~(\ref{eq:FFccfit}),
 but the estimates of these uncertainties (as discussed above) were added to the data covariance matrix used in the fit.
\end{itemize}
Figures \ref{fig:systLbp} and \ref{fig:systLbLc} show that near $q^2=q^2_{\rm max}$, the finite-volume and chiral-extrapolation uncertainties are the largest, but as the momentum
$|\mathbf{p}^\prime|$ increases (corresponding to decreasing $q^2$), the $z$-expansion and continuum extrapolation uncertainties grow and become dominant. The continuum extrapolation
uncertainty should not be interpreted as the actual size of lattice discretization errors; the reason for the large continuum extrapolation uncertainty is primarily that we have only two lattice
spacings and our data do not tightly constrain all of the extrapolation coefficients.

Discretization errors associated with the relativistic heavy quark actions used for the $b$ and $c$ quarks are not necessarily well approximated by the leading terms in an
expansion in $a\Lambda_{\rm QCD}$ and $a \mathbf{p}^\prime$. These errors can be described by mismatches of the coefficients of higher-dimensional operators in
the heavy-quark expansions of the lattice theory and continuum QCD \cite{ElKhadra:1996mp, Kronfeld:2000ck, Harada:2001fi, Harada:2001fj}. In Ref.~\cite{Flynn:2015mha},
the resulting heavy-quark discretization errors in the $B \to \pi$ form factors were estimated using power counting
to be of order 2\% for the same lattice actions and parameters as used in the present work. For most of the kinematic range, our estimate of the total continuum-extrapolation
uncertainty in the $\Lambda_b \to p$ form factors is larger than this power-counting estimate, see Fig.~\ref{fig:systLbp}. Similarly,
a comparison with the analysis of $B \to D$ form factors in Ref.~\cite{Lattice:2015rga} suggests that heavy-quark discretization errors
in the $\Lambda_b \to \Lambda_c$ form factors are smaller than or compatible with our estimates of the total continuum-extrapolation uncertainties in the entire kinematic range.

\begin{table}[b]
\begin{tabular}{lccccccccccc}
\hline\hline
$\Lambda_b\to p$ \hspace{2ex}  & $a_0^{f_+}$ & $a_1^{f_+}$ & $a_0^{f_0}$ & $a_1^{f_0}$ & $a_0^{f_\perp}$ & $a_1^{f_\perp}$ & $a_0^{g_\perp,g_+}$ & $a_1^{g_+}$ & $a_0^{g_0}$ & $a_1^{g_0}$ & $a_1^{g_\perp}$ \\ 
\hline
$a_0^{f_+}$ & $\wm 1$ & $-0.9058$ & $\wm 0.5081$ & $-0.4403$ & $\wm 0.5299$ & $-0.3987$ & $\wm 0.5362$ & $-0.4112$ & $\wm 0.6302$ & $-0.5305$ & $-0.3898$ \\ 
$a_1^{f_+}$ & $-0.9058$ & $\wm 1$ & $-0.4280$ & $\wm 0.4312$ & $-0.4238$ & $\wm 0.3739$ & $-0.4402$ & $\wm 0.3912$ & $-0.4901$ & $\wm 0.4839$ & $\wm 0.3668$ \\ 
$a_0^{f_0}$ & $\wm 0.5081$ & $-0.4280$ & $\wm 1$ & $-0.8533$ & $\wm 0.4251$ & $-0.3226$ & $\wm 0.6963$ & $-0.5274$ & $\wm 0.5504$ & $-0.4694$ & $-0.4886$ \\ 
$a_1^{f_0}$ & $-0.4403$ & $\wm 0.4312$ & $-0.8533$ & $\wm 1$ & $-0.3525$ & $\wm 0.3008$ & $-0.5963$ & $\wm 0.5917$ & $-0.4661$ & $\wm 0.4554$ & $\wm 0.5667$ \\ 
$a_0^{f_\perp}$ & $\wm 0.5299$ & $-0.4238$ & $\wm 0.4251$ & $-0.3525$ & $\wm 1$ & $-0.8930$ & $\wm 0.4748$ & $-0.3554$ & $\wm 0.5975$ & $-0.5111$ & $-0.3348$ \\ 
$a_1^{f_\perp}$ & $-0.3987$ & $\wm 0.3739$ & $-0.3226$ & $\wm 0.3008$ & $-0.8930$ & $\wm 1$ & $-0.3664$ & $\wm 0.3156$ & $-0.4349$ & $\wm 0.4542$ & $\wm 0.2900$ \\ 
$a_0^{g_\perp,g_+}$ & $\wm 0.5362$ & $-0.4402$ & $\wm 0.6963$ & $-0.5963$ & $\wm 0.4748$ & $-0.3664$ & $\wm 1$ & $-0.8434$ & $\wm 0.6238$ & $-0.5500$ & $-0.7905$ \\ 
$a_1^{g_+}$ & $-0.4112$ & $\wm 0.3912$ & $-0.5274$ & $\wm 0.5917$ & $-0.3554$ & $\wm 0.3156$ & $-0.8434$ & $\wm 1$ & $-0.4761$ & $\wm 0.5011$ & $\wm 0.8778$ \\ 
$a_0^{g_0}$ & $\wm 0.6302$ & $-0.4901$ & $\wm 0.5504$ & $-0.4661$ & $\wm 0.5975$ & $-0.4349$ & $\wm 0.6238$ & $-0.4761$ & $\wm 1$ & $-0.9039$ & $-0.4497$ \\ 
$a_1^{g_0}$ & $-0.5305$ & $\wm 0.4839$ & $-0.4694$ & $\wm 0.4554$ & $-0.5111$ & $\wm 0.4542$ & $-0.5500$ & $\wm 0.5011$ & $-0.9039$ & $\wm 1$ & $\wm 0.4632$ \\ 
$a_1^{g_\perp}$ & $-0.3898$ & $\wm 0.3668$ & $-0.4886$ & $\wm 0.5667$ & $-0.3348$ & $\wm 0.2900$ & $-0.7905$ & $\wm 0.8778$ & $-0.4497$ & $\wm 0.4632$ & $\wm 1$ \\ 
\hline\hline
$\Lambda_b\to \Lambda_c$ \hspace{2ex}  & $a_0^{f_+}$ & $a_1^{f_+}$ & $a_0^{f_0}$ & $a_1^{f_0}$ & $a_0^{f_\perp}$ & $a_1^{f_\perp}$ & $a_0^{g_\perp,g_+}$ & $a_1^{g_+}$ & $a_0^{g_0}$ & $a_1^{g_0}$ & $a_1^{g_\perp}$ \\ 
\hline
$a_0^{f_+}$ & $\wm 1$ & $-0.6644$ & $\wm 0.6827$ & $-0.4853$ & $\wm 0.6218$ & $-0.3906$ & $\wm 0.4828$ & $-0.3152$ & $\wm 0.5636$ & $-0.4317$ & $-0.3763$ \\ 
$a_1^{f_+}$ & $-0.6644$ & $\wm 1$ & $-0.6515$ & $\wm 0.9445$ & $-0.3853$ & $\wm 0.5109$ & $-0.3831$ & $\wm 0.4915$ & $-0.2979$ & $\wm 0.4916$ & $\wm 0.4764$ \\ 
$a_0^{f_0}$ & $\wm 0.6827$ & $-0.6515$ & $\wm 1$ & $-0.7040$ & $\wm 0.4208$ & $-0.3620$ & $\wm 0.6174$ & $-0.4822$ & $\wm 0.4320$ & $-0.4726$ & $-0.4756$ \\ 
$a_1^{f_0}$ & $-0.4853$ & $\wm 0.9445$ & $-0.7040$ & $\wm 1$ & $-0.2738$ & $\wm 0.4739$ & $-0.3888$ & $\wm 0.5261$ & $-0.2164$ & $\wm 0.4779$ & $\wm 0.4877$ \\ 
$a_0^{f_\perp}$ & $\wm 0.6218$ & $-0.3853$ & $\wm 0.4208$ & $-0.2738$ & $\wm 1$ & $-0.6637$ & $\wm 0.3933$ & $-0.2369$ & $\wm 0.5161$ & $-0.3639$ & $-0.2926$ \\ 
$a_1^{f_\perp}$ & $-0.3906$ & $\wm 0.5109$ & $-0.3620$ & $\wm 0.4739$ & $-0.6637$ & $\wm 1$ & $-0.2903$ & $\wm 0.3509$ & $-0.2443$ & $\wm 0.3640$ & $\wm 0.3400$ \\ 
$a_0^{g_\perp,g_+}$ & $\wm 0.4828$ & $-0.3831$ & $\wm 0.6174$ & $-0.3888$ & $\wm 0.3933$ & $-0.2903$ & $\wm 1$ & $-0.7304$ & $\wm 0.6365$ & $-0.6743$ & $-0.7301$ \\ 
$a_1^{g_+}$ & $-0.3152$ & $\wm 0.4915$ & $-0.4822$ & $\wm 0.5261$ & $-0.2369$ & $\wm 0.3509$ & $-0.7304$ & $\wm 1$ & $-0.3829$ & $\wm 0.8725$ & $\wm 0.9171$ \\ 
$a_0^{g_0}$ & $\wm 0.5636$ & $-0.2979$ & $\wm 0.4320$ & $-0.2164$ & $\wm 0.5161$ & $-0.2443$ & $\wm 0.6365$ & $-0.3829$ & $\wm 1$ & $-0.6843$ & $-0.4846$ \\ 
$a_1^{g_0}$ & $-0.4317$ & $\wm 0.4916$ & $-0.4726$ & $\wm 0.4779$ & $-0.3639$ & $\wm 0.3640$ & $-0.6743$ & $\wm 0.8725$ & $-0.6843$ & $\wm 1$ & $\wm 0.8456$ \\ 
$a_1^{g_\perp}$ & $-0.3763$ & $\wm 0.4764$ & $-0.4756$ & $\wm 0.4877$ & $-0.2926$ & $\wm 0.3400$ & $-0.7301$ & $\wm 0.9171$ & $-0.4846$ & $\wm 0.8456$ & $\wm 1$ \\ \hline\hline
\end{tabular}
\caption{\label{tab:nominalcorr}Correlation matrices of the nominal form factor parameters for $\Lambda_b \to p$ (top) and $\Lambda_b \to \Lambda_c$ (bottom).}
\end{table}

\begin{table}
\begin{tabular}{lllll}
\hline \hline
 Parameter         &  & \hspace{5ex} $\Lambda_b \to p$  & \hspace{1ex}  & \hspace{5ex}  $\Lambda_b \to \Lambda_c$    \\
\hline
$a_0^{f_+}$ && $\wm 0.4251\pm 0.0388$ && $\wm 0.8103\pm 0.0276$ \\ 
$a_1^{f_+}$ && $-0.7088\pm 0.3361$ && $-4.7480\pm 0.9429$ \\ 
$a_2^{f_+}$ && $\wm 0.8925\pm 0.8869$ && $\wm 0.7862\pm 8.8020$ \\ 
$a_0^{f_0}$ && $\wm 0.4144\pm 0.0321$ && $\wm 0.7389\pm 0.0225$ \\ 
$a_1^{f_0}$ && $-1.0420\pm 0.3142$ && $-4.5630\pm 0.9426$ \\ 
$a_2^{f_0}$ && $\wm 1.9260\pm 0.9190$ && $\wm 2.7050\pm 8.4430$ \\ 
$a_0^{f_\perp}$ && $\wm 0.5214\pm 0.0520$ && $\wm 1.0940\pm 0.0435$ \\ 
$a_1^{f_\perp}$ && $-0.8247\pm 0.4424$ && $-6.4410\pm 1.5010$ \\ 
$a_2^{f_\perp}$ && $\wm 0.7609\pm 1.2770$ && $\wm 2.3160\pm 11.320$ \\ 
$a_0^{g_\perp,g_+}$ && $\wm 0.3889\pm 0.0260$ && $\wm 0.6848\pm 0.0184$ \\ 
$a_1^{g_+}$ && $-1.0730\pm 0.2617$ && $-4.3790\pm 0.6954$ \\ 
$a_2^{g_+}$ && $\wm 1.9860\pm 0.8247$ && $\wm 1.2810\pm 7.3650$ \\ 
$a_0^{g_0}$ && $\wm 0.4419\pm 0.0388$ && $\wm 0.7408\pm 0.0258$ \\ 
$a_1^{g_0}$ && $-0.8649\pm 0.3481$ && $-4.3860\pm 0.8774$ \\ 
$a_2^{g_0}$ && $\wm 0.9969\pm 0.8955$ && $\wm 1.3380\pm 8.0440$ \\ 
$a_1^{g_\perp}$ && $-1.0840\pm 0.2732$ && $-4.6270\pm 0.7088$ \\ 
$a_2^{g_\perp}$ && $\wm 1.4520\pm 1.0680$ && $\wm 1.6140\pm 7.4530$ \\ 
\hline\hline
\end{tabular}
\caption{\label{tab:HO}Central values and uncertainties of the higher-order form factor parameters for $\Lambda_b \to p$ and $\Lambda_b \to \Lambda_c$.
See Table \protect\ref{tab:HOcorr} for the correlation matrices.}
\end{table}

\begin{turnpage}

\begin{table}
\small
\begin{tabular}{lccccccccccccccccc}
\hline\hline
$\Lambda_b \to p$ & $a_0^{f_+}$ & $a_1^{f_+}$ & $a_2^{f_+}$ & $a_0^{f_0}$ & $a_1^{f_0}$ & $a_2^{f_0}$ & $a_0^{f_\perp}$ & $a_1^{f_\perp}$ & $a_2^{f_\perp}$ & $a_0^{g_\perp,g_+}$ & $a_1^{g_+}$ & $a_2^{g_+}$ & $a_0^{g_0}$ & $a_1^{g_0}$ & $a_2^{g_0}$ & $a_1^{g_\perp}$ & $a_2^{g_\perp}$ \\ 
\hline
$a_0^{f_+}$ & $\wm 1$ & $-0.7671$ & $\wm 0.2482$ & $\wm 0.5337$ & $-0.2670$ & $-0.0922$ & $\wm 0.5121$ & $-0.2469$ & $-0.0180$ & $\wm 0.3774$ & $-0.2148$ & $-0.0472$ & $\wm 0.4420$ & $-0.2680$ & $\wm 0.0018$ & $-0.2284$ & $-0.0231$ \\ 
$a_1^{f_+}$ & $-0.7671$ & $\wm 1$ & $-0.6611$ & $-0.2486$ & $\wm 0.1617$ & $\wm 0.0653$ & $-0.2526$ & $\wm 0.1671$ & $\wm 0.0056$ & $-0.2177$ & $\wm 0.1480$ & $\wm 0.0287$ & $-0.2496$ & $\wm 0.1849$ & $-0.0169$ & $\wm 0.1534$ & $\wm 0.0147$ \\ 
$a_2^{f_+}$ & $\wm 0.2482$ & $-0.6611$ & $\wm 1$ & $-0.0792$ & $\wm 0.0267$ & $\wm 0.2795$ & $-0.0035$ & $-0.0120$ & $\wm 0.0425$ & $-0.0562$ & $\wm 0.0382$ & $\wm 0.0559$ & $-0.0279$ & $-0.0074$ & $\wm 0.0870$ & $\wm 0.0370$ & $\wm 0.0469$ \\ 
$a_0^{f_0}$ & $\wm 0.5337$ & $-0.2486$ & $-0.0792$ & $\wm 1$ & $-0.7202$ & $\wm 0.2599$ & $\wm 0.4581$ & $-0.2052$ & $-0.0146$ & $\wm 0.4734$ & $-0.2798$ & $-0.0031$ & $\wm 0.3860$ & $-0.2266$ & $-0.0115$ & $-0.2781$ & $\wm 0.0048$ \\ 
$a_1^{f_0}$ & $-0.2670$ & $\wm 0.1617$ & $\wm 0.0267$ & $-0.7202$ & $\wm 1$ & $-0.6947$ & $-0.2404$ & $\wm 0.1415$ & $\wm 0.0128$ & $-0.2964$ & $\wm 0.2603$ & $-0.0377$ & $-0.2410$ & $\wm 0.1694$ & $\wm 0.0090$ & $\wm 0.2610$ & $-0.0279$ \\ 
$a_2^{f_0}$ & $-0.0922$ & $\wm 0.0653$ & $\wm 0.2795$ & $\wm 0.2599$ & $-0.6947$ & $\wm 1$ & $\wm 0.0190$ & $-0.0056$ & $\wm 0.0297$ & $-0.0019$ & $-0.0529$ & $\wm 0.1086$ & $-0.0081$ & $-0.0097$ & $\wm 0.0664$ & $-0.0568$ & $\wm 0.0874$ \\ 
$a_0^{f_\perp}$ & $\wm 0.5121$ & $-0.2526$ & $-0.0035$ & $\wm 0.4581$ & $-0.2404$ & $\wm 0.0190$ & $\wm 1$ & $-0.7672$ & $\wm 0.1031$ & $\wm 0.3418$ & $-0.1831$ & $-0.0539$ & $\wm 0.4313$ & $-0.2713$ & $\wm 0.0163$ & $-0.1994$ & $-0.0127$ \\ 
$a_1^{f_\perp}$ & $-0.2469$ & $\wm 0.1671$ & $-0.0120$ & $-0.2052$ & $\wm 0.1415$ & $-0.0056$ & $-0.7672$ & $\wm 1$ & $-0.5040$ & $-0.1983$ & $\wm 0.1259$ & $\wm 0.0378$ & $-0.2429$ & $\wm 0.1907$ & $-0.0274$ & $\wm 0.1347$ & $\wm 0.0083$ \\ 
$a_2^{f_\perp}$ & $-0.0180$ & $\wm 0.0056$ & $\wm 0.0425$ & $-0.0146$ & $\wm 0.0128$ & $\wm 0.0297$ & $\wm 0.1031$ & $-0.5040$ & $\wm 1$ & $-0.0271$ & $\wm 0.0045$ & $\wm 0.0524$ & $-0.0286$ & $\wm 0.0090$ & $\wm 0.0530$ & $\wm 0.0120$ & $\wm 0.0187$ \\ 
$a_0^{g_\perp,g_+}$ & $\wm 0.3774$ & $-0.2177$ & $-0.0562$ & $\wm 0.4734$ & $-0.2964$ & $-0.0019$ & $\wm 0.3418$ & $-0.1983$ & $-0.0271$ & $\wm 1$ & $-0.6751$ & $\wm 0.2299$ & $\wm 0.5903$ & $-0.2849$ & $-0.0084$ & $-0.6325$ & $\wm 0.1314$ \\ 
$a_1^{g_+}$ & $-0.2148$ & $\wm 0.1480$ & $\wm 0.0382$ & $-0.2798$ & $\wm 0.2603$ & $-0.0529$ & $-0.1831$ & $\wm 0.1259$ & $\wm 0.0045$ & $-0.6751$ & $\wm 1$ & $-0.6972$ & $-0.2576$ & $\wm 0.1666$ & $-0.0268$ & $\wm 0.6832$ & $-0.1976$ \\ 
$a_2^{g_+}$ & $-0.0472$ & $\wm 0.0287$ & $\wm 0.0559$ & $-0.0031$ & $-0.0377$ & $\wm 0.1086$ & $-0.0539$ & $\wm 0.0378$ & $\wm 0.0524$ & $\wm 0.2299$ & $-0.6972$ & $\wm 1$ & $-0.0760$ & $\wm 0.0463$ & $\wm 0.2693$ & $-0.3207$ & $\wm 0.2419$ \\ 
$a_0^{g_0}$ & $\wm 0.4420$ & $-0.2496$ & $-0.0279$ & $\wm 0.3860$ & $-0.2410$ & $-0.0081$ & $\wm 0.4313$ & $-0.2429$ & $-0.0286$ & $\wm 0.5903$ & $-0.2576$ & $-0.0760$ & $\wm 1$ & $-0.7868$ & $\wm 0.3673$ & $-0.2892$ & $-0.0105$ \\ 
$a_1^{g_0}$ & $-0.2680$ & $\wm 0.1849$ & $-0.0074$ & $-0.2266$ & $\wm 0.1694$ & $-0.0097$ & $-0.2713$ & $\wm 0.1907$ & $\wm 0.0090$ & $-0.2849$ & $\wm 0.1666$ & $\wm 0.0463$ & $-0.7868$ & $\wm 1$ & $-0.7393$ & $\wm 0.1798$ & $\wm 0.0107$ \\ 
$a_2^{g_0}$ & $\wm 0.0018$ & $-0.0169$ & $\wm 0.0870$ & $-0.0115$ & $\wm 0.0090$ & $\wm 0.0664$ & $\wm 0.0163$ & $-0.0274$ & $\wm 0.0530$ & $-0.0084$ & $-0.0268$ & $\wm 0.2693$ & $\wm 0.3673$ & $-0.7393$ & $\wm 1$ & $\wm 0.0302$ & $\wm 0.0637$ \\ 
$a_1^{g_\perp}$ & $-0.2284$ & $\wm 0.1534$ & $\wm 0.0370$ & $-0.2781$ & $\wm 0.2610$ & $-0.0568$ & $-0.1994$ & $\wm 0.1347$ & $\wm 0.0120$ & $-0.6325$ & $\wm 0.6832$ & $-0.3207$ & $-0.2892$ & $\wm 0.1798$ & $\wm 0.0302$ & $\wm 1$ & $-0.6223$ \\ 
$a_2^{g_\perp}$ & $-0.0231$ & $\wm 0.0147$ & $\wm 0.0469$ & $\wm 0.0048$ & $-0.0279$ & $\wm 0.0874$ & $-0.0127$ & $\wm 0.0083$ & $\wm 0.0187$ & $\wm 0.1314$ & $-0.1976$ & $\wm 0.2419$ & $-0.0105$ & $\wm 0.0107$ & $\wm 0.0637$ & $-0.6223$ & $\wm 1$ \\ 
\hline\hline
$\Lambda_b \to \Lambda_c$ & $a_0^{f_+}$ & $a_1^{f_+}$ & $a_2^{f_+}$ & $a_0^{f_0}$ & $a_1^{f_0}$ & $a_2^{f_0}$ & $a_0^{f_\perp}$ & $a_1^{f_\perp}$ & $a_2^{f_\perp}$ & $a_0^{g_\perp,g_+}$ & $a_1^{g_+}$ & $a_2^{g_+}$ & $a_0^{g_0}$ & $a_1^{g_0}$ & $a_2^{g_0}$ & $a_1^{g_\perp}$ & $a_2^{g_\perp}$ \\ 
\hline
$a_0^{f_+}$ & $\wm 1$ & $-0.5220$ & $\wm 0.1623$ & $\wm 0.7106$ & $-0.2661$ & $-0.0293$ & $\wm 0.6259$ & $-0.2683$ & $\wm 0.0077$ & $\wm 0.1992$ & $-0.1307$ & $-0.0277$ & $\wm 0.2833$ & $-0.1838$ & $\wm 0.0436$ & $-0.1611$ & $\wm 0.0088$ \\ 
$a_1^{f_+}$ & $-0.5220$ & $\wm 1$ & $-0.6595$ & $-0.3199$ & $\wm 0.4277$ & $\wm 0.0649$ & $-0.2548$ & $\wm 0.2618$ & $-0.0102$ & $-0.1403$ & $\wm 0.1878$ & $\wm 0.0413$ & $-0.1575$ & $\wm 0.1932$ & $-0.0364$ & $\wm 0.1703$ & $\wm 0.0030$ \\ 
$a_2^{f_+}$ & $\wm 0.1623$ & $-0.6595$ & $\wm 1$ & $-0.0350$ & $\wm 0.1309$ & $\wm 0.0939$ & $\wm 0.0181$ & $-0.0149$ & $\wm 0.0300$ & $-0.0111$ & $\wm 0.0190$ & $\wm 0.0007$ & $\wm 0.0005$ & $-0.0041$ & $\wm 0.0186$ & $\wm 0.0246$ & $\wm 0.0088$ \\ 
$a_0^{f_0}$ & $\wm 0.7106$ & $-0.3199$ & $-0.0350$ & $\wm 1$ & $-0.5132$ & $\wm 0.1123$ & $\wm 0.5190$ & $-0.2037$ & $-0.0014$ & $\wm 0.2531$ & $-0.2100$ & $\wm 0.0128$ & $\wm 0.2012$ & $-0.1481$ & $\wm 0.0096$ & $-0.2057$ & $-0.0079$ \\ 
$a_1^{f_0}$ & $-0.2661$ & $\wm 0.4277$ & $\wm 0.1309$ & $-0.5132$ & $\wm 1$ & $-0.5243$ & $-0.1791$ & $\wm 0.2285$ & $\wm 0.0094$ & $-0.1770$ & $\wm 0.2589$ & $\wm 0.0134$ & $-0.1266$ & $\wm 0.1854$ & $-0.0086$ & $\wm 0.2339$ & $\wm 0.0127$ \\ 
$a_2^{f_0}$ & $-0.0293$ & $\wm 0.0649$ & $\wm 0.0939$ & $\wm 0.1123$ & $-0.5243$ & $\wm 1$ & $-0.0222$ & $\wm 0.0275$ & $\wm 0.0138$ & $\wm 0.0044$ & $-0.0148$ & $\wm 0.0300$ & $-0.0074$ & $\wm 0.0112$ & $-0.0034$ & $-0.0218$ & $\wm 0.0075$ \\ 
$a_0^{f_\perp}$ & $\wm 0.6259$ & $-0.2548$ & $\wm 0.0181$ & $\wm 0.5190$ & $-0.1791$ & $-0.0222$ & $\wm 1$ & $-0.5829$ & $\wm 0.1142$ & $\wm 0.1754$ & $-0.1255$ & $-0.0168$ & $\wm 0.2874$ & $-0.1811$ & $\wm 0.0416$ & $-0.1320$ & $-0.0086$ \\ 
$a_1^{f_\perp}$ & $-0.2683$ & $\wm 0.2618$ & $-0.0149$ & $-0.2037$ & $\wm 0.2285$ & $\wm 0.0275$ & $-0.5829$ & $\wm 1$ & $-0.4656$ & $-0.1154$ & $\wm 0.1472$ & $\wm 0.0360$ & $-0.1487$ & $\wm 0.1650$ & $-0.0341$ & $\wm 0.1319$ & $\wm 0.0096$ \\ 
$a_2^{f_\perp}$ & $\wm 0.0077$ & $-0.0102$ & $\wm 0.0300$ & $-0.0014$ & $\wm 0.0094$ & $\wm 0.0138$ & $\wm 0.1142$ & $-0.4656$ & $\wm 1$ & $-0.0006$ & $-0.0003$ & $\wm 0.0057$ & $\wm 0.0049$ & $-0.0059$ & $\wm 0.0087$ & $-0.0006$ & $\wm 0.0033$ \\ 
$a_0^{g_\perp,g_+}$ & $\wm 0.1992$ & $-0.1403$ & $-0.0111$ & $\wm 0.2531$ & $-0.1770$ & $\wm 0.0044$ & $\wm 0.1754$ & $-0.1154$ & $-0.0006$ & $\wm 1$ & $-0.4436$ & $\wm 0.0876$ & $\wm 0.7054$ & $-0.2594$ & $\wm 0.0128$ & $-0.4268$ & $\wm 0.0479$ \\ 
$a_1^{g_+}$ & $-0.1307$ & $\wm 0.1878$ & $\wm 0.0190$ & $-0.2100$ & $\wm 0.2589$ & $-0.0148$ & $-0.1255$ & $\wm 0.1472$ & $-0.0003$ & $-0.4436$ & $\wm 1$ & $-0.5465$ & $-0.2790$ & $\wm 0.3438$ & $\wm 0.0541$ & $\wm 0.4776$ & $-0.1381$ \\ 
$a_2^{g_+}$ & $-0.0277$ & $\wm 0.0413$ & $\wm 0.0007$ & $\wm 0.0128$ & $\wm 0.0134$ & $\wm 0.0300$ & $-0.0168$ & $\wm 0.0360$ & $\wm 0.0057$ & $\wm 0.0876$ & $-0.5465$ & $\wm 1$ & $-0.0447$ & $\wm 0.1194$ & $\wm 0.0577$ & $-0.1482$ & $\wm 0.2692$ \\ 
$a_0^{g_0}$ & $\wm 0.2833$ & $-0.1575$ & $\wm 0.0005$ & $\wm 0.2012$ & $-0.1266$ & $-0.0074$ & $\wm 0.2874$ & $-0.1487$ & $\wm 0.0049$ & $\wm 0.7054$ & $-0.2790$ & $-0.0447$ & $\wm 1$ & $-0.5511$ & $\wm 0.2196$ & $-0.3015$ & $\wm 0.0059$ \\ 
$a_1^{g_0}$ & $-0.1838$ & $\wm 0.1932$ & $-0.0041$ & $-0.1481$ & $\wm 0.1854$ & $\wm 0.0112$ & $-0.1811$ & $\wm 0.1650$ & $-0.0059$ & $-0.2594$ & $\wm 0.3438$ & $\wm 0.1194$ & $-0.5511$ & $\wm 1$ & $-0.7687$ & $\wm 0.2440$ & $\wm 0.0190$ \\ 
$a_2^{g_0}$ & $\wm 0.0436$ & $-0.0364$ & $\wm 0.0186$ & $\wm 0.0096$ & $-0.0086$ & $-0.0034$ & $\wm 0.0416$ & $-0.0341$ & $\wm 0.0087$ & $\wm 0.0128$ & $\wm 0.0541$ & $\wm 0.0577$ & $\wm 0.2196$ & $-0.7687$ & $\wm 1$ & $\wm 0.0004$ & $\wm 0.0405$ \\ 
$a_1^{g_\perp}$ & $-0.1611$ & $\wm 0.1703$ & $\wm 0.0246$ & $-0.2057$ & $\wm 0.2339$ & $-0.0218$ & $-0.1320$ & $\wm 0.1319$ & $-0.0006$ & $-0.4268$ & $\wm 0.4776$ & $-0.1482$ & $-0.3015$ & $\wm 0.2440$ & $\wm 0.0004$ & $\wm 1$ & $-0.5028$ \\ 
$a_2^{g_\perp}$ & $\wm 0.0088$ & $\wm 0.0030$ & $\wm 0.0088$ & $-0.0079$ & $\wm 0.0127$ & $\wm 0.0075$ & $-0.0086$ & $\wm 0.0096$ & $\wm 0.0033$ & $\wm 0.0479$ & $-0.1381$ & $\wm 0.2692$ & $\wm 0.0059$ & $\wm 0.0190$ & $\wm 0.0405$ & $-0.5028$ & $\wm 1$ \\ 
\hline\hline
\end{tabular}
\caption{\label{tab:HOcorr}Correlation matrices of the higher-order form factor parameters for $\Lambda_b \to p$ (top) and $\Lambda_b \to \Lambda_c$ (bottom).}
\end{table}

\end{turnpage}

\FloatBarrier
\section{\label{sec:dGamma}Predictions for the $\Lambda_b \to p\,\ell^-\bar{\nu}_\ell$ and $\Lambda_b \to \Lambda_c\,\ell^-\bar{\nu}_\ell$ decay rates}
\FloatBarrier

In this section, we present predictions for the $\Lambda_b \to p\,\ell^-\bar{\nu}_\ell$ and $\Lambda_b \to \Lambda_c\,\ell^-\bar{\nu}_\ell$ differential
and integrated decay rates using our form factor results. Including possible right-handed currents with real-valued $\epsilon_q^{R}$, the effective Hamiltonian in Eq.~(\ref{eq:Heff})
leads to the following expression for the differential decay rate in terms of the helicity form factors,
\begin{eqnarray}
\nonumber \frac{\mathrm{d}\Gamma}{\mathrm{d} q^2} &=&
\frac{G_F^2 |V_{qb}^L|^2 \sqrt{s_+s_-}
    }{768 \pi ^3 m_{\Lambda_b}^3 } \left(1-\frac{m_\ell^2}{q^2}\right)^2 \\
\nonumber     &&\times\Bigg\{4 \left(m_\ell^2+2 q^2\right) \left(
     s_+ \left[ (1-\epsilon_q^{R}) g_\perp  \right]^2 + s_- \left[(1+\epsilon_q^{R})f_\perp\right]^2 \right)  \\
\nonumber    && \hspace{2ex} +2 \frac{m_\ell^2+2 q^2}{q^2} \left(s_+
   \left[\left(m_{\Lambda_b}-m_X\right)(1-\epsilon_q^{R}) g_+ \right]^2+s_- \left[\left(m_{\Lambda_b}+m_X\right)(1+\epsilon_q^{R})f_+ \right]^2\right)\\
   &&\hspace{2ex} +\frac{6 m_\ell^2}{q^2} \left(s_+ \left[ \left(m_{\Lambda_b}-m_X\right) (1+\epsilon_q^{R})f_0
   \right]^2 + s_-
   \left[ \left(m_{\Lambda_b}+m_X\right)(1-\epsilon_q^{R}) g_0  \right]^2\right)\Bigg\},
\end{eqnarray}
where, as before, $X=p,\Lambda_c$ denotes the final-state baryon, and
\begin{equation}
 s_\pm =(m_{\Lambda_b} \pm m_X)^2-q^2.
\end{equation}
Expressions for the individual helicity amplitudes and the angular distributions can be found in Refs.~\cite{Gutsche:2014zna, Gutsche:2015mxa, Shivashankara:2015cta}.
By combining experimental data with our form factor results, novel constraints in the $(V_{qb}^L,\:\epsilon_q^{R})$ plane can be obtained.

\begin{figure}
 \includegraphics[width=0.7\linewidth]{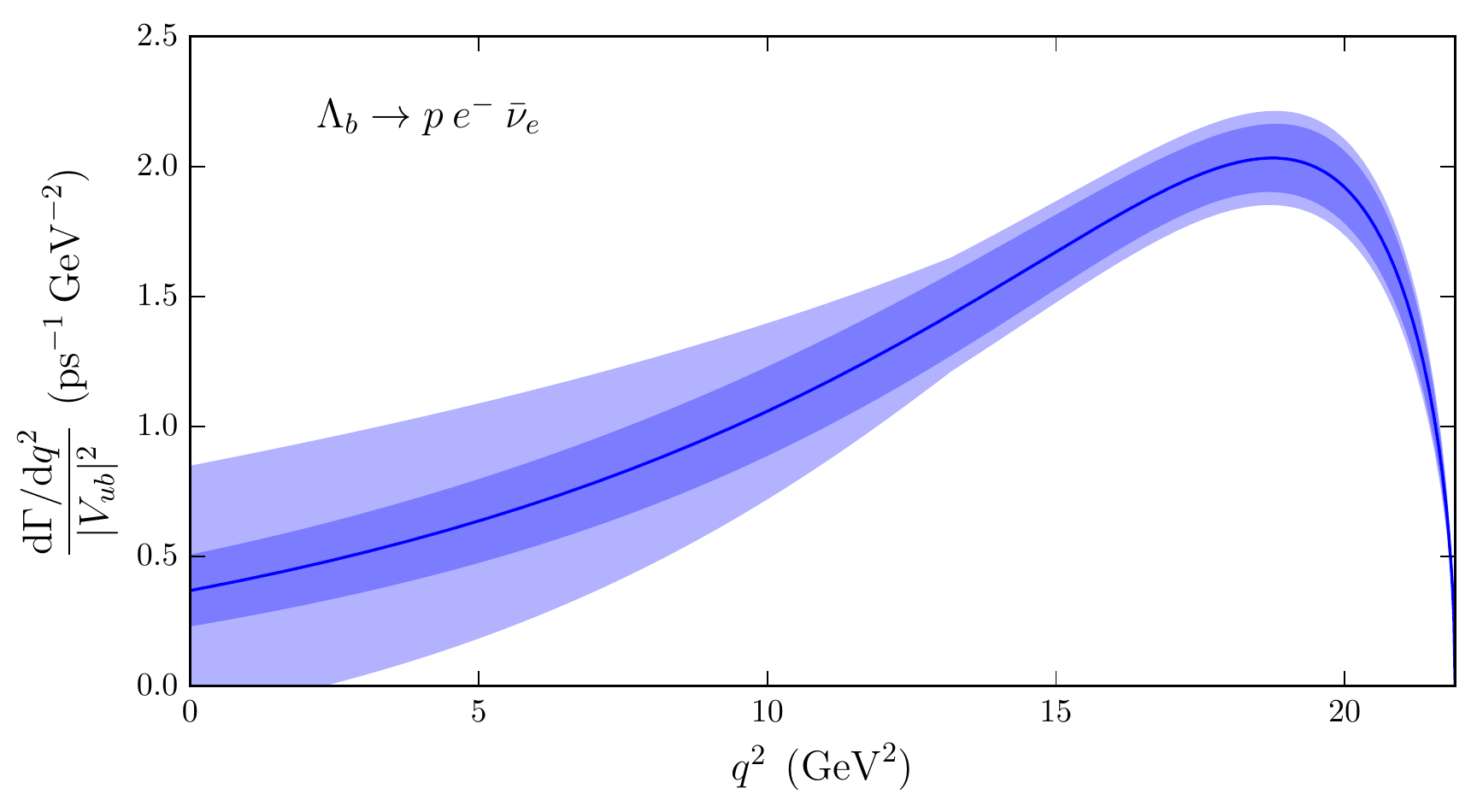} 

 \includegraphics[width=0.7\linewidth]{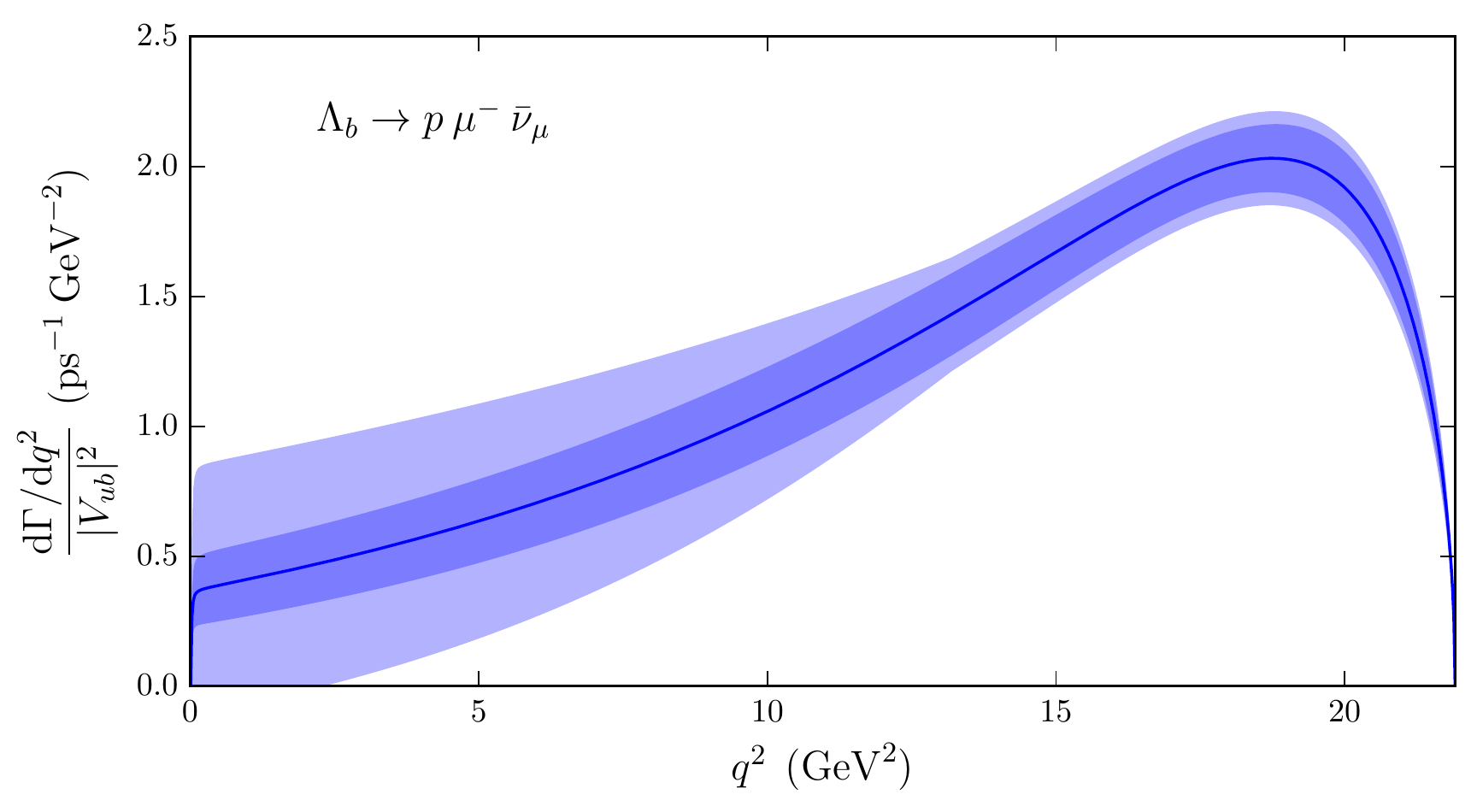} 

 \includegraphics[width=0.7\linewidth]{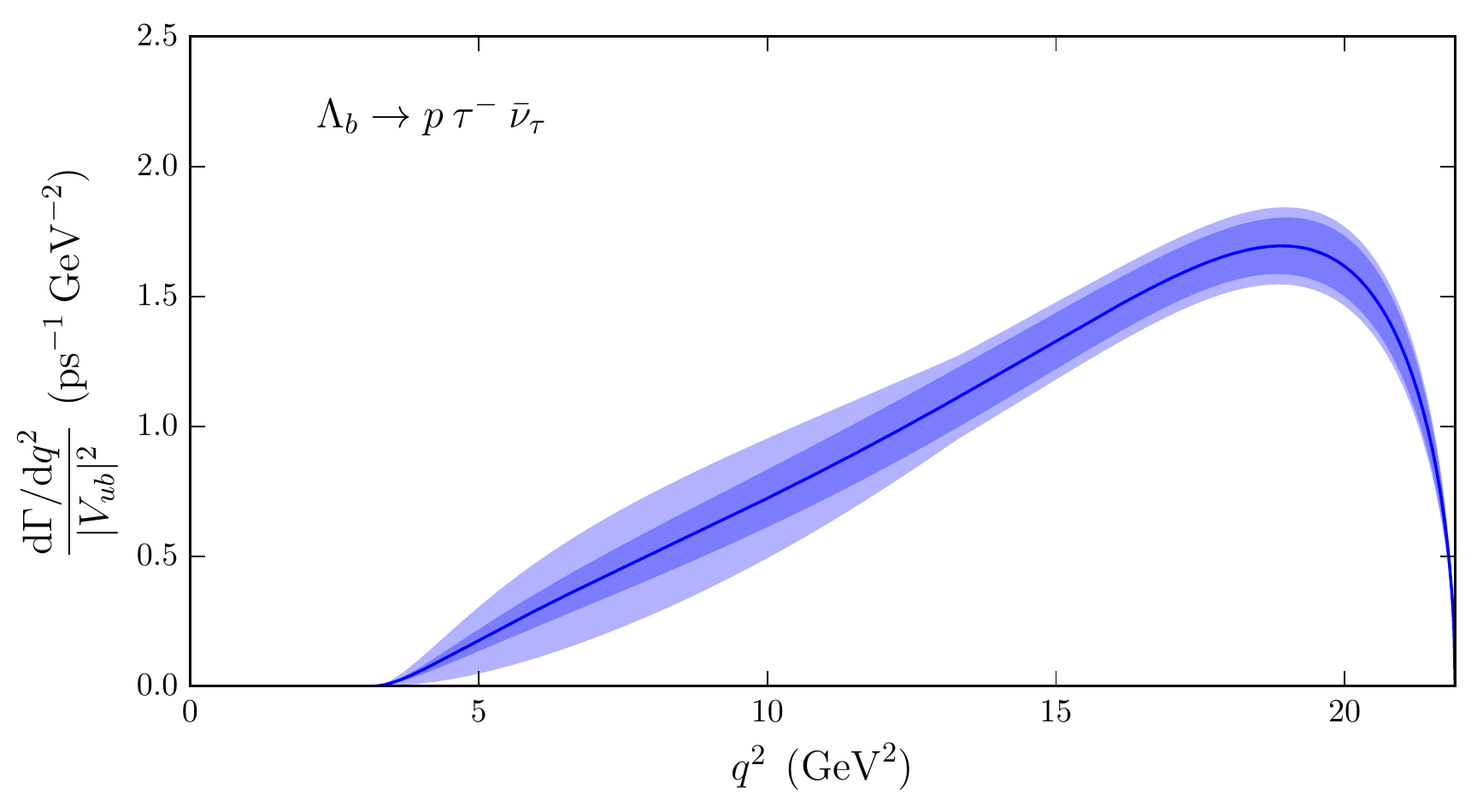} 

 \caption{\label{fig:LbpdGamma}Predictions for the $\Lambda_b \to p\,\ell^-\bar{\nu}_\ell$ differential decay rates for $\ell=e,\mu,\tau$ in the Standard Model.
 The inner bands show the statistical uncertainty and the outer bands show the total uncertainty, calculated using Eq.~(\ref{eq:finalres}).}
\end{figure}

\begin{figure}

 \includegraphics[width=0.7\linewidth]{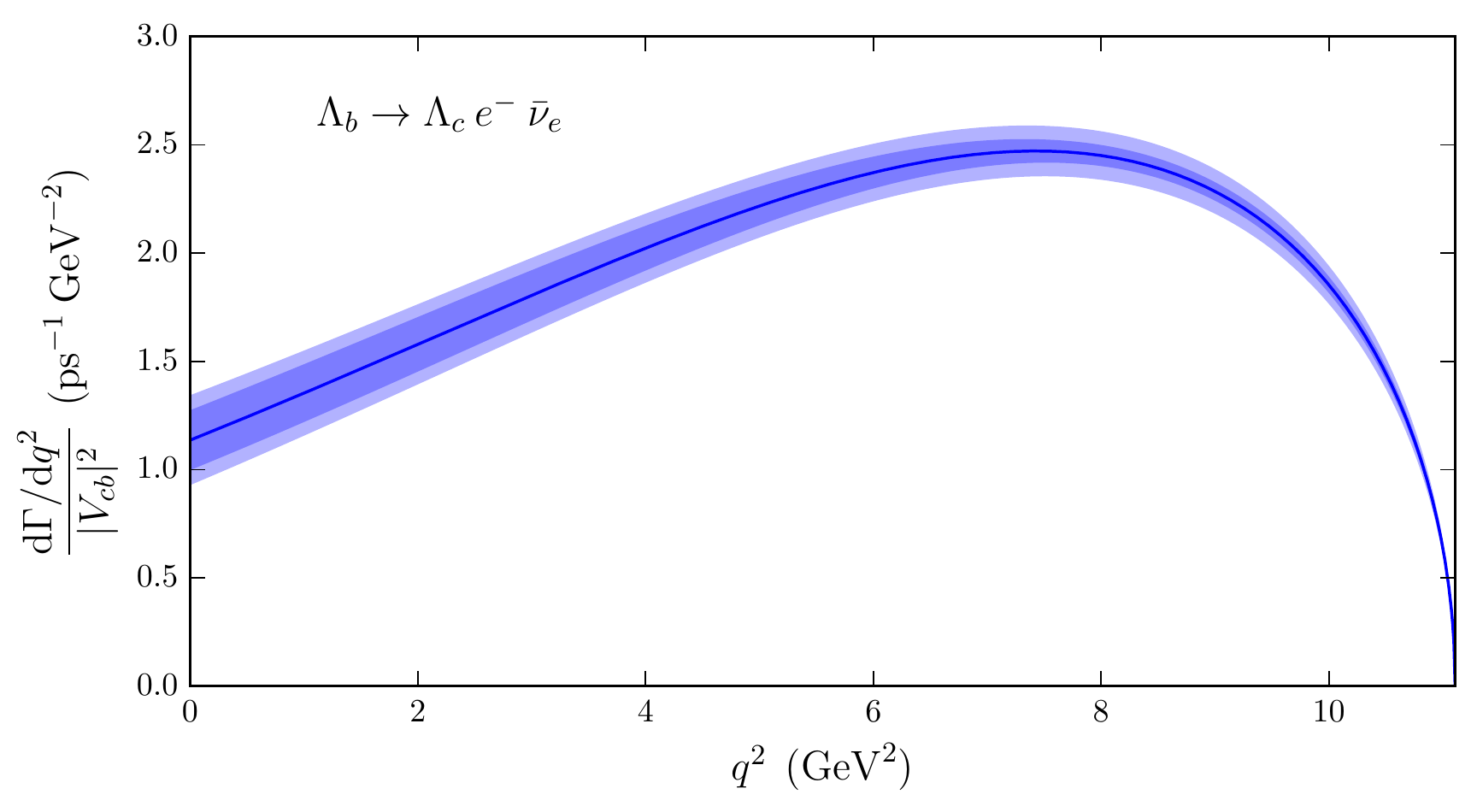}

 \includegraphics[width=0.7\linewidth]{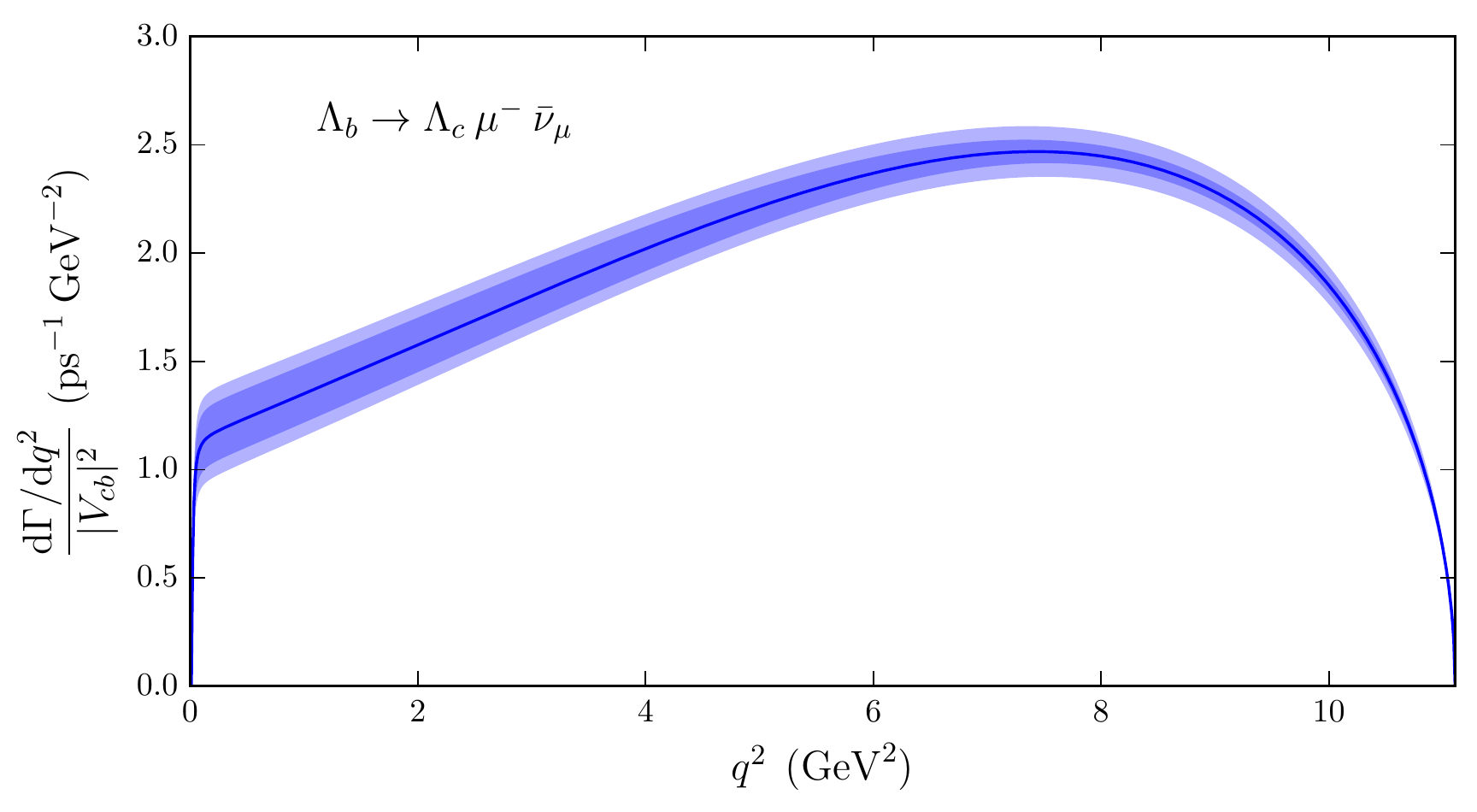}

 \includegraphics[width=0.7\linewidth]{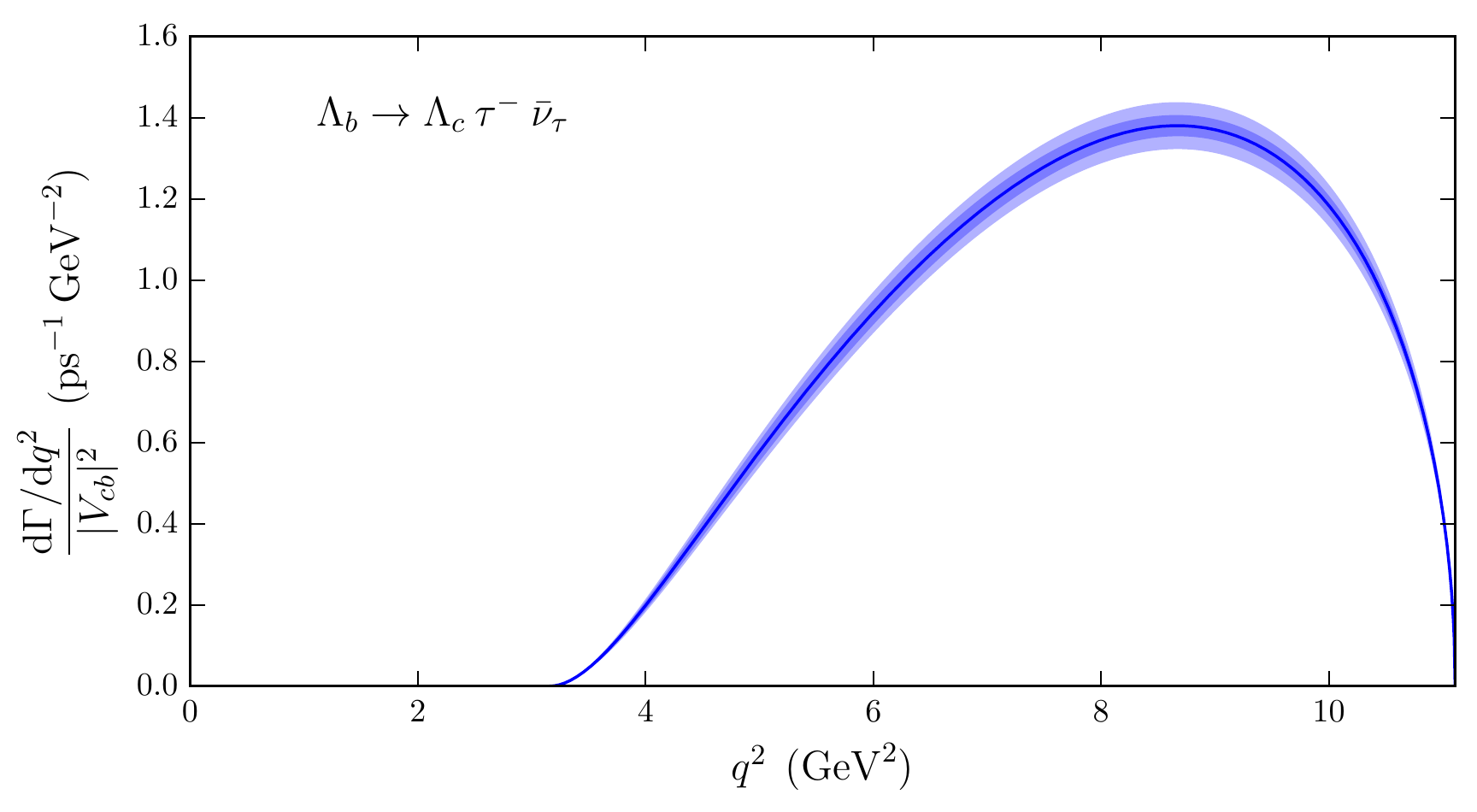}

 \caption{\label{fig:LbLcdGamma}Predictions for the $\Lambda_b \to \Lambda_c\,\ell^-\bar{\nu}_\ell$ differential decay rates for $\ell=e,\mu,\tau$ in the Standard Model.
 The inner bands show the statistical uncertainty and the outer bands show the total uncertainty, calculated using Eq.~(\ref{eq:finalres}).}
\end{figure}

In the following, we consider the Standard Model with $V_{qb}^L=V_{qb}$ and $\epsilon_q^{R}=0$. Our predictions of the $\Lambda_b \to p\,\ell^-\bar{\nu}_\ell$
and $\Lambda_b \to \Lambda_c\,\ell^-\bar{\nu}_\ell$ differential decay rates for $\ell=e,\mu,\tau$ are shown in Figs.~\ref{fig:LbpdGamma} and \ref{fig:LbLcdGamma}.
The central values, statistical uncertainties, and systematic uncertainties have been calculated using Eq.~(\ref{eq:finalres}); all baryon and lepton masses
were taken from Ref.~\cite{Agashe:2014kda}. Our results are most precise in the high-$q^2$ region, where the form factor shapes are most tightly constrained by the lattice QCD data.
We obtain the following partially integrated decay rates
\begin{eqnarray}
 \zeta_{p\mu\bar{\nu}}(15\:{\rm GeV}^2)\equiv\frac{1}{|V_{ub}|^2}\int_{15\:{\rm GeV}^2}^{q^2_{\rm max}}
\frac{\mathrm{d}\Gamma (\Lambda_b \to p\: \mu^- \bar{\nu}_\mu)}{\mathrm{d}q^2} \mathrm{d} q^2 &=& (12.31 \pm 0.76 \pm 0.77)\:\:{\rm ps}^{-1}, \label{eq:Vub} \\
 \zeta_{\Lambda_c\mu\bar{\nu}}(7\:{\rm GeV}^2)\equiv\frac{1}{|V_{cb}|^2}\int_{7\:{\rm GeV}^2}^{q^2_{\rm max}}
\frac{\mathrm{d}\Gamma (\Lambda_b \to \Lambda_c\: \mu^- \bar{\nu}_\mu)}{\mathrm{d}q^2} \mathrm{d} q^2 &=& (8.37 \pm 0.16 \pm 0.34)\:\:{\rm ps}^{-1}, \label{eq:Vcb}
\end{eqnarray}
and their ratio
\begin{eqnarray}
 \frac{\zeta_{p\mu\bar{\nu}}(15\:{\rm GeV}^2)}{\zeta_{\Lambda_c\mu\bar{\nu}}(7\:{\rm GeV}^2)} &=& 1.471 \pm 0.095 \pm 0.109, \label{eq:VubVcb}
\end{eqnarray}
where the first uncertainty is statistical and the second uncertainty is systematic.
Together with experimental data, Eqs.~(\ref{eq:Vub}), (\ref{eq:Vcb}), and (\ref{eq:VubVcb}) will allow determinations of $|V_{ub}|$, $|V_{cb}|$, and
$|V_{ub}/V_{cb}|$ with theory uncertainties of $4.4\%$, $2.2\%$, and $4.9\%$, respectively. A breakdown of the uncertainties into the individual sources,
obtained by applying Eq.~(\ref{eq:finalres}) to the various additional form factor fits discussed at the end of Sec.~\ref{sec:ccextrap}, is given in Table \ref{tab:errorbreakdown}.

The predicted total decay rates
for all possible lepton flavors are
\begin{eqnarray}
  \Gamma (\Lambda_b \to p\: e^- \bar{\nu}_e)/|V_{ub}|^2 &=& (25.7 \pm 2.6 \pm 4.6)\:\:{\rm ps}^{-1} \\
  \Gamma (\Lambda_b \to p\: \mu^- \bar{\nu}_\mu)/|V_{ub}|^2 &=& (25.7 \pm 2.6 \pm 4.6)\:\:{\rm ps}^{-1}, \\
  \Gamma (\Lambda_b \to p\: \tau^- \bar{\nu}_\mu)/|V_{ub}|^2 &=& (17.7 \pm 1.3 \pm 1.6)\:\:{\rm ps}^{-1}, \\
  \Gamma (\Lambda_b \to \Lambda_c\: e^- \bar{\nu}_e)/|V_{cb}|^2 &=& (21.5 \pm 0.8 \pm 1.1)\:\:{\rm ps}^{-1}, \\
  \Gamma (\Lambda_b \to \Lambda_c\: \mu^- \bar{\nu}_\mu)/|V_{cb}|^2 &=& (21.5 \pm 0.8 \pm 1.1)\:\:{\rm ps}^{-1}, \\
  \Gamma (\Lambda_b \to \Lambda_c\: \tau^- \bar{\nu}_\mu)/|V_{cb}|^2 &=& (7.15 \pm 0.15 \pm  0.27)\:\:{\rm ps}^{-1}.
\end{eqnarray}
Motivated by the $\mathcal{R}(D^{(*)})$ puzzle \cite{Lees:2012xj}, we also provide predictions for the following ratios:
\begin{eqnarray}
 \frac{\Gamma (\Lambda_b \to \Lambda_c\: \tau^- \bar{\nu}_\mu)}{\Gamma (\Lambda_b \to \Lambda_c\: e^- \bar{\nu}_\mu)} &=& 0.3318 \pm 0.0074 \pm 0.0070, \\
 \frac{\Gamma (\Lambda_b \to \Lambda_c\: \tau^- \bar{\nu}_\mu)}{\Gamma (\Lambda_b \to \Lambda_c\: \mu^- \bar{\nu}_\mu)} &=& 0.3328 \pm 0.0074 \pm  0.0070.
\end{eqnarray}
QED corrections to the decay rates, which may be relevant at this level of precision, have been neglected here.

\begin{table}
\begin{tabular}{lcccccc}
\hline\hline
 & \hspace{1ex} & $\zeta_{p\mu\bar{\nu}}(15\:{\rm GeV}^2)$  & \hspace{1ex} & $\zeta_{\Lambda_c\mu\bar{\nu}}(7\:{\rm GeV}^2)$  & \hspace{1ex} & $\displaystyle\frac{\zeta_{p\mu\bar{\nu}}(15\:{\rm GeV}^2)}{\zeta_{\Lambda_c\mu\bar{\nu}}(7\:{\rm GeV}^2)}$  \\
\hline
Statistics                    &&  $6.2$  &&  $1.9$  &&  $6.5$  \\
Finite volume                 &&  $5.0$  &&  $2.5$  &&  $4.9$  \\
Continuum extrapolation       &&  $3.0$  &&  $1.4$  &&  $2.8$  \\
Chiral extrapolation          &&  $2.6$  &&  $1.8$  &&  $2.6$  \\
RHQ parameters                &&  $1.4$  &&  $1.7$  &&  $2.3$  \\
Matching \& improvement       &&  $1.7$  &&  $0.9$  &&  $2.1$  \\
Missing isospin breaking/QED  &&  $1.2$  &&  $1.4$  &&  $2.0$  \\
Scale setting                 &&  $1.7$  &&  $0.3$  &&  $1.8$  \\
$z$ expansion                 &&  $1.2$  &&  $0.2$  &&  $1.3$  \\
\hline
Total                         &&  $8.8$  &&  $4.5$  &&  $9.8$  \\
\hline\hline
\end{tabular}
\caption{\label{tab:errorbreakdown} Approximate breakdown of relative uncertainties (in \%) in the partially integrated $\Lambda_b \to p\: \mu^- \bar{\nu}_\mu$
and $\Lambda_b \to \Lambda_c\: \mu^- \bar{\nu}_\mu$ decay rates and their ratio, defined in Eqs.~(\ref{eq:Vub}), (\ref{eq:Vcb}), and (\ref{eq:VubVcb}). As explained in the main text,
 the combined uncertainty is not simply the quadratic sum of the individual uncertainties. }
\end{table}

\FloatBarrier

\FloatBarrier
\section{\label{sec:discussion}Summary}
\FloatBarrier

We have presented a high-precision lattice QCD calculation of the complete set of relativistic form factors describing the $\Lambda_b \to p$ and $\Lambda_b \to \Lambda_c$
matrix elements of the vector and axial vector $b \to u$ and $b \to c$ currents. The form factors and their uncertainties in the physical limit are shown in Figs.~\ref{fig:finalFFsLbp} and \ref{fig:finalFFsLbLc}.
Any observable depending on the form factors can be calculated using Eq.~(\ref{eq:finalres}), which is based on two different sets of form factor
parameters. The ``nominal'' form factors are used to calculate the central value and statistical uncertainty of the observable, and are given by the functions (\ref{eq:nominalfitphys})
with parameters and correlation matrices from Tables \ref{tab:nominal} and \ref{tab:nominalcorr}, together with
the pole masses from Table \ref{tab:polemasses}. The ``higher order'' form factors are additionally needed to calculate the systematic uncertainty of the observable,
and are given by Eq.~(\ref{eq:HOfitphys}) with the parameters from Tables \ref{tab:HO} and \ref{tab:HOcorr}. The higher-order fit was performed in such a way that the
systematic uncertainty obtained from Eq.~(\ref{eq:finalres}) includes the continuum extrapolation uncertainty, the chiral extrapolation uncertainty,
the kinematic ($q^2$) extrapolation uncertainty, the perturbative matching/improvement uncertainty, the uncertainty due to the finite lattice volume, and the uncertainty from the
missing isospin breaking effects. The individual contributions to the systematic uncertainties in the form factors
are shown in Figs.~\ref{fig:systLbp} and \ref{fig:systLbLc}.

Our predictions of the $\Lambda_b \to p\,\ell^-\,\bar{\nu}_\ell$ and $\Lambda_b \to \Lambda_c\,\ell^-\,\bar{\nu}_\ell$ differential decay rates using the new form factors are presented in Sec.~\ref{sec:dGamma}.
The results (\ref{eq:Vub}), (\ref{eq:Vcb}), and (\ref{eq:VubVcb}) for the $\Lambda_b \to p\: \mu^- \bar{\nu}_\mu$ and $\Lambda_b \to \Lambda_c\: \mu^- \bar{\nu}_\mu$ differential decay rates in the high-$q^2$ region
can be combined with forthcoming experimental data to determine $|V_{ub}|$, $|V_{cb}|$, and $|V_{ub}/V_{cb}|$ with theory uncertainties of $4.4\%$, $2.2\%$, and $4.9\%$, respectively. These uncertainties are
competitive with the total uncertainties in the 2014 PDG values based on exclusive $B$ meson decays [see Eq.~(\ref{eq:VubVcbRPP2014})].
Compared to Ref.~\cite{Detmold:2013nia}, we have reduced the uncertainty in the $\Lambda_b \to p\,\ell^-\bar{\nu}_\ell$ decay rate at high $q^2$ by a factor of 3.
This reduction in uncertainty mainly resulted from the elimination of the static approximation for the $b$ quark. Combined with experimental data, our form factor results
will also provide novel constraints on right-handed couplings beyond the Standard Model \cite{Chen:2008se, Crivellin:2009sd, Buras:2010pz, Crivellin:2014zpa}.
The constraints from the baryonic decays nicely complement existing constraints from mesonic decays due to the unique dependence of the baryonic
decays on $\epsilon_R$. Using our $\Lambda_b \to \Lambda_c$ form factors, very precise predictions can also be made for the decay $\Lambda_b \to \Lambda_c \,\tau^-\,\bar{\nu}_\tau$,
which may provide new insights into the $\mathcal{R}(D^{(*)})$ puzzle \cite{Gutsche:2015mxa, Shivashankara:2015cta}.

\FloatBarrier
\section*{Acknowledgments}
\FloatBarrier

We thank William Sutcliffe, Ulrik Egede, and Patrick Owen for numerous discussions about the analysis of the decays $\Lambda_b \to p\, \mu\,\bar{\nu}_\mu$ and $\Lambda_b \to \Lambda_c\, \mu\,\bar{\nu}_\mu$
using LHCb data, and Gil Paz for comments regarding the $z$ expansion. We are grateful to the RBC and UKQCD collaborations for making their gauge field configurations available. The lattice calculations were carried out using
the Chroma software \cite{Edwards:2004sx} on high-performance computing resources provided by XSEDE (supported by National Science Foundation Grant Number OCI-1053575) and NERSC (supported by U.S.~Department of Energy Grant Number DE-AC02-05CH11231).
SM is supported by the RHIC Physics Fellow Program of the RIKEN BNL Research Center. WD is supported by the U.S.~Department of Energy Early Career Research Award D{E}-S{C001}0495 and the Solomon Buchsbaum Fund at MIT.
CL is supported by the U.S. Department of Energy contract D{E}-AC02-98CH10886(BNL).

\FloatBarrier
\section*{Note added}
\FloatBarrier

After the completion of this work, a measurement of the ratio of partially integrated $\Lambda_b \to p\: \mu^- \bar{\nu}_\mu$ and $\Lambda_b \to \Lambda_c\: \mu^- \bar{\nu}_\mu$
decay rates was published by the LHCb Collaboration, with the result \cite{Aaij:2015bfa}
\begin{eqnarray}
\frac{\int_{15\:{\rm GeV}^2}^{q^2_{\rm max}}
\frac{\mathrm{d}\Gamma (\Lambda_b \to p\: \mu^- \bar{\nu}_\mu)}{\mathrm{d}q^2} \mathrm{d} q^2}{\int_{7\:{\rm GeV}^2}^{q^2_{\rm max}}
\frac{\mathrm{d}\Gamma (\Lambda_b \to \Lambda_c\: \mu^- \bar{\nu}_\mu)}{\mathrm{d}q^2} \mathrm{d} q^2} &=& (1.00 \pm 0.04 \pm 0.08)\times 10^{-2},
\end{eqnarray}
where the first uncertainty is statistical and the second uncertainty is systematic. Combined with our lattice QCD result in Eq.~(\ref{eq:VubVcb}), this gives \cite{Aaij:2015bfa}
\begin{equation}
 \frac{|V_{ub}|}{|V_{cb}|} = 0.083 \pm 0.004 ({\rm expt}) \pm 0.004 ({\rm lattice}),
\end{equation}
and, taking the value of $|V_{cb}|$ extracted from exclusive $B$ decays \cite{Aaij:2015bfa},
\begin{equation}
 |V_{ub}| = \left( 3.27 \pm 0.15({\rm expt}) \pm 0.16 ({\rm lattice}) \pm 0.06 (|V_{cb}|) \right)\times 10^{-3}.
\end{equation}

\newpage
\appendix

\FloatBarrier
\section{\label{sec:fftables}Tables of lattice form factor data}
\FloatBarrier

\begin{table}
\vspace{-5ex}
\footnotesize
\begin{tabular}{ccccllllllllllll}
\hline\hline
 $f(\Lambda_b \to p)$ & & $|\mathbf{p'}|^2/(2\pi/L)^2$ && \hspace{2.5ex} \texttt{C14} & \hspace{2ex} & \hspace{2.5ex} \texttt{C24} & \hspace{2ex} & \hspace{2.5ex} \texttt{C54} & \hspace{2ex} & \hspace{2.5ex} \texttt{F23} & \hspace{2ex} & \hspace{2.5ex} \texttt{F43} & \hspace{2ex} & \hspace{2.5ex} \texttt{F63} \\
\hline
 $f_+$                 && 1 &&  $1.436(60)$ &&  $1.417(52)$ &&  $1.429(54)$ &&  $1.422(60)$ &&  $1.419(51)$ &&  $1.436(41)$  \\ 
                       && 2 &&  $1.209(66)$ &&  $1.202(59)$ &&  $1.213(55)$ &&  $1.210(83)$ &&  $1.218(57)$ &&  $1.236(44)$  \\ 
                       && 3 &&  $1.037(73)$ &&  $1.037(52)$ &&  $1.050(50)$ &&  $1.032(56)$ &&  $1.051(36)$ &&  $1.083(34)$  \\ 
                       && 4 &&  $0.912(64)$ &&  $0.925(27)$ &&  $0.938(27)$ &&  $0.968(31)$ &&  $0.964(22)$ &&  $0.969(19)$  \\ 
                       && 5 &&  $0.809(34)$ &&  $0.823(26)$ &&  $0.836(26)$ &&  $0.856(30)$ &&  $0.857(22)$ &&  $0.875(19)$  \\ 
                       && 6 &&  $0.740(34)$ &&  $0.754(26)$ &&  $0.768(26)$ &&  $0.780(29)$ &&  $0.788(22)$ &&  $0.810(19)$  \\ 
                       && 8 &&  $0.614(35)$ &&  $0.651(27)$ &&  $0.664(27)$ &&  $0.682(29)$ &&  $0.690(21)$ &&  $0.710(19)$  \\ 
                       && 9 &&  $0.590(36)$ &&  $0.608(28)$ &&  $0.623(27)$ &&  $0.655(29)$ &&  $0.649(22)$ &&  $0.672(19)$  \\ 
\hline
 $f_\perp$             && 1 &&  $1.767(88)$ &&  $1.762(64)$ &&  $1.802(66)$ &&  $1.780(81)$ &&  $1.771(73)$ &&  $1.804(59)$  \\ 
                       && 2 &&  $1.526(85)$ &&  $1.523(68)$ &&  $1.558(64)$ &&  $1.547(97)$ &&  $1.554(71)$ &&  $1.582(50)$  \\ 
                       && 3 &&  $1.32(11)$ &&  $1.325(85)$ &&  $1.353(81)$ &&  $1.31(11)$ &&  $1.333(76)$ &&  $1.390(57)$  \\ 
                       && 4 &&  $1.136(95)$ &&  $1.156(61)$ &&  $1.185(54)$ &&  $1.201(57)$ &&  $1.203(34)$ &&  $1.224(46)$  \\ 
                       && 5 &&  $1.009(50)$ &&  $1.024(39)$ &&  $1.056(39)$ &&  $1.068(45)$ &&  $1.079(32)$ &&  $1.112(29)$  \\ 
                       && 6 &&  $0.923(50)$ &&  $0.941(39)$ &&  $0.968(39)$ &&  $0.964(44)$ &&  $0.988(32)$ &&  $1.029(29)$  \\ 
                       && 8 &&  $0.756(50)$ &&  $0.805(40)$ &&  $0.830(39)$ &&  $0.839(45)$ &&  $0.857(32)$ &&  $0.893(30)$  \\ 
                       && 9 &&  $0.726(51)$ &&  $0.754(41)$ &&  $0.781(40)$ &&  $0.808(46)$ &&  $0.811(33)$ &&  $0.849(31)$  \\ 
\hline
 $f_0$                 && 1 &&  $1.056(24)$ &&  $1.011(26)$ &&  $1.008(26)$ &&  $1.051(29)$ &&  $1.040(24)$ &&  $1.025(21)$  \\ 
                       && 2 &&  $0.887(37)$ &&  $0.878(34)$ &&  $0.874(37)$ &&  $0.889(43)$ &&  $0.894(36)$ &&  $0.891(36)$  \\ 
                       && 3 &&  $0.775(39)$ &&  $0.777(24)$ &&  $0.777(24)$ &&  $0.788(29)$ &&  $0.796(24)$ &&  $0.798(23)$  \\ 
                       && 4 &&  $0.718(40)$ &&  $0.727(25)$ &&  $0.724(29)$ &&  $0.757(49)$ &&  $0.748(48)$ &&  $0.746(20)$  \\ 
                       && 5 &&  $0.654(24)$ &&  $0.674(20)$ &&  $0.672(20)$ &&  $0.666(26)$ &&  $0.669(21)$ &&  $0.681(17)$  \\ 
                       && 6 &&  $0.606(24)$ &&  $0.631(20)$ &&  $0.633(19)$ &&  $0.626(25)$ &&  $0.627(21)$ &&  $0.645(17)$  \\ 
                       && 8 &&  $0.516(26)$ &&  $0.549(22)$ &&  $0.555(20)$ &&  $0.550(29)$ &&  $0.555(24)$ &&  $0.581(19)$  \\ 
                       && 9 &&  $0.502(30)$ &&  $0.529(28)$ &&  $0.535(20)$ &&  $0.530(29)$ &&  $0.530(26)$ &&  $0.559(21)$  \\ 
\hline
 $g_+$                 && 1 &&  $0.952(17)$ &&  $0.922(17)$ &&  $0.920(14)$ &&  $0.947(24)$ &&  $0.939(23)$ &&  $0.925(20)$  \\ 
                       && 2 &&  $0.828(19)$ &&  $0.815(17)$ &&  $0.813(16)$ &&  $0.823(27)$ &&  $0.826(24)$ &&  $0.821(21)$  \\ 
                       && 3 &&  $0.721(28)$ &&  $0.721(19)$ &&  $0.722(19)$ &&  $0.721(33)$ &&  $0.729(25)$ &&  $0.735(22)$  \\ 
                       && 4 &&  $0.648(24)$ &&  $0.664(16)$ &&  $0.663(12)$ &&  $0.665(27)$ &&  $0.659(24)$ &&  $0.663(21)$  \\ 
                       && 5 &&  $0.582(20)$ &&  $0.606(15)$ &&  $0.606(11)$ &&  $0.598(27)$ &&  $0.600(27)$ &&  $0.607(23)$  \\ 
                       && 6 &&  $0.540(20)$ &&  $0.567(15)$ &&  $0.569(11)$ &&  $0.559(30)$ &&  $0.561(31)$ &&  $0.576(26)$  \\ 
                       && 8 &&  $0.463(22)$ &&  $0.497(16)$ &&  $0.505(12)$ &&  $0.494(38)$ &&  $0.496(36)$ &&  $0.517(30)$  \\ 
                       && 9 &&  $0.458(25)$ &&  $0.481(24)$ &&  $0.489(12)$ &&  $0.487(30)$ &&  $0.483(30)$ &&  $0.507(25)$  \\ 
\hline
 $g_\perp$             && 1 &&  $0.952(23)$ &&  $0.920(20)$ &&  $0.919(16)$ &&  $0.947(30)$ &&  $0.939(28)$ &&  $0.924(25)$  \\ 
                       && 2 &&  $0.827(25)$ &&  $0.812(20)$ &&  $0.811(18)$ &&  $0.818(32)$ &&  $0.822(30)$ &&  $0.817(26)$  \\ 
                       && 3 &&  $0.719(32)$ &&  $0.720(22)$ &&  $0.720(22)$ &&  $0.715(38)$ &&  $0.723(31)$ &&  $0.729(27)$  \\ 
                       && 4 &&  $0.643(29)$ &&  $0.659(20)$ &&  $0.658(16)$ &&  $0.657(33)$ &&  $0.651(30)$ &&  $0.655(26)$  \\ 
                       && 5 &&  $0.578(26)$ &&  $0.603(19)$ &&  $0.602(14)$ &&  $0.586(34)$ &&  $0.589(33)$ &&  $0.598(28)$  \\ 
                       && 6 &&  $0.535(27)$ &&  $0.566(20)$ &&  $0.566(15)$ &&  $0.546(38)$ &&  $0.549(37)$ &&  $0.566(32)$  \\ 
                       && 8 &&  $0.456(36)$ &&  $0.495(24)$ &&  $0.503(17)$ &&  $0.476(51)$ &&  $0.480(48)$ &&  $0.506(40)$  \\ 
                       && 9 &&  $0.454(39)$ &&  $0.482(36)$ &&  $0.491(17)$ &&  $0.473(45)$ &&  $0.470(45)$ &&  $0.499(36)$  \\ 
\hline
 $g_0$                 && 1 &&  $1.475(60)$ &&  $1.469(47)$ &&  $1.477(41)$ &&  $1.496(75)$ &&  $1.469(57)$ &&  $1.476(43)$  \\ 
                       && 2 &&  $1.237(44)$ &&  $1.229(39)$ &&  $1.242(36)$ &&  $1.274(50)$ &&  $1.262(42)$ &&  $1.275(26)$  \\ 
                       && 3 &&  $1.055(24)$ &&  $1.048(21)$ &&  $1.065(21)$ &&  $1.082(28)$ &&  $1.089(22)$ &&  $1.113(17)$  \\ 
                       && 4 &&  $0.912(22)$ &&  $0.924(20)$ &&  $0.935(19)$ &&  $0.982(25)$ &&  $0.972(19)$ &&  $0.972(15)$  \\ 
                       && 5 &&  $0.808(21)$ &&  $0.818(18)$ &&  $0.833(19)$ &&  $0.872(22)$ &&  $0.867(17)$ &&  $0.881(14)$  \\ 
                       && 6 &&  $0.734(21)$ &&  $0.746(18)$ &&  $0.759(19)$ &&  $0.782(21)$ &&  $0.789(17)$ &&  $0.811(14)$  \\ 
                       && 8 &&  $0.603(30)$ &&  $0.631(18)$ &&  $0.646(19)$ &&  $0.668(20)$ &&  $0.675(17)$ &&  $0.695(14)$  \\ 
                       && 9 &&  $0.576(24)$ &&  $0.592(20)$ &&  $0.607(20)$ &&  $0.640(23)$ &&  $0.638(18)$ &&  $0.660(16)$  \\ 
\hline\hline
\end{tabular}
\normalsize
\caption{$\Lambda_b \to p$ helicity form factors.}
\end{table}

\begin{table}
\footnotesize
\begin{tabular}{ccccllllllllllll}
\hline\hline
 $f(\Lambda_b \to p)$ & & $|\mathbf{p'}|^2/(2\pi/L)^2$ && \hspace{4ex} \texttt{C14} & \hspace{2ex} & \hspace{4ex} \texttt{C24} & \hspace{2ex} & \hspace{4ex} \texttt{C54} & \hspace{2ex} & \hspace{4ex} \texttt{F23} & \hspace{2ex} & \hspace{4ex} \texttt{F43} & \hspace{2ex} & \hspace{4ex} \texttt{F63} \\
\hline
 $f_1^V$               && 1 &&  $\wm1.168(42)$ &&  $\wm1.144(46)$ &&  $\wm1.152(47)$ &&  $\wm1.123(48)$ &&  $\wm1.138(37)$ &&  $\wm1.164(31)$  \\ 
                       && 2 &&  $\wm0.974(55)$ &&  $\wm0.968(54)$ &&  $\wm0.977(51)$ &&  $\wm0.952(75)$ &&  $\wm0.973(49)$ &&  $\wm1.000(41)$  \\ 
                       && 3 &&  $\wm0.846(49)$ &&  $\wm0.844(31)$ &&  $\wm0.858(31)$ &&  $\wm0.838(24)$ &&  $\wm0.860(20)$ &&  $\wm0.889(21)$  \\ 
                       && 4 &&  $\wm0.771(44)$ &&  $\wm0.782(19)$ &&  $\wm0.792(19)$ &&  $\wm0.816(24)$ &&  $\wm0.814(25)$ &&  $\wm0.819(15)$  \\ 
                       && 5 &&  $\wm0.692(25)$ &&  $\wm0.707(20)$ &&  $\wm0.715(20)$ &&  $\wm0.728(26)$ &&  $\wm0.729(21)$ &&  $\wm0.745(15)$  \\ 
                       && 6 &&  $\wm0.641(25)$ &&  $\wm0.654(19)$ &&  $\wm0.666(20)$ &&  $\wm0.677(23)$ &&  $\wm0.681(20)$ &&  $\wm0.698(15)$  \\ 
                       && 8 &&  $\wm0.547(26)$ &&  $\wm0.579(20)$ &&  $\wm0.590(20)$ &&  $\wm0.606(23)$ &&  $\wm0.611(19)$ &&  $\wm0.629(15)$  \\ 
                       && 9 &&  $\wm0.530(27)$ &&  $\wm0.545(21)$ &&  $\wm0.557(20)$ &&  $\wm0.587(23)$ &&  $\wm0.578(20)$ &&  $\wm0.597(15)$  \\ 
\hline
 $f_2^V$               && 1 &&  $\wm0.505(46)$ &&  $\wm0.520(29)$ &&  $\wm0.543(25)$ &&  $\wm0.556(45)$ &&  $\wm0.532(39)$ &&  $\wm0.534(30)$  \\ 
                       && 2 &&  $\wm0.465(32)$ &&  $\wm0.466(28)$ &&  $\wm0.485(25)$ &&  $\wm0.503(39)$ &&  $\wm0.489(30)$ &&  $\wm0.485(23)$  \\ 
                       && 3 &&  $\wm0.397(53)$ &&  $\wm0.405(48)$ &&  $\wm0.413(43)$ &&  $\wm0.397(78)$ &&  $\wm0.397(58)$ &&  $\wm0.417(32)$  \\ 
                       && 4 &&  $\wm0.307(44)$ &&  $\wm0.315(47)$ &&  $\wm0.328(46)$ &&  $\wm0.325(54)$ &&  $\wm0.327(40)$ &&  $\wm0.337(40)$  \\ 
                       && 5 &&  $\wm0.267(24)$ &&  $\wm0.267(20)$ &&  $\wm0.285(18)$ &&  $\wm0.288(29)$ &&  $\wm0.295(25)$ &&  $\wm0.306(18)$  \\ 
                       && 6 &&  $\wm0.238(24)$ &&  $\wm0.241(20)$ &&  $\wm0.251(18)$ &&  $\wm0.243(28)$ &&  $\wm0.258(24)$ &&  $\wm0.276(18)$  \\ 
                       && 8 &&  $\wm0.176(24)$ &&  $\wm0.190(20)$ &&  $\wm0.200(18)$ &&  $\wm0.198(29)$ &&  $\wm0.207(25)$ &&  $\wm0.220(19)$  \\ 
                       && 9 &&  $\wm0.165(25)$ &&  $\wm0.175(21)$ &&  $\wm0.187(19)$ &&  $\wm0.187(30)$ &&  $\wm0.196(26)$ &&  $\wm0.210(20)$  \\ 
\hline
 $f_3^V$               && 1 &&  $-0.145(59)$ &&  $-0.172(46)$ &&  $-0.188(42)$ &&  $-0.092(64)$ &&  $-0.127(43)$ &&  $-0.181(29)$  \\ 
                       && 2 &&  $-0.118(38)$ &&  $-0.123(41)$ &&  $-0.141(33)$ &&  $-0.085(59)$ &&  $-0.106(33)$ &&  $-0.149(24)$  \\ 
                       && 3 &&  $-0.101(37)$ &&  $-0.095(39)$ &&  $-0.116(34)$ &&  $-0.072(41)$ &&  $-0.091(31)$ &&  $-0.131(25)$  \\ 
                       && 4 &&  $-0.079(39)$ &&  $-0.081(43)$ &&  $-0.102(41)$ &&  $-0.088(61)$ &&  $-0.099(40)$ &&  $-0.111(27)$  \\ 
                       && 5 &&  $-0.059(33)$ &&  $-0.051(34)$ &&  $-0.068(31)$ &&  $-0.097(39)$ &&  $-0.094(29)$ &&  $-0.100(28)$  \\ 
                       && 6 &&  $-0.056(34)$ &&  $-0.037(34)$ &&  $-0.055(29)$ &&  $-0.083(40)$ &&  $-0.087(30)$ &&  $-0.087(23)$  \\ 
                       && 8 &&  $-0.056(44)$ &&  $-0.054(42)$ &&  $-0.064(33)$ &&  $-0.101(48)$ &&  $-0.101(38)$ &&  $-0.086(29)$  \\ 
                       && 9 &&  $-0.053(58)$ &&  $-0.031(44)$ &&  $-0.043(34)$ &&  $-0.107(50)$ &&  $-0.092(46)$ &&  $-0.072(35)$  \\ 
\hline
 $f_1^A$               && 1 &&  $\wm0.959(18)$ &&  $\wm0.957(19)$ &&  $\wm0.950(17)$ &&  $\wm0.953(21)$ &&  $\wm0.947(17)$ &&  $\wm0.953(15)$  \\ 
                       && 2 &&  $\wm0.846(18)$ &&  $\wm0.837(20)$ &&  $\wm0.838(17)$ &&  $\wm0.868(24)$ &&  $\wm0.868(22)$ &&  $\wm0.865(20)$  \\ 
                       && 3 &&  $\wm0.731(13)$ &&  $\wm0.726(12)$ &&  $\wm0.738(13)$ &&  $\wm0.762(18)$ &&  $\wm0.769(16)$ &&  $\wm0.773(16)$  \\ 
                       && 4 &&  $\wm0.668(14)$ &&  $\wm0.685(15)$ &&  $\wm0.687(13)$ &&  $\wm0.701(15)$ &&  $\wm0.696(13)$ &&  $\wm0.701(11)$  \\ 
                       && 5 &&  $\wm0.598(12)$ &&  $\wm0.614(12)$ &&  $\wm0.621(12)$ &&  $\wm0.642(14)$ &&  $\wm0.639(12)$ &&  $\wm0.643(11)$  \\ 
                       && 6 &&  $\wm0.554(12)$ &&  $\wm0.568(11)$ &&  $\wm0.576(11)$ &&  $\wm0.597(14)$ &&  $\wm0.599(15)$ &&  $\wm0.606(12)$  \\ 
                       && 8 &&  $\wm0.480(21)$ &&  $\wm0.500(11)$ &&  $\wm0.508(14)$ &&  $\wm0.532(16)$ &&  $\wm0.531(13)$ &&  $\wm0.543(12)$  \\ 
                       && 9 &&  $\wm0.467(16)$ &&  $\wm0.477(13)$ &&  $\wm0.487(16)$ &&  $\wm0.512(15)$ &&  $\wm0.508(14)$ &&  $\wm0.522(12)$  \\ 
\hline
 $f_2^A$               && 1 &&  $\wm0.008(39)$ &&  $\wm0.046(33)$ &&  $\wm0.039(29)$ &&  $\wm0.007(41)$ &&  $\wm0.010(37)$ &&  $\wm0.036(34)$  \\ 
                       && 2 &&  $\wm0.024(38)$ &&  $\wm0.030(33)$ &&  $\wm0.034(34)$ &&  $\wm0.061(45)$ &&  $\wm0.057(40)$ &&  $\wm0.059(41)$  \\ 
                       && 3 &&  $\wm0.015(39)$ &&  $\wm0.007(30)$ &&  $\wm0.023(33)$ &&  $\wm0.058(52)$ &&  $\wm0.057(42)$ &&  $\wm0.055(41)$  \\ 
                       && 4 &&  $\wm0.030(40)$ &&  $\wm0.032(34)$ &&  $\wm0.036(29)$ &&  $\wm0.053(42)$ &&  $\wm0.056(37)$ &&  $\wm0.057(37)$  \\ 
                       && 5 &&  $\wm0.024(38)$ &&  $\wm0.012(29)$ &&  $\wm0.023(27)$ &&  $\wm0.068(40)$ &&  $\wm0.061(37)$ &&  $\wm0.056(34)$  \\ 
                       && 6 &&  $\wm0.023(39)$ &&  $\wm0.003(30)$ &&  $\wm0.012(27)$ &&  $\wm0.062(42)$ &&  $\wm0.061(38)$ &&  $\wm0.050(35)$  \\ 
                       && 8 &&  $\wm0.030(64)$ &&  $\wm0.006(35)$ &&  $\wm0.007(31)$ &&  $\wm0.068(55)$ &&  $\wm0.064(52)$ &&  $\wm0.046(42)$  \\ 
                       && 9 &&  $\wm0.017(53)$ &&  $-0.006(45)$ &&  $-0.004(35)$ &&  $\wm0.048(57)$ &&  $\wm0.048(57)$ &&  $\wm0.028(42)$  \\ 
\hline
 $f_3^A$               && 1 &&  $-0.98(12)$ &&  $-0.976(89)$ &&  $-1.033(80)$ &&  $-1.01(15)$ &&  $-0.99(12)$ &&  $-1.027(86)$  \\ 
                       && 2 &&  $-0.77(10)$ &&  $-0.784(98)$ &&  $-0.829(94)$ &&  $-0.79(12)$ &&  $-0.78(11)$ &&  $-0.844(79)$  \\ 
                       && 3 &&  $-0.672(44)$ &&  $-0.676(39)$ &&  $-0.702(32)$ &&  $-0.654(53)$ &&  $-0.667(43)$ &&  $-0.732(33)$  \\ 
                       && 4 &&  $-0.532(43)$ &&  $-0.526(37)$ &&  $-0.558(31)$ &&  $-0.605(53)$ &&  $-0.601(42)$ &&  $-0.610(32)$  \\ 
                       && 5 &&  $-0.479(40)$ &&  $-0.471(35)$ &&  $-0.501(29)$ &&  $-0.520(48)$ &&  $-0.521(40)$ &&  $-0.561(31)$  \\ 
                       && 6 &&  $-0.429(40)$ &&  $-0.428(34)$ &&  $-0.452(29)$ &&  $-0.437(45)$ &&  $-0.457(40)$ &&  $-0.504(31)$  \\ 
                       && 8 &&  $-0.323(41)$ &&  $-0.350(36)$ &&  $-0.372(30)$ &&  $-0.353(47)$ &&  $-0.379(42)$ &&  $-0.412(33)$  \\ 
                       && 9 &&  $-0.299(43)$ &&  $-0.318(37)$ &&  $-0.340(31)$ &&  $-0.348(50)$ &&  $-0.358(44)$ &&  $-0.391(35)$  \\ 
\hline\hline
\end{tabular}
\normalsize
\caption{\label{LbpWeinberg}$\Lambda_b \to p$ Weinberg form factors.}
\end{table}

\begin{table}
\footnotesize
\begin{tabular}{ccccllllllllllll}
\hline\hline
 $f(\Lambda_b \to \Lambda_c)$ & & $|\mathbf{p'}|^2/(2\pi/L)^2$ && \hspace{2.5ex} \texttt{C14} & \hspace{2ex} & \hspace{2.5ex} \texttt{C24} & \hspace{2ex} & \hspace{2.5ex} \texttt{C54} & \hspace{2ex} & \hspace{2.5ex} \texttt{F23} & \hspace{2ex} & \hspace{2.5ex} \texttt{F43} & \hspace{2ex} & \hspace{2.5ex} \texttt{F63} \\
\hline
$f_+$                  && 1 &&  $1.0401(82)$ &&  $1.0114(81)$ &&  $1.0170(77)$ &&  $1.068(15)$ &&  $1.062(13)$ &&  $1.047(10)$  \\ 
                       && 2 &&  $0.9887(74)$ &&  $0.9613(76)$ &&  $0.9676(72)$ &&  $1.016(14)$ &&  $1.011(12)$ &&  $0.998(10)$  \\ 
                       && 3 &&  $0.9418(69)$ &&  $0.9167(73)$ &&  $0.9229(69)$ &&  $0.968(13)$ &&  $0.963(12)$ &&  $0.9528(95)$  \\ 
                       && 4 &&  $0.9018(92)$ &&  $0.8826(77)$ &&  $0.8852(71)$ &&  $0.936(13)$ &&  $0.927(12)$ &&  $0.9133(88)$  \\ 
                       && 5 &&  $0.864(12)$ &&  $0.8515(73)$ &&  $0.8528(67)$ &&  $0.894(17)$ &&  $0.887(17)$ &&  $0.876(14)$  \\ 
                       && 6 &&  $0.8327(67)$ &&  $0.820(11)$ &&  $0.8202(81)$ &&  $0.859(15)$ &&  $0.852(14)$ &&  $0.8455(96)$  \\ 
                       && 8 &&  $0.7692(81)$ &&  $0.757(12)$ &&  $0.758(10)$ &&  $0.799(15)$ &&  $0.791(14)$ &&  $0.788(11)$  \\ 
                       && 9 &&  $0.7406(98)$ &&  $0.727(14)$ &&  $0.728(12)$ &&  $0.766(15)$ &&  $0.758(14)$ &&  $0.757(12)$  \\ 
                       && 10 &&  $0.7098(89)$ &&  $0.709(16)$ &&  $0.708(15)$ &&  $0.748(18)$ &&  $0.737(16)$ &&  $0.735(14)$  \\ 
\hline
$f_\perp$              && 1 &&  $1.467(17)$ &&  $1.431(13)$ &&  $1.450(17)$ &&  $1.464(23)$ &&  $1.458(16)$ &&  $1.453(14)$  \\ 
                       && 2 &&  $1.400(13)$ &&  $1.368(12)$ &&  $1.386(13)$ &&  $1.398(19)$ &&  $1.394(14)$ &&  $1.390(13)$  \\ 
                       && 3 &&  $1.339(12)$ &&  $1.308(12)$ &&  $1.326(13)$ &&  $1.335(17)$ &&  $1.333(13)$ &&  $1.330(12)$  \\ 
                       && 4 &&  $1.268(17)$ &&  $1.244(12)$ &&  $1.257(15)$ &&  $1.282(18)$ &&  $1.276(17)$ &&  $1.270(17)$  \\ 
                       && 5 &&  $1.219(20)$ &&  $1.204(12)$ &&  $1.215(12)$ &&  $1.228(18)$ &&  $1.224(21)$ &&  $1.218(21)$  \\ 
                       && 6 &&  $1.180(12)$ &&  $1.163(13)$ &&  $1.172(12)$ &&  $1.185(15)$ &&  $1.181(14)$ &&  $1.176(11)$  \\ 
                       && 8 &&  $1.094(14)$ &&  $1.078(18)$ &&  $1.086(15)$ &&  $1.103(19)$ &&  $1.099(17)$ &&  $1.102(16)$  \\ 
                       && 9 &&  $1.056(14)$ &&  $1.039(19)$ &&  $1.047(16)$ &&  $1.055(18)$ &&  $1.054(16)$ &&  $1.061(16)$  \\ 
                       && 10 &&  $0.997(14)$ &&  $1.000(23)$ &&  $1.004(21)$ &&  $1.021(22)$ &&  $1.015(20)$ &&  $1.027(20)$  \\ 
\hline
$f_0$                  && 1 &&  $0.9025(45)$ &&  $0.8952(57)$ &&  $0.8937(54)$ &&  $0.945(13)$ &&  $0.9392(98)$ &&  $0.9206(62)$  \\ 
                       && 2 &&  $0.8674(41)$ &&  $0.8598(54)$ &&  $0.8586(51)$ &&  $0.906(12)$ &&  $0.8996(88)$ &&  $0.8846(57)$  \\ 
                       && 3 &&  $0.8336(38)$ &&  $0.8273(50)$ &&  $0.8258(48)$ &&  $0.867(11)$ &&  $0.8619(81)$ &&  $0.8508(51)$  \\ 
                       && 4 &&  $0.8032(46)$ &&  $0.7935(61)$ &&  $0.7920(58)$ &&  $0.842(13)$ &&  $0.8337(86)$ &&  $0.8231(75)$  \\ 
                       && 5 &&  $0.7748(41)$ &&  $0.7714(46)$ &&  $0.7692(43)$ &&  $0.805(12)$ &&  $0.8012(99)$ &&  $0.7939(68)$  \\ 
                       && 6 &&  $0.7458(46)$ &&  $0.7429(53)$ &&  $0.7409(53)$ &&  $0.772(10)$ &&  $0.7691(88)$ &&  $0.7669(76)$  \\ 
                       && 8 &&  $0.6970(70)$ &&  $0.6928(75)$ &&  $0.6924(78)$ &&  $0.729(17)$ &&  $0.723(13)$ &&  $0.723(11)$  \\ 
                       && 9 &&  $0.6716(69)$ &&  $0.6655(68)$ &&  $0.6655(70)$ &&  $0.696(14)$ &&  $0.691(11)$ &&  $0.693(10)$  \\ 
                       && 10 &&  $0.6582(84)$ &&  $0.659(12)$ &&  $0.658(12)$ &&  $0.694(22)$ &&  $0.685(18)$ &&  $0.685(16)$  \\ 
\hline
$g_+$                  && 1 &&  $0.8397(32)$ &&  $0.8334(62)$ &&  $0.8318(56)$ &&  $0.8724(73)$ &&  $0.8673(57)$ &&  $0.8512(52)$  \\ 
                       && 2 &&  $0.8069(33)$ &&  $0.8024(70)$ &&  $0.7998(56)$ &&  $0.8426(71)$ &&  $0.8361(54)$ &&  $0.8173(30)$  \\ 
                       && 3 &&  $0.7777(30)$ &&  $0.7738(58)$ &&  $0.7718(53)$ &&  $0.8099(66)$ &&  $0.8031(49)$ &&  $0.7887(29)$  \\ 
                       && 4 &&  $0.7527(26)$ &&  $0.7498(68)$ &&  $0.7476(54)$ &&  $0.7783(58)$ &&  $0.7728(43)$ &&  $0.7633(28)$  \\ 
                       && 5 &&  $0.7268(28)$ &&  $0.7232(56)$ &&  $0.7217(50)$ &&  $0.7503(56)$ &&  $0.7442(42)$ &&  $0.7378(30)$  \\ 
                       && 6 &&  $0.7023(34)$ &&  $0.6978(59)$ &&  $0.6965(53)$ &&  $0.7208(57)$ &&  $0.7155(49)$ &&  $0.7147(32)$  \\ 
                       && 8 &&  $0.6595(45)$ &&  $0.6537(65)$ &&  $0.6536(56)$ &&  $0.6731(65)$ &&  $0.6680(54)$ &&  $0.6732(45)$  \\ 
                       && 9 &&  $0.6402(52)$ &&  $0.6311(72)$ &&  $0.6319(70)$ &&  $0.650(11)$ &&  $0.6452(91)$ &&  $0.6547(62)$  \\ 
                       && 10 &&  $0.6240(71)$ &&  $0.624(11)$ &&  $0.625(11)$ &&  $0.641(10)$ &&  $0.6352(87)$ &&  $0.642(11)$  \\ 
\hline
$g_\perp$              && 1 &&  $0.8389(35)$ &&  $0.8332(50)$ &&  $0.8315(48)$ &&  $0.8720(72)$ &&  $0.8663(57)$ &&  $0.8510(54)$  \\ 
                       && 2 &&  $0.8054(37)$ &&  $0.8016(67)$ &&  $0.7989(52)$ &&  $0.8415(72)$ &&  $0.8347(57)$ &&  $0.8167(34)$  \\ 
                       && 3 &&  $0.7756(35)$ &&  $0.7729(51)$ &&  $0.7708(44)$ &&  $0.8081(68)$ &&  $0.8017(50)$ &&  $0.7878(33)$  \\ 
                       && 4 &&  $0.7511(32)$ &&  $0.7500(64)$ &&  $0.7472(55)$ &&  $0.7775(56)$ &&  $0.7714(42)$ &&  $0.7634(32)$  \\ 
                       && 5 &&  $0.7244(34)$ &&  $0.7226(47)$ &&  $0.7206(48)$ &&  $0.7482(54)$ &&  $0.7426(41)$ &&  $0.7373(33)$  \\ 
                       && 6 &&  $0.6983(41)$ &&  $0.6958(50)$ &&  $0.6944(45)$ &&  $0.7159(56)$ &&  $0.7116(47)$ &&  $0.7120(33)$  \\ 
                       && 8 &&  $0.6540(57)$ &&  $0.6511(57)$ &&  $0.6510(49)$ &&  $0.6664(56)$ &&  $0.6626(48)$ &&  $0.6691(41)$  \\ 
                       && 9 &&  $0.6335(58)$ &&  $0.6275(67)$ &&  $0.6284(56)$ &&  $0.6408(94)$ &&  $0.6374(81)$ &&  $0.6495(52)$  \\ 
                       && 10 &&  $0.6206(90)$ &&  $0.6230(91)$ &&  $0.6233(95)$ &&  $0.637(11)$ &&  $0.6325(88)$ &&  $0.641(12)$  \\ 
\hline
$g_0$                  && 1 &&  $0.9771(97)$ &&  $0.959(12)$ &&  $0.9608(99)$ &&  $1.007(15)$ &&  $0.998(13)$ &&  $0.9801(93)$  \\ 
                       && 2 &&  $0.9296(66)$ &&  $0.913(11)$ &&  $0.9151(88)$ &&  $0.958(13)$ &&  $0.951(11)$ &&  $0.9292(75)$  \\ 
                       && 3 &&  $0.8866(63)$ &&  $0.873(11)$ &&  $0.8740(86)$ &&  $0.916(12)$ &&  $0.908(10)$ &&  $0.8891(67)$  \\ 
                       && 4 &&  $0.8478(61)$ &&  $0.838(11)$ &&  $0.8383(86)$ &&  $0.873(13)$ &&  $0.866(13)$ &&  $0.853(11)$  \\ 
                       && 5 &&  $0.8141(62)$ &&  $0.804(11)$ &&  $0.8044(86)$ &&  $0.838(12)$ &&  $0.831(10)$ &&  $0.8197(88)$  \\ 
                       && 6 &&  $0.7854(66)$ &&  $0.774(12)$ &&  $0.7737(95)$ &&  $0.807(12)$ &&  $0.800(11)$ &&  $0.7950(76)$  \\ 
                       && 8 &&  $0.7281(86)$ &&  $0.716(14)$ &&  $0.717(11)$ &&  $0.745(15)$ &&  $0.738(13)$ &&  $0.7409(95)$  \\ 
                       && 9 &&  $0.704(11)$ &&  $0.688(16)$ &&  $0.690(14)$ &&  $0.718(19)$ &&  $0.712(16)$ &&  $0.717(12)$  \\ 
                       && 10 &&  $0.6732(91)$ &&  $0.676(18)$ &&  $0.675(16)$ &&  $0.698(18)$ &&  $0.691(15)$ &&  $0.695(13)$  \\ 
\hline\hline
\end{tabular}
\normalsize
\caption{$\Lambda_b \to \Lambda_c$ helicity form factors.}
\end{table}

\begin{table}
\footnotesize
\begin{tabular}{ccccllllllllllll}
\hline\hline
 $f(\Lambda_b \to \Lambda_c)$ & & $|\mathbf{p'}|^2/(2\pi/L)^2$ && \hspace{4ex} \texttt{C14} & \hspace{2ex} & \hspace{4ex} \texttt{C24} & \hspace{2ex} & \hspace{4ex} \texttt{C54} & \hspace{2ex} & \hspace{4ex} \texttt{F23} & \hspace{2ex} & \hspace{4ex} \texttt{F43} & \hspace{2ex} & \hspace{4ex} \texttt{F63} \\
\hline
 $f_1^V$               && 1 &&  $\wm0.9534(80)$ &&  $\wm0.9273(80)$ &&  $\wm0.9316(71)$ &&  $\wm0.988(15)$ &&  $\wm0.981(14)$ &&  $\wm0.965(11)$  \\ 
                       && 2 &&  $\wm0.9096(72)$ &&  $\wm0.8842(74)$ &&  $\wm0.8894(66)$ &&  $\wm0.943(14)$ &&  $\wm0.937(13)$ &&  $\wm0.923(11)$  \\ 
                       && 3 &&  $\wm0.8697(66)$ &&  $\wm0.8465(71)$ &&  $\wm0.8517(63)$ &&  $\wm0.902(13)$ &&  $\wm0.896(12)$ &&  $\wm0.8852(95)$  \\ 
                       && 4 &&  $\wm0.8390(82)$ &&  $\wm0.8214(75)$ &&  $\wm0.8230(65)$ &&  $\wm0.877(14)$ &&  $\wm0.867(12)$ &&  $\wm0.8527(88)$  \\ 
                       && 5 &&  $\wm0.807(11)$ &&  $\wm0.7950(70)$ &&  $\wm0.7957(60)$ &&  $\wm0.840(18)$ &&  $\wm0.832(16)$ &&  $\wm0.821(13)$  \\ 
                       && 6 &&  $\wm0.7798(63)$ &&  $\wm0.768(11)$ &&  $\wm0.7679(80)$ &&  $\wm0.810(16)$ &&  $\wm0.802(14)$ &&  $\wm0.7954(98)$  \\ 
                       && 8 &&  $\wm0.7255(78)$ &&  $\wm0.714(12)$ &&  $\wm0.7154(98)$ &&  $\wm0.758(15)$ &&  $\wm0.749(15)$ &&  $\wm0.746(11)$  \\ 
                       && 9 &&  $\wm0.7009(96)$ &&  $\wm0.688(13)$ &&  $\wm0.689(11)$ &&  $\wm0.730(15)$ &&  $\wm0.721(14)$ &&  $\wm0.719(12)$  \\ 
                       && 10 &&  $\wm0.6761(87)$ &&  $\wm0.676(15)$ &&  $\wm0.674(14)$ &&  $\wm0.716(19)$ &&  $\wm0.704(16)$ &&  $\wm0.700(14)$  \\ 
\hline
$f_2^V$                && 1 &&  $\wm0.365(11)$ &&  $\wm0.3569(91)$ &&  $\wm0.367(11)$ &&  $\wm0.338(18)$ &&  $\wm0.338(13)$ &&  $\wm0.3455(96)$  \\ 
                       && 2 &&  $\wm0.3481(80)$ &&  $\wm0.3427(62)$ &&  $\wm0.3516(72)$ &&  $\wm0.323(13)$ &&  $\wm0.3241(87)$ &&  $\wm0.3308(73)$  \\ 
                       && 3 &&  $\wm0.3330(67)$ &&  $\wm0.3273(56)$ &&  $\wm0.3354(60)$ &&  $\wm0.308(11)$ &&  $\wm0.3107(83)$ &&  $\wm0.3149(71)$  \\ 
                       && 4 &&  $\wm0.3045(82)$ &&  $\wm0.2997(70)$ &&  $\wm0.3072(83)$ &&  $\wm0.287(11)$ &&  $\wm0.2903(86)$ &&  $\wm0.296(11)$  \\ 
                       && 5 &&  $\wm0.2927(78)$ &&  $\wm0.2903(52)$ &&  $\wm0.2968(56)$ &&  $\wm0.2756(10)$ &&  $\wm0.2784(76)$ &&  $\wm0.2813(75)$  \\ 
                       && 6 &&  $\wm0.2839(60)$ &&  $\wm0.2802(51)$ &&  $\wm0.2861(55)$ &&  $\wm0.2665(98)$ &&  $\wm0.2690(74)$ &&  $\wm0.2700(63)$  \\ 
                       && 8 &&  $\wm0.2618(76)$ &&  $\wm0.2578(66)$ &&  $\wm0.2624(65)$ &&  $\wm0.244(11)$ &&  $\wm0.248(11)$ &&  $\wm0.2523(94)$  \\ 
                       && 9 &&  $\wm0.2524(77)$ &&  $\wm0.2489(66)$ &&  $\wm0.2531(66)$ &&  $\wm0.231(11)$ &&  $\wm0.2363(10)$ &&  $\wm0.2422(90)$  \\ 
                       && 10 &&  $\wm0.2279(75)$ &&  $\wm0.2300(88)$ &&  $\wm0.2330(78)$ &&  $\wm0.216(13)$ &&  $\wm0.221(11)$ &&  $\wm0.231(12)$  \\ 
\hline
$f_3^V$                && 1 &&  $-0.090(15)$ &&  $-0.057(13)$ &&  $-0.068(11)$ &&  $-0.075(32)$ &&  $-0.074(29)$ &&  $-0.079(20)$  \\ 
                       && 2 &&  $-0.078(14)$ &&  $-0.045(12)$ &&  $-0.057(10)$ &&  $-0.070(31)$ &&  $-0.070(28)$ &&  $-0.072(20)$  \\ 
                       && 3 &&  $-0.070(13)$ &&  $-0.037(12)$ &&  $-0.051(10)$ &&  $-0.067(31)$ &&  $-0.065(29)$ &&  $-0.067(20)$  \\ 
                       && 4 &&  $-0.073(14)$ &&  $-0.057(13)$ &&  $-0.064(11)$ &&  $-0.072(31)$ &&  $-0.067(28)$ &&  $-0.060(19)$  \\ 
                       && 5 &&  $-0.068(22)$ &&  $-0.051(12)$ &&  $-0.0572(97)$ &&  $-0.074(41)$ &&  $-0.065(40)$ &&  $-0.058(31)$  \\ 
                       && 6 &&  $-0.076(14)$ &&  $-0.057(17)$ &&  $-0.061(11)$ &&  $-0.084(31)$ &&  $-0.073(28)$ &&  $-0.064(21)$  \\ 
                       && 8 &&  $-0.072(17)$ &&  $-0.054(21)$ &&  $-0.058(14)$ &&  $-0.075(50)$ &&  $-0.066(44)$ &&  $-0.057(28)$  \\ 
                       && 9 &&  $-0.078(25)$ &&  $-0.060(28)$ &&  $-0.063(21)$ &&  $-0.091(44)$ &&  $-0.080(38)$ &&  $-0.067(27)$  \\ 
                       && 10 &&  $-0.051(17)$ &&  $-0.049(20)$ &&  $-0.048(15)$ &&  $-0.063(54)$ &&  $-0.053(46)$ &&  $-0.042(29)$  \\ 
\hline
$f_1^A$                && 1 &&  $\wm0.8581(72)$ &&  $\wm0.838(12)$ &&  $\wm0.8386(87)$ &&  $\wm0.881(12)$ &&  $\wm0.890(10)$ &&  $\wm0.8552(65)$  \\ 
                       && 2 &&  $\wm0.8235(57)$ &&  $\wm0.810(12)$ &&  $\wm0.8089(85)$ &&  $\wm0.853(10)$ &&  $\wm0.8514(86)$ &&  $\wm0.8242(58)$  \\ 
                       && 3 &&  $\wm0.7918(53)$ &&  $\wm0.780(11)$ &&  $\wm0.7791(84)$ &&  $\wm0.822(10)$ &&  $\wm0.8126(84)$ &&  $\wm0.7948(58)$  \\ 
                       && 4 &&  $\wm0.7608(51)$ &&  $\wm0.749(11)$ &&  $\wm0.7498(83)$ &&  $\wm0.7824(94)$ &&  $\wm0.7795(83)$ &&  $\wm0.7627(69)$  \\ 
                       && 5 &&  $\wm0.7357(55)$ &&  $\wm0.726(11)$ &&  $\wm0.7255(84)$ &&  $\wm0.7579(96)$ &&  $\wm0.7502(81)$ &&  $\wm0.7397(67)$  \\ 
                       && 6 &&  $\wm0.7144(56)$ &&  $\wm0.704(12)$ &&  $\wm0.7030(96)$ &&  $\wm0.7355(95)$ &&  $\wm0.7271(82)$ &&  $\wm0.7232(68)$  \\ 
                       && 8 &&  $\wm0.6709(71)$ &&  $\wm0.659(13)$ &&  $\wm0.659(10)$ &&  $\wm0.687(12)$ &&  $\wm0.6791(98)$ &&  $\wm0.6818(88)$  \\ 
                       && 9 &&  $\wm0.6519(90)$ &&  $\wm0.637(15)$ &&  $\wm0.638(14)$ &&  $\wm0.666(17)$ &&  $\wm0.659(14)$ &&  $\wm0.664(11)$  \\ 
                       && 10 &&  $\wm0.6289(80)$ &&  $\wm0.626(16)$ &&  $\wm0.627(14)$ &&  $\wm0.647(14)$ &&  $\wm0.639(11)$ &&  $\wm0.643(12)$  \\ 
\hline
$f_2^A$                && 1 &&  $\wm0.032(12)$ &&  $\wm0.008(15)$ &&  $\wm0.012(10)$ &&  $\wm0.016(18)$ &&  $\wm0.040(15)$ &&  $\wm0.007(14)$  \\ 
                       && 2 &&  $\wm0.031(11)$ &&  $\wm0.014(15)$ &&  $\wm0.017(10)$ &&  $\wm0.020(15)$ &&  $\wm0.028(13)$ &&  $\wm0.013(12)$  \\ 
                       && 3 &&  $\wm0.027(11)$ &&  $\wm0.013(15)$ &&  $\wm0.014(10)$ &&  $\wm0.024(15)$ &&  $\wm0.018(12)$ &&  $\wm0.012(11)$  \\ 
                       && 4 &&  $\wm0.016(10)$ &&  $-0.002(14)$ &&  $\wm0.0045(98)$ &&  $\wm0.008(16)$ &&  $\wm0.014(14)$ &&  $-0.001(15)$  \\ 
                       && 5 &&  $\wm0.019(11)$ &&  $\wm0.005(15)$ &&  $\wm0.008(10)$ &&  $\wm0.016(15)$ &&  $\wm0.013(13)$ &&  $\wm0.004(14)$  \\ 
                       && 6 &&  $\wm0.027(12)$ &&  $\wm0.014(18)$ &&  $\wm0.015(13)$ &&  $\wm0.033(15)$ &&  $\wm0.026(13)$ &&  $\wm0.019(13)$  \\ 
                       && 8 &&  $\wm0.029(15)$ &&  $\wm0.013(24)$ &&  $\wm0.014(17)$ &&  $\wm0.034(18)$ &&  $\wm0.028(16)$ &&  $\wm0.022(15)$  \\ 
                       && 9 &&  $\wm0.031(17)$ &&  $\wm0.017(27)$ &&  $\wm0.016(21)$ &&  $\wm0.043(22)$ &&  $\wm0.037(20)$ &&  $\wm0.025(18)$  \\ 
                       && 10 &&  $\wm0.014(16)$ &&  $\wm0.006(20)$ &&  $\wm0.005(13)$ &&  $\wm0.016(20)$ &&  $\wm0.011(17)$ &&  $\wm0.003(15)$  \\ 
\hline
$f_3^A$                && 1 &&  $-0.501(24)$ &&  $-0.513(46)$ &&  $-0.525(43)$ &&  $-0.530(42)$ &&  $-0.457(34)$ &&  $-0.531(30)$  \\ 
                       && 2 &&  $-0.467(16)$ &&  $-0.459(14)$ &&  $-0.478(14)$ &&  $-0.462(29)$ &&  $-0.438(24)$ &&  $-0.467(16)$  \\ 
                       && 3 &&  $-0.438(15)$ &&  $-0.431(13)$ &&  $-0.447(12)$ &&  $-0.433(27)$ &&  $-0.439(23)$ &&  $-0.439(15)$  \\ 
                       && 4 &&  $-0.422(17)$ &&  $-0.437(17)$ &&  $-0.438(15)$ &&  $-0.439(37)$ &&  $-0.421(32)$ &&  $-0.440(26)$  \\ 
                       && 5 &&  $-0.399(16)$ &&  $-0.402(14)$ &&  $-0.410(13)$ &&  $-0.408(28)$ &&  $-0.411(25)$ &&  $-0.410(18)$  \\ 
                       && 6 &&  $-0.381(16)$ &&  $-0.378(14)$ &&  $-0.387(13)$ &&  $-0.384(27)$ &&  $-0.388(24)$ &&  $-0.387(16)$  \\ 
                       && 8 &&  $-0.343(26)$ &&  $-0.342(15)$ &&  $-0.352(14)$ &&  $-0.348(33)$ &&  $-0.348(30)$ &&  $-0.354(17)$  \\ 
                       && 9 &&  $-0.330(31)$ &&  $-0.329(17)$ &&  $-0.337(16)$ &&  $-0.331(29)$ &&  $-0.331(26)$ &&  $-0.339(19)$  \\ 
                       && 10 &&  $-0.299(31)$ &&  $-0.339(35)$ &&  $-0.332(31)$ &&  $-0.348(59)$ &&  $-0.345(52)$ &&  $-0.355(43)$  \\ 
\hline\hline
\end{tabular}
\normalsize
\caption{\label{LbLcWeinberg}$\Lambda_b \to \Lambda_c$ Weinberg form factors.}
\end{table}

\FloatBarrier

\providecommand{\href}[2]{#2}\begingroup\raggedright

\endgroup


\begin{thebibliography}{10}

\bibitem{Agashe:2014kda}
{\bfseries Particle Data Group} Collaboration, K.~Olive {\em et~al.}, ``{Review
  of Particle Physics},''
\href{http://dx.doi.org/10.1088/1674-1137/38/9/090001}{Chin.Phys. {\bfseries
  C38} (2014) 090001}.

\bibitem{Bailey:2008wp}
J.~A. Bailey, C.~Bernard, C.~E. DeTar, M.~Di~Pierro, A.~El-Khadra, {\em
  et~al.}, ``{The $B \to \pi \ell \nu$ semileptonic form factor from
  three-flavor lattice QCD: A Model-independent determination of $|V_{ub}|$},''
  \href{http://dx.doi.org/10.1103/PhysRevD.79.054507}{Phys.Rev. {\bfseries D79}
  (2009) 054507},
\href{http://arxiv.org/abs/0811.3640}{{\ttfamily arXiv:0811.3640 [hep-lat]}}.

\bibitem{Bernard:2008dn}
C.~Bernard, C.~E. DeTar, M.~Di~Pierro, A.~El-Khadra, R.~Evans, {\em et~al.},
  ``{The $\bar{B} \to D^{*} \ell \bar{\nu}$ form factor at zero recoil from
  three-flavor lattice QCD: A Model independent determination of $|V_{cb}|$},''
  \href{http://dx.doi.org/10.1103/PhysRevD.79.014506}{Phys.Rev. {\bfseries D79}
  (2009) 014506},
\href{http://arxiv.org/abs/0808.2519}{{\ttfamily arXiv:0808.2519 [hep-lat]}}.

\bibitem{Kowalewski:2010zz}
{\bfseries BaBar} Collaboration, R.~Kowalewski, ``{Status of $|V_{ub}|$ and
  $|V_{cb}|$ determinations},''
PoS {\bfseries FPCP2010} (2010) 028.

\bibitem{Mannel:2010zz}
T.~Mannel, ``{Determination of $|V_{ub}|$ and $|V_{cb}|$: A Theory
  Perspective},''
PoS {\bfseries FPCP2010} (2010) 029.

\bibitem{Ricciardi:2014aya}
G.~Ricciardi, ``{Status of $|V_{cb}|$ and $|V_{ub}|$ CKM matrix elements},''
\href{http://arxiv.org/abs/1412.4288}{{\ttfamily arXiv:1412.4288 [hep-ph]}}.

\bibitem{Chen:2008se}
C.-H. Chen and S.-h. Nam, ``{Left-right mixing on leptonic and semileptonic $b
  \to u$ decays},''
  \href{http://dx.doi.org/10.1016/j.physletb.2008.07.095}{Phys.Lett. {\bfseries
  B666} (2008) 462--466},
\href{http://arxiv.org/abs/0807.0896}{{\ttfamily arXiv:0807.0896 [hep-ph]}}.

\bibitem{Crivellin:2009sd}
A.~Crivellin, ``{Effects of right-handed charged currents on the determinations
  of $|V_{ub}|$ and $|V_{cb}|$},''
  \href{http://dx.doi.org/10.1103/PhysRevD.81.031301}{Phys.Rev. {\bfseries D81}
  (2010) 031301},
\href{http://arxiv.org/abs/0907.2461}{{\ttfamily arXiv:0907.2461 [hep-ph]}}.

\bibitem{Buras:2010pz}
A.~J. Buras, K.~Gemmler, and G.~Isidori, ``{Quark flavour mixing with
  right-handed currents: an effective theory approach},''
  \href{http://dx.doi.org/10.1016/j.nuclphysb.2010.09.021}{Nucl.Phys.
  {\bfseries B843} (2011) 107--142},
\href{http://arxiv.org/abs/1007.1993}{{\ttfamily arXiv:1007.1993 [hep-ph]}}.

\bibitem{Crivellin:2014zpa}
A.~Crivellin and S.~Pokorski, ``{Can the differences in the determinations of
  $V_{ub}$ and $V_{cb}$ be explained by New Physics?},''
  \href{http://dx.doi.org/10.1103/PhysRevLett.114.011802}{Phys.Rev.Lett.
  {\bfseries 114} (2015) 011802},
\href{http://arxiv.org/abs/1407.1320}{{\ttfamily arXiv:1407.1320 [hep-ph]}}.

\bibitem{Lattice:2015tia}
{\bfseries Fermilab Lattice, MILC} Collaboration, J.~A. Bailey {\em et~al.},
  ``{$|V_{ub}|$ from $B\to\pi\ell\nu$ decays and (2+1)-flavor lattice QCD},''
\href{http://arxiv.org/abs/1503.07839}{{\ttfamily arXiv:1503.07839 [hep-lat]}}.

\bibitem{Flynn:2015mha}
J.~Flynn, T.~Izubuchi, T.~Kawanai, C.~Lehner, A.~Soni, {\em et~al.}, ``{The
  $B\to \pi \ell \bar{\nu}$ and $B_s\to K \ell \bar{\nu}$ form factors and
  $|V_{ub}|$ from 2+1-flavor lattice QCD with domain-wall light quarks and
  relativistic heavy quarks},''
\href{http://arxiv.org/abs/1501.05373}{{\ttfamily arXiv:1501.05373 [hep-lat]}}.

\bibitem{Bailey:2014tva}
J.~A. Bailey, A.~Bazavov, C.~Bernard, C.~Bouchard, C.~DeTar, {\em et~al.},
  ``{Update of $|V_{cb}|$ from the $\bar{B}\to D^*\ell\bar{\nu}$ form factor at
  zero recoil with three-flavor lattice QCD},''
  \href{http://dx.doi.org/10.1103/PhysRevD.89.114504}{Phys.Rev. {\bfseries D89}
  no.~11, (2014) 114504},
\href{http://arxiv.org/abs/1403.0635}{{\ttfamily arXiv:1403.0635 [hep-lat]}}.

\bibitem{Lees:2012xj}
{\bfseries BaBar} Collaboration, J.~Lees {\em et~al.}, ``{Evidence for an
  excess of $\bar{B} \to D^{(*)} \tau^-\bar{\nu}_\tau$ decays},''
  \href{http://dx.doi.org/10.1103/PhysRevLett.109.101802}{Phys.Rev.Lett.
  {\bfseries 109} (2012) 101802},
\href{http://arxiv.org/abs/1205.5442}{{\ttfamily arXiv:1205.5442 [hep-ex]}}.

\bibitem{Adinolfi:2012qfa}
{\bfseries LHCb RICH Group} Collaboration, M.~Adinolfi {\em et~al.},
  ``{Performance of the LHCb RICH detector at the LHC},''
  \href{http://dx.doi.org/10.1140/epjc/s10052-013-2431-9}{Eur.Phys.J.
  {\bfseries C73} (2013) 2431},
\href{http://arxiv.org/abs/1211.6759}{{\ttfamily arXiv:1211.6759
  [physics.ins-det]}}.

\bibitem{Aaij:2014jyk}
{\bfseries LHCb} Collaboration, R.~Aaij {\em et~al.}, ``{Study of the kinematic
  dependences of $\Lambda_{b}^{0}$ production in pp collisions and a
  measurement of the $\Lambda_{b}^{0} \to \Lambda_{c}^{+}$ $\pi^{-}$ branching
  fraction},'' \href{http://dx.doi.org/10.1007/JHEP08(2014)143}{JHEP {\bfseries
  1408} (2014) 143},
\href{http://arxiv.org/abs/1405.6842}{{\ttfamily arXiv:1405.6842 [hep-ex]}}.

\bibitem{Cardarelli:1997sx}
F.~Cardarelli and S.~Simula, ``{Isgur-Wise form-factors of heavy baryons within
  a light front constituent quark model},''
  \href{http://dx.doi.org/10.1016/S0370-2693(97)01581-5}{Phys.Lett. {\bfseries
  B421} (1998) 295--302},
\href{http://arxiv.org/abs/hep-ph/9711207}{{\ttfamily arXiv:hep-ph/9711207
  [hep-ph]}}.

\bibitem{Dosch:1997zx}
H.~G. Dosch, E.~Ferreira, M.~Nielsen, and R.~Rosenfeld, ``{Evidence from QCD
  sum rules for large violation of heavy quark symmetry in $\Lambda_b$
  semileptonic decay},''
  \href{http://dx.doi.org/10.1016/S0370-2693(98)00566-8}{Phys.Lett. {\bfseries
  B431} (1998) 173--178},
\href{http://arxiv.org/abs/hep-ph/9712350}{{\ttfamily arXiv:hep-ph/9712350
  [hep-ph]}}.

\bibitem{Huang:1998rq}
C.-S. Huang, C.-F. Qiao, and H.-G. Yan, ``{Decay $\Lambda_b \to p \ell
  \bar{\nu}$ in QCD sum rules},''
  \href{http://dx.doi.org/10.1016/S0370-2693(98)00909-5}{Phys.Lett. {\bfseries
  B437} (1998) 403--407},
\href{http://arxiv.org/abs/hep-ph/9805452}{{\ttfamily arXiv:hep-ph/9805452
  [hep-ph]}}.

\bibitem{Carvalho:1999ia}
R.~Marques~de Carvalho, F.~Navarra, M.~Nielsen, E.~Ferreira, and H.~G. Dosch,
  ``{Form-factors and decay rates for heavy $\Lambda$ semileptonic decays from
  QCD sum rules},''
  \href{http://dx.doi.org/10.1103/PhysRevD.60.034009}{Phys.Rev. {\bfseries D60}
  (1999) 034009},
\href{http://arxiv.org/abs/hep-ph/9903326}{{\ttfamily arXiv:hep-ph/9903326
  [hep-ph]}}.

\bibitem{Huang:2004vf}
M.-q. Huang and D.-W. Wang, ``{Light cone QCD sum rules for the semileptonic
  decay $\Lambda_b \to p \ell \bar{\nu}$},''
  \href{http://dx.doi.org/10.1103/PhysRevD.69.094003}{Phys.Rev. {\bfseries D69}
  (2004) 094003},
\href{http://arxiv.org/abs/hep-ph/0401094}{{\ttfamily arXiv:hep-ph/0401094
  [hep-ph]}}.

\bibitem{Pervin:2005ve}
M.~Pervin, W.~Roberts, and S.~Capstick, ``{Semileptonic decays of heavy
  $\Lambda$ baryons in a quark model},''
  \href{http://dx.doi.org/10.1103/PhysRevC.72.035201}{Phys.Rev. {\bfseries C72}
  (2005) 035201},
\href{http://arxiv.org/abs/nucl-th/0503030}{{\ttfamily arXiv:nucl-th/0503030
  [nucl-th]}}.

\bibitem{Ke:2007tg}
H.-W. Ke, X.-Q. Li, and Z.-T. Wei, ``{Diquarks and $\Lambda_b \to \Lambda_c$
  weak decays},'' \href{http://dx.doi.org/10.1103/PhysRevD.77.014020}{Phys.Rev.
  {\bfseries D77} (2008) 014020},
\href{http://arxiv.org/abs/0710.1927}{{\ttfamily arXiv:0710.1927 [hep-ph]}}.

\bibitem{Wang:2009hra}
Y.-M. Wang, Y.-L. Shen, and C.-D. Lu, ``{$\Lambda_b \to p, \Lambda$ transition
  form factors from QCD light-cone sum rules},''
  \href{http://dx.doi.org/10.1103/PhysRevD.80.074012}{Phys.Rev. {\bfseries D80}
  (2009) 074012},
\href{http://arxiv.org/abs/0907.4008}{{\ttfamily arXiv:0907.4008 [hep-ph]}}.

\bibitem{Azizi:2009wn}
K.~Azizi, M.~Bayar, Y.~Sarac, and H.~Sundu, ``{Semileptonic $\Lambda_{b,c}$ to
  Nucleon Transitions in Full QCD at Light Cone},''
  \href{http://dx.doi.org/10.1103/PhysRevD.80.096007}{Phys.Rev. {\bfseries D80}
  (2009) 096007},
\href{http://arxiv.org/abs/0908.1758}{{\ttfamily arXiv:0908.1758 [hep-ph]}}.

\bibitem{Khodjamirian:2011jp}
A.~Khodjamirian, C.~Klein, T.~Mannel, and Y.-M. Wang, ``{Form Factors and
  Strong Couplings of Heavy Baryons from QCD Light-Cone Sum Rules},''
  \href{http://dx.doi.org/10.1007/JHEP09(2011)106}{JHEP {\bfseries 1109} (2011)
  106},
\href{http://arxiv.org/abs/1108.2971}{{\ttfamily arXiv:1108.2971 [hep-ph]}}.

\bibitem{Gutsche:2014zna}
T.~Gutsche, M.~A. Ivanov, J.~G. K$\ddot{\mathrm{o}}$rner, V.~E. Lyubovitskij,
  and P.~Santorelli, ``{Heavy-to-light semileptonic decays of $\Lambda_b$ and
  $\Lambda_c$ baryons in the covariant confined quark model},''
\href{http://arxiv.org/abs/1410.6043}{{\ttfamily arXiv:1410.6043 [hep-ph]}}.

\bibitem{Gutsche:2015mxa}
T.~Gutsche, M.~A. Ivanov, J.~G. K$\ddot{\mathrm{o}}$rner, V.~E. Lyubovitskij,
  P.~Santorelli, {\em et~al.}, ``{The semileptonic decay $\Lambda_b \to
  \Lambda_c \tau^- \bar{\nu}_\tau$ in the covariant confined quark model},''
\href{http://arxiv.org/abs/1502.04864}{{\ttfamily arXiv:1502.04864 [hep-ph]}}.

\bibitem{Detmold:2013nia}
W.~Detmold, C.-J.~D. Lin, S.~Meinel, and M.~Wingate, ``{$\Lambda_b \to p l^-
  \bar{\nu}_\ell$ form factors from lattice QCD with static b quarks},''
  \href{http://dx.doi.org/10.1103/PhysRevD.88.014512}{Phys.Rev. {\bfseries D88}
  (2013) 014512},
\href{http://arxiv.org/abs/1306.0446}{{\ttfamily arXiv:1306.0446 [hep-lat]}}.

\bibitem{Mannel:1990vg}
T.~Mannel, W.~Roberts, and Z.~Ryzak, ``{Baryons in the heavy quark effective
  theory},''
\href{http://dx.doi.org/10.1016/0550-3213(91)90301-D}{Nucl.Phys. {\bfseries
  B355} (1991) 38--53}.

\bibitem{Hussain:1990uu}
F.~Hussain, J.~K$\ddot{\mathrm{o}}$rner, M.~Kramer, and G.~Thompson, ``{On
  heavy baryon decay form-factors},''
\href{http://dx.doi.org/10.1007/BF01475799}{Z.Phys. {\bfseries C51} (1991)
  321--328}.

\bibitem{Hussain:1992rb}
F.~Hussain, D.-S. Liu, M.~Kramer, J.~K$\ddot{\mathrm{o}}$rner, and S.~Tawfiq,
  ``{General analysis of weak decay form-factors in heavy to heavy and heavy to
  light baryon transitions},''
\href{http://dx.doi.org/10.1016/0550-3213(92)90286-K}{Nucl.Phys. {\bfseries
  B370} (1992) 259--277}.

\bibitem{Bowler:1997ej}
{\bfseries UKQCD} Collaboration, K.~Bowler {\em et~al.}, ``{First lattice study
  of semileptonic decays of $\Lambda_b$ and $\Xi_b$ baryons},''
  \href{http://dx.doi.org/10.1103/PhysRevD.57.6948}{Phys.Rev. {\bfseries D57}
  (1998) 6948--6974},
\href{http://arxiv.org/abs/hep-lat/9709028}{{\ttfamily arXiv:hep-lat/9709028
  [hep-lat]}}.

\bibitem{Gottlieb:2003yb}
S.~A. Gottlieb and S.~Tamhankar, ``{A Lattice study of $\Lambda_b$ semileptonic
  decay},''
  \href{http://dx.doi.org/10.1016/S0920-5632(03)01612-8}{Nucl.Phys.Proc.Suppl.
  {\bfseries 119} (2003) 644--646},
\href{http://arxiv.org/abs/hep-lat/0301022}{{\ttfamily arXiv:hep-lat/0301022
  [hep-lat]}}.

\bibitem{Meinel:2014wua}
S.~Meinel, ``{Flavor physics with $\Lambda_b$ baryons},'' PoS {\bfseries
  LATTICE2013} (2014) 024,
\href{http://arxiv.org/abs/1401.2685}{{\ttfamily arXiv:1401.2685 [hep-lat]}}.

\bibitem{Aoki:2010dy}
{\bfseries RBC, UKQCD} Collaboration, Y.~Aoki {\em et~al.}, ``{Continuum Limit
  Physics from 2+1 Flavor Domain Wall QCD},''
  \href{http://dx.doi.org/10.1103/PhysRevD.83.074508}{Phys.Rev. {\bfseries D83}
  (2011) 074508},
\href{http://arxiv.org/abs/1011.0892}{{\ttfamily arXiv:1011.0892 [hep-lat]}}.

\bibitem{Bourrely:2008za}
C.~Bourrely, I.~Caprini, and L.~Lellouch, ``{Model-independent description of
  $B \to \pi \ell \nu$ decays and a determination of $|V_{ub}|$},''
  \href{http://dx.doi.org/10.1103/PhysRevD.82.099902,
  10.1103/PhysRevD.79.013008}{Phys.Rev. {\bfseries D79} (2009) 013008},
\href{http://arxiv.org/abs/0807.2722}{{\ttfamily arXiv:0807.2722 [hep-ph]}}.

\bibitem{Feldmann:2011xf}
T.~Feldmann and M.~W. Yip, ``{Form Factors for $\Lambda_b \to \Lambda$
  Transitions in SCET},'' \href{http://dx.doi.org/10.1103/PhysRevD.85.014035,
  10.1103/PhysRevD.86.079901}{Phys.Rev. {\bfseries D85} (2012) 014035},
\href{http://arxiv.org/abs/1111.1844}{{\ttfamily arXiv:1111.1844 [hep-ph]}}.

\bibitem{Weinberg:1958ut}
S.~Weinberg, ``{Charge symmetry of weak interactions},''
\href{http://dx.doi.org/10.1103/PhysRev.112.1375}{Phys.Rev. {\bfseries 112}
  (1958) 1375--1379}.

\bibitem{Sasaki:2008ha}
S.~Sasaki and T.~Yamazaki, ``{Lattice study of flavor SU(3) breaking in hyperon
  beta decay},'' \href{http://dx.doi.org/10.1103/PhysRevD.79.074508}{Phys.Rev.
  {\bfseries D79} (2009) 074508},
\href{http://arxiv.org/abs/0811.1406}{{\ttfamily arXiv:0811.1406 [hep-ph]}}.

\bibitem{Iwasaki:1983ck}
Y.~Iwasaki,
``{Renormalization group analysis of lattice theories and improved lattice
  action. 2. Four-dimensional nonabelian $SU(N)$ gauge model},''.

\bibitem{Iwasaki:1984cj}
Y.~Iwasaki and T.~Yoshie, ``{Renormalization group improved action for $SU(3)$
  lattice gauge theory and the string tension},''
\href{http://dx.doi.org/10.1016/0370-2693(84)91500-4}{Phys.Lett. {\bfseries
  B143} (1984) 449}.

\bibitem{Kaplan:1992bt}
D.~B. Kaplan, ``{A Method for simulating chiral fermions on the lattice},''
  \href{http://dx.doi.org/10.1016/0370-2693(92)91112-M}{Phys.Lett. {\bfseries
  B288} (1992) 342--347},
\href{http://arxiv.org/abs/hep-lat/9206013}{{\ttfamily arXiv:hep-lat/9206013}}.

\bibitem{Furman:1994ky}
V.~Furman and Y.~Shamir, ``{Axial symmetries in lattice QCD with Kaplan
  fermions},'' \href{http://dx.doi.org/10.1016/0550-3213(95)00031-M}{Nucl.Phys.
  {\bfseries B439} (1995) 54--78},
\href{http://arxiv.org/abs/hep-lat/9405004}{{\ttfamily arXiv:hep-lat/9405004}}.

\bibitem{Shamir:1993zy}
Y.~Shamir, ``{Chiral fermions from lattice boundaries},''
  \href{http://dx.doi.org/10.1016/0550-3213(93)90162-I}{Nucl.Phys. {\bfseries
  B406} (1993) 90--106},
\href{http://arxiv.org/abs/hep-lat/9303005}{{\ttfamily arXiv:hep-lat/9303005}}.

\bibitem{Eichten:1989kb}
E.~Eichten and B.~R. Hill, ``{Renormalization of Heavy - Light Bilinears and
  $f_B$ for Wilson Fermions},''
\href{http://dx.doi.org/10.1016/0370-2693(90)90432-6}{Phys.Lett. {\bfseries
  B240} (1990) 193}.

\bibitem{ElKhadra:1996mp}
A.~X. El-Khadra, A.~S. Kronfeld, and P.~B. Mackenzie, ``{Massive fermions in
  lattice gauge theory},''
  \href{http://dx.doi.org/10.1103/PhysRevD.55.3933}{Phys.Rev. {\bfseries D55}
  (1997) 3933--3957},
\href{http://arxiv.org/abs/hep-lat/9604004}{{\ttfamily arXiv:hep-lat/9604004
  [hep-lat]}}.

\bibitem{Aoki:2001ra}
S.~Aoki, Y.~Kuramashi, and S.-i. Tominaga, ``{Relativistic heavy quarks on the
  lattice},'' \href{http://dx.doi.org/10.1143/PTP.109.383}{Prog.Theor.Phys.
  {\bfseries 109} (2003) 383--413},
\href{http://arxiv.org/abs/hep-lat/0107009}{{\ttfamily arXiv:hep-lat/0107009
  [hep-lat]}}.

\bibitem{Aoki:2003dg}
S.~Aoki, Y.~Kayaba, and Y.~Kuramashi, ``{A Perturbative determination of mass
  dependent $O(a)$ improvement coefficients in a relativistic heavy quark
  action},''
  \href{http://dx.doi.org/10.1016/j.nuclphysb.2004.07.017}{Nucl.Phys.
  {\bfseries B697} (2004) 271--301},
\href{http://arxiv.org/abs/hep-lat/0309161}{{\ttfamily arXiv:hep-lat/0309161
  [hep-lat]}}.

\bibitem{Lin:2006ur}
H.-W. Lin and N.~Christ, ``{Non-perturbatively Determined Relativistic Heavy
  Quark Action},''
  \href{http://dx.doi.org/10.1103/PhysRevD.76.074506}{Phys.Rev. {\bfseries D76}
  (2007) 074506},
\href{http://arxiv.org/abs/hep-lat/0608005}{{\ttfamily arXiv:hep-lat/0608005
  [hep-lat]}}.

\bibitem{Aoki:2012xaa}
{\bfseries RBC, UKQCD} Collaboration, Y.~Aoki {\em et~al.}, ``{Nonperturbative
  tuning of an improved relativistic heavy-quark action with application to
  bottom spectroscopy},''
  \href{http://dx.doi.org/10.1103/PhysRevD.86.116003}{Phys.Rev. {\bfseries D86}
  (2012) 116003},
\href{http://arxiv.org/abs/1206.2554}{{\ttfamily arXiv:1206.2554 [hep-lat]}}.

\bibitem{Brown:2014ena}
Z.~S. Brown, W.~Detmold, S.~Meinel, and K.~Orginos, ``{Charmed bottom baryon
  spectroscopy from lattice QCD},''
  \href{http://dx.doi.org/10.1103/PhysRevD.90.094507}{Phys.Rev. {\bfseries D90}
  (2014) 094507},
\href{http://arxiv.org/abs/1409.0497}{{\ttfamily arXiv:1409.0497 [hep-lat]}}.

\bibitem{Detmold:2012vy}
W.~Detmold, C.-J.~D. Lin, S.~Meinel, and M.~Wingate, ``{$\Lambda_b \to \Lambda
  \ell^+ \ell^-$ form factors and differential branching fraction from lattice
  QCD},'' \href{http://dx.doi.org/10.1103/PhysRevD.87.074502}{Phys. Rev. D 87,
  {\bfseries 074502} (2013) },
\href{http://arxiv.org/abs/1212.4827}{{\ttfamily arXiv:1212.4827 [hep-lat]}}.

\bibitem{Meinel:2010pv}
S.~Meinel, ``{Bottomonium spectrum at order $v^6$ from domain-wall lattice QCD:
  Precise results for hyperfine splittings},''
  \href{http://dx.doi.org/10.1103/PhysRevD.82.114502}{Phys.Rev. {\bfseries D82}
  (2010) 114502},
\href{http://arxiv.org/abs/1007.3966}{{\ttfamily arXiv:1007.3966 [hep-lat]}}.

\bibitem{Hashimoto:1999yp}
S.~Hashimoto, A.~X. El-Khadra, A.~S. Kronfeld, P.~B. Mackenzie, S.~M. Ryan,
  {\em et~al.}, ``{Lattice QCD calculation of $\bar{B} \to D \ell \bar{\nu}$
  decay form-factors at zero recoil},''
  \href{http://dx.doi.org/10.1103/PhysRevD.61.014502}{Phys.Rev. {\bfseries D61}
  (1999) 014502},
\href{http://arxiv.org/abs/hep-ph/9906376}{{\ttfamily arXiv:hep-ph/9906376
  [hep-ph]}}.

\bibitem{ElKhadra:2001rv}
A.~X. El-Khadra, A.~S. Kronfeld, P.~B. Mackenzie, S.~M. Ryan, and J.~N. Simone,
  ``{The Semileptonic decays $B \to \pi \ell \nu$ and $D \to \pi \ell \nu$ from
  lattice QCD},'' \href{http://dx.doi.org/10.1103/PhysRevD.64.014502}{Phys.Rev.
  {\bfseries D64} (2001) 014502},
\href{http://arxiv.org/abs/hep-ph/0101023}{{\ttfamily arXiv:hep-ph/0101023
  [hep-ph]}}.

\bibitem{Lehner:2012bt}
C.~Lehner, ``{Automated lattice perturbation theory and relativistic heavy
  quarks in the Columbia formulation},'' PoS {\bfseries LATTICE2012} (2012)
  126,
\href{http://arxiv.org/abs/1211.4013}{{\ttfamily arXiv:1211.4013 [hep-lat]}}.

\bibitem{PhySyHCAl}
C.~Lehner, ``{PhySyHCAl}.''
\newblock \url{http://www.lhnr.de/physyhcal/}.

\bibitem{Lattice:2015rga}
{\bfseries Fermilab Lattice, MILC} Collaboration, J.~A. Bailey {\em et~al.},
  ``{The $B \to D \ell \nu$ form factors at nonzero recoil and $|V_{cb}|$ from
  $2+1$-flavor lattice QCD},''
\href{http://arxiv.org/abs/1503.07237}{{\ttfamily arXiv:1503.07237 [hep-lat]}}.

\bibitem{Christ:2014uea}
N.~H. Christ, J.~M. Flynn, T.~Izubuchi, T.~Kawanai, C.~Lehner, {\em et~al.},
  ``{B-meson decay constants from 2+1-flavor lattice QCD with domain-wall light
  quarks and relativistic heavy quarks},''
\href{http://arxiv.org/abs/1404.4670}{{\ttfamily arXiv:1404.4670 [hep-lat]}}.

\bibitem{Detmold:2012ge}
W.~Detmold, C.~D. Lin, and S.~Meinel, ``{Calculation of the heavy-hadron axial
  couplings $g_1$, $g_2$, and $g_3$ using lattice QCD},''
  \href{http://dx.doi.org/10.1103/PhysRevD.85.114508}{Phys.Rev. {\bfseries D85}
  (2012) 114508},
\href{http://arxiv.org/abs/1203.3378}{{\ttfamily arXiv:1203.3378 [hep-lat]}}.

\bibitem{Morningstar:2003gk}
C.~Morningstar and M.~J. Peardon, ``{Analytic smearing of $SU(3)$ link
  variables in lattice QCD},''
  \href{http://dx.doi.org/10.1103/PhysRevD.69.054501}{Phys.Rev. {\bfseries D69}
  (2004) 054501},
\href{http://arxiv.org/abs/hep-lat/0311018}{{\ttfamily arXiv:hep-lat/0311018
  [hep-lat]}}.

\bibitem{Colangelo:2010et}
G.~Colangelo, S.~Durr, A.~Juttner, L.~Lellouch, H.~Leutwyler, {\em et~al.},
  ``{Review of lattice results concerning low energy particle physics},''
  \href{http://dx.doi.org/10.1140/epjc/s10052-011-1695-1}{Eur.Phys.J.
  {\bfseries C71} (2011) 1695},
\href{http://arxiv.org/abs/1011.4408}{{\ttfamily arXiv:1011.4408 [hep-lat]}}.

\bibitem{Beane:2004tw}
S.~R. Beane, ``{Nucleon masses and magnetic moments in a finite volume},''
  \href{http://dx.doi.org/10.1103/PhysRevD.70.034507}{Phys.Rev. {\bfseries D70}
  (2004) 034507},
\href{http://arxiv.org/abs/hep-lat/0403015}{{\ttfamily arXiv:hep-lat/0403015
  [hep-lat]}}.

\bibitem{Beane:2004rf}
S.~R. Beane and M.~J. Savage, ``{Baryon axial charge in a finite volume},''
  \href{http://dx.doi.org/10.1103/PhysRevD.70.074029}{Phys.Rev. {\bfseries D70}
  (2004) 074029},
\href{http://arxiv.org/abs/hep-ph/0404131}{{\ttfamily arXiv:hep-ph/0404131
  [hep-ph]}}.

\bibitem{Detmold:2011rb}
W.~Detmold, C.-J.~D. Lin, and S.~Meinel, ``{Axial couplings in heavy hadron
  chiral perturbation theory at the next-to-leading order},''
  \href{http://dx.doi.org/10.1103/PhysRevD.84.094502}{Phys.Rev. {\bfseries D84}
  (2011) 094502},
\href{http://arxiv.org/abs/1108.5594}{{\ttfamily arXiv:1108.5594 [hep-lat]}}.

\bibitem{Kronfeld:2000ck}
A.~S. Kronfeld, ``{Application of heavy quark effective theory to lattice QCD.
  1. Power corrections},''
  \href{http://dx.doi.org/10.1103/PhysRevD.62.014505}{Phys.Rev. {\bfseries D62}
  (2000) 014505},
\href{http://arxiv.org/abs/hep-lat/0002008}{{\ttfamily arXiv:hep-lat/0002008
  [hep-lat]}}.

\bibitem{Harada:2001fi}
J.~Harada, S.~Hashimoto, K.-I. Ishikawa, A.~S. Kronfeld, T.~Onogi, {\em
  et~al.}, ``{Application of heavy quark effective theory to lattice QCD. 2.
  Radiative corrections to heavy light currents},''
  \href{http://dx.doi.org/10.1103/PhysRevD.71.019903,
  10.1103/PhysRevD.65.094513}{Phys.Rev. {\bfseries D65} (2002) 094513},
\href{http://arxiv.org/abs/hep-lat/0112044}{{\ttfamily arXiv:hep-lat/0112044
  [hep-lat]}}.

\bibitem{Harada:2001fj}
J.~Harada, S.~Hashimoto, A.~S. Kronfeld, and T.~Onogi, ``{Application of heavy
  quark effective theory to lattice QCD. 3. Radiative corrections to
  heavy-heavy currents},''
  \href{http://dx.doi.org/10.1103/PhysRevD.65.094514}{Phys.Rev. {\bfseries D65}
  (2002) 094514},
\href{http://arxiv.org/abs/hep-lat/0112045}{{\ttfamily arXiv:hep-lat/0112045
  [hep-lat]}}.

\bibitem{Shivashankara:2015cta}
S.~Shivashankara, W.~Wu, and A.~Datta, ``{$\Lambda_b \to \Lambda_c \tau
  \bar{\nu}_{\tau}$ Decay in the Standard Model and with New Physics},''
\href{http://arxiv.org/abs/1502.07230}{{\ttfamily arXiv:1502.07230 [hep-ph]}}.

\bibitem{Edwards:2004sx}
{\bfseries SciDAC, LHPC, UKQCD} Collaboration, R.~G. Edwards and B.~Joo, ``{The
  Chroma software system for lattice QCD},''
  \href{http://dx.doi.org/10.1016/j.nuclphysbps.2004.11.254}{Nucl.Phys.Proc.Suppl.
  {\bfseries 140} (2005) 832},
\href{http://arxiv.org/abs/hep-lat/0409003}{{\ttfamily arXiv:hep-lat/0409003
  [hep-lat]}}.

\bibitem{Aaij:2015bfa}
{\bfseries LHCb} Collaboration, R.~Aaij {\em et~al.}, ``{Determination of the
  quark coupling strength $|V_{ub}|$ using baryonic decays},''
\href{http://arxiv.org/abs/1504.01568}{{\ttfamily arXiv:1504.01568 [hep-ex]}}.

\end{thebibliography}
\end{document}